\pdfoutput=1

\documentclass[11pt,twoside,a4paper,cmspaper,final,collab]{cms-tdr}
\def\svnVersion{491456P}\def\cmsCernNoTag{CERN-EP-2018-065}\def\cmsCernDate{\today}\def\cmsMessage{Published in the Journal of High Energy Physics as \href{http://dx.doi.org/10.1007/JHEP03(2019)026}{\doi{10.1007/JHEP03(2019)026.}}}

\begin{document}\cmsNoteHeader{HIG-17-026}

\hyphenation{had-ron-i-za-tion}
\hyphenation{cal-or-i-me-ter}
\hyphenation{de-vices}
\RCS$HeadURL: svn+ssh://svn.cern.ch/reps/tdr2/papers/HIG-17-026/trunk/HIG-17-026.tex $
\RCS$Id: HIG-17-026.tex 489037 2019-02-14 10:09:15Z mschrode $

\providecommand{\cmsLeft}{left\xspace}
\providecommand{\cmsRight}{right\xspace}
\newlength\cmsTabSkip\setlength{\cmsTabSkip}{1ex}
\providecommand{\cmsTable}[1]{\resizebox{\textwidth}{!}{#1}}
\newcommand{\lumivalue}{\ensuremath{35.9\fbinv}\xspace}

\newcommand{\ljSixThreeIncl}{\ensuremath{{(\geq6\,\text{jets}, \geq 3\,\text{\cPqb\ tags})}}\xspace}
\newcommand{\ljFiveThreeIncl}{\ensuremath{{(5\,\text{jets}, \geq 3\,\text{\cPqb\ tags})}}\xspace}
\newcommand{\ljFourThreeIncl}{\ensuremath{{(4\,\text{jets}, \geq 3\,\text{\cPqb\ tags})}}\xspace}
\newcommand{\dlFourThree}{\ensuremath{{(\geq4\,\text{jets}, 3\,\text{\cPqb\ tags})}}\xspace}
\newcommand{\dlFourFour}{\ensuremath{{(\geq4\,\text{jets}, \geq4\,\text{\cPqb\ tags})}}\xspace}

\newcommand{\btagger}{CSVv2\xspace}

\newcommand{\protonproton}{\ensuremath{\Pp\Pp}\xspace}

\newcommand{\ttH}{\ensuremath{\ttbar\PH}\xspace}

\newcommand{\ttbb}{\ensuremath{\ttbar\text{+}\bbbar}\xspace}
\newcommand{\ttb}{\ensuremath{\ttbar\text{+}\cPqb}\xspace}
\newcommand{\tttwob}{\ensuremath{\ttbar\text{+}2\cPqb}\xspace}
\newcommand{\ttcc}{\ensuremath{\ttbar\text{+}\ccbar}\xspace}
\newcommand{\ttlf}{\ensuremath{\ttbar\text{+}\text{lf}}\xspace}
\newcommand{\tthf}{\ensuremath{\ttbar\text{+}\text{hf}}\xspace}
\newcommand{\ttjets}{\ensuremath{\ttbar\text{+}\text{jets}}\xspace}
\newcommand{\singlet}{\ensuremath{\text{Single \cPqt}}}
\newcommand{\ttW}{\ensuremath{\ttbar\text{+}\PW}\xspace}
\newcommand{\ttZ}{\ensuremath{\ttbar\text{+}\PZ}\xspace}
\newcommand{\ttV}{\ensuremath{\ttbar\text{+}\text{V}}\xspace}
\newcommand{\Wjets}{\ensuremath{\PW\text{+jets}}\xspace}
\newcommand{\Zjets}{\ensuremath{\PZ\text{+jets}}\xspace}
\newcommand{\Vjets}{\ensuremath{\text{V+jets}}\xspace}
\newcommand{\diboson}{\ensuremath{\text{Diboson}}\xspace}

\newcommand{\Hbb}{\ensuremath{\PH\to\bbbar}\xspace}

\newcommand{\mumu}{\ensuremath{\PGmp\PGmm}\xspace}
\newcommand{\ee}{\ensuremath{\Pep\Pem}\xspace}
\newcommand{\mue}{\ensuremath{\PGmpm\Pemp}\xspace}

\newcommand{\varHT}{\ensuremath{H_{\text{T}}^{\text{j}}}\xspace}
\newcommand{\varHTtagged}{\ensuremath{H_{\text{T}}^{\text{b}}}\xspace}
\newcommand{\varfirstjetpt}{\ensuremath{\pt(\text{jet 1})}\xspace}
\newcommand{\varsecondjetpt}{\ensuremath{\pt(\text{jet 2})}\xspace}
\newcommand{\varthirdjetpt}{\ensuremath{\pt(\text{jet 3})}\xspace}
\newcommand{\varfourthjetpt}{\ensuremath{\pt(\text{jet 4})}\xspace}
\newcommand{\varEvtDetaTaggedJetsAverageAlt}{\ensuremath{\Delta\eta_{\text{b},\text{b}}^{\text{avg}}}\xspace}
\newcommand{\varavgdrtaggedjetsAlt}{\ensuremath{\Delta R_{\text{b},\text{b}}^{\text{avg}}}\xspace}
\newcommand{\varavgdrtaggeduntaggedjetsAlt}{\ensuremath{\Delta R_{\text{j},\text{b}}^{\text{avg}}}\xspace}
\newcommand{\varmindrjets}{\ensuremath{\Delta R_{\text{j},\text{j}}^{\text{min}}}\xspace}
\newcommand{\varmindrtaggedjets}{\ensuremath{\Delta R_{\text{b},\text{b}}^{\text{min}}}\xspace}
\newcommand{\varmaxdrtaggedjets}{\ensuremath{\Delta R_{\text{b},\text{b}}^{\text{max}}}\xspace}
\newcommand{\varmaxdrjets}{\ensuremath{\Delta R_{\text{j},\text{j}}^{\text{max}}}\xspace}
\newcommand{\varmaxdetataggedjetsAlt}{\ensuremath{\Delta \eta_{\text{b},\text{b}}^{\text{max}}}\xspace}
\newcommand{\varmaxdetajetsAlt}{\ensuremath{\Delta \eta_{\text{j},\text{j}}^{\text{max}}}\xspace}
\newcommand{\vardrbetweenlepandclosesttagAlt}{\ensuremath{\Delta R_{\text{lep},\text{b}}^{\text{min}\Delta R}}\xspace}
\newcommand{\vardrbetweenlepandclosestjetAlt}{\ensuremath{\Delta R_{\text{lep},\text{j}}^{\text{min}\Delta R}}\xspace}
\newcommand{\vartaggeddijetmassclosesttoMHAlt}{\ensuremath{m_{\text{b},\text{b}}^{ \text{closest to 125}}}\xspace}
\newcommand{\vartaggeddijetmassMaxMass}{\ensuremath{m_{\text{b},\text{b}}^{ \text{max mass}}}\xspace}
\newcommand{\varclosesttaggeddijetmassAlt}{\ensuremath{m_{\text{b},\text{b}}^{\text{min}\Delta R}}\xspace}
\newcommand{\varclosesttaggeduntaggeddijetmassAlt}{\ensuremath{m_{\text{j},\text{b}}^{\text{min}\Delta R}}\xspace}
\newcommand{\varclosesttaggeddijetpTAlt}{\ensuremath{\pt {}_{\text{b},\text{b}}^{\text{min}\Delta R}}\xspace}
\newcommand{\varclosesttaggeduntaggeddijetpTAlt}{\ensuremath{\pt {}_{\text{j},\text{b}}^{\text{min}\Delta R}}\xspace}
\newcommand{\varMlbAlt}{\ensuremath{m_{\text{lep},\text{b}}^{\text{min}\Delta R}}\xspace}
\newcommand{\varMjets}{\ensuremath{m_{\text{j}}^{\text{avg}}}\xspace}
\newcommand{\varMsquaredtaggedjets}{\ensuremath{(m^2){}_{\text{b}}^{\text{avg}}}\xspace}
\newcommand{\varavgbtagdiscbtagsForm}{\ensuremath{d^{\text{avg}}_{\text{b}}}\xspace}
\newcommand{\varavgbtagdiscjetsForm}{\ensuremath{d^{\text{avg}}_{\text{j}}}\xspace}
\newcommand{\varavgbtagdiscuntaggedjetsForm}{\ensuremath{d^{\text{avg}}_{\text{non-b}}}\xspace}
\newcommand{\vardevfromavgdiscbtagsAlt}{\ensuremath{\tfrac{1}{N_{\text{b}}}\sum^{N_{\text{b}}}_{b}\left( d - d_{\text{b}}^{\text{avg}}\right)^{2}}\xspace}
\newcommand{\varhzero}{\ensuremath{H_{0}}\xspace}
\newcommand{\varhtwo}{\ensuremath{H_{2}}\xspace}
\newcommand{\varhthree}{\ensuremath{H_{3}}\xspace}
\newcommand{\varhthreetagged}{\ensuremath{H_{3}^{\text{b}}}\xspace}
\newcommand{\varhfour}{\ensuremath{H_{4}}\xspace}
\newcommand{\varRthree}{\ensuremath{R_{3}}\xspace}
\newcommand{\varNjj}{\ensuremath{N^{\text{j},\text{b}}}\xspace}

\cmsNoteHeader{HIG-17-026}

\title{Search for \ttH production in the \Hbb decay channel with leptonic \ttbar decays in proton-proton collisions at $\sqrt{s}=13\TeV$}

\date{\today}

\abstract{
A search is presented for the associated production of a standard model Higgs boson with a top quark-antiquark pair (\ttH), in which the Higgs boson decays into a \PQb quark-antiquark pair, in proton-proton collisions at a centre-of-mass energy $\sqrt{s}=13\TeV$. The data correspond to an integrated luminosity of 35.9\fbinv recorded with the CMS detector at the CERN LHC. Candidate \ttH events are selected that contain either one or two electrons or muons from the \ttbar decays and are categorised according to the number of jets. Multivariate techniques are employed to further classify the events and eventually discriminate between signal and background. The results are characterised by an observed \ttH signal strength relative to the standard model cross section, $\mu = \sigma/\sigma_{\mathrm{SM}}$, under the assumption of a Higgs boson mass of 125\GeV. A combined fit of multivariate discriminant distributions in all categories results in an observed (expected) upper limit on $\mu$ of 1.5\,(0.9) at 95\% confidence level, and a best fit value of $0.72 \pm 0.24\stat \pm 0.38\syst$, corresponding to an observed (expected) signal significance of 1.6 (2.2) standard deviations above the background-only hypothesis.
}

\hypersetup{
pdfauthor={CMS Collaboration},
pdftitle={Search for ttH production in the Hbb decay channel with leptonic tt decays in proton-proton collisions at sqrt(s) = 13 TeV},
pdfsubject={CMS},
pdfkeywords={CMS, physics, Higgs, top quarks, bottom quarks}}

\maketitle

\section{Introduction}
\label{sec:intro}
The observation~\cite{Aad:2012tfa,Chatrchyan:2012xdj,CMS:2012nga} of a Higgs boson with a mass of approximately 125\GeV~\cite{Aad:2015zhl,Sirunyan:2017exp} at the CERN LHC marked the starting point of a broad experimental
programme to determine the properties of the newly discovered particle. Decays into $\gamma\gamma$, $\PZ\PZ$, $\PW\PW$, and $\Pgt\Pgt$ final states
have been observed, and there is evidence for the direct decay of the
particle to the bottom quark-antiquark (\bbbar) final
state~\cite{Chatrchyan:2014vua,Aad:2015vsa,Sirunyan:2017khh,Aaboud:2017xsd,Sirunyan:2017elk}. The
measured rates for various production and decay channels are consistent with the standard model (SM) expectations~\cite{Aad:2013wqa,Khachatryan:2014jba}, and the
hypothesis of a spin-0 particle is favoured over other
hypotheses~\cite{Aad:2013xqa,Khachatryan:2014kca}.

\begin{figure}[hbtp]
 \centering
   \begin{tabular}{m{0.45\textwidth}m{0.45\textwidth}}
   \includegraphics[height=0.23\textheight]{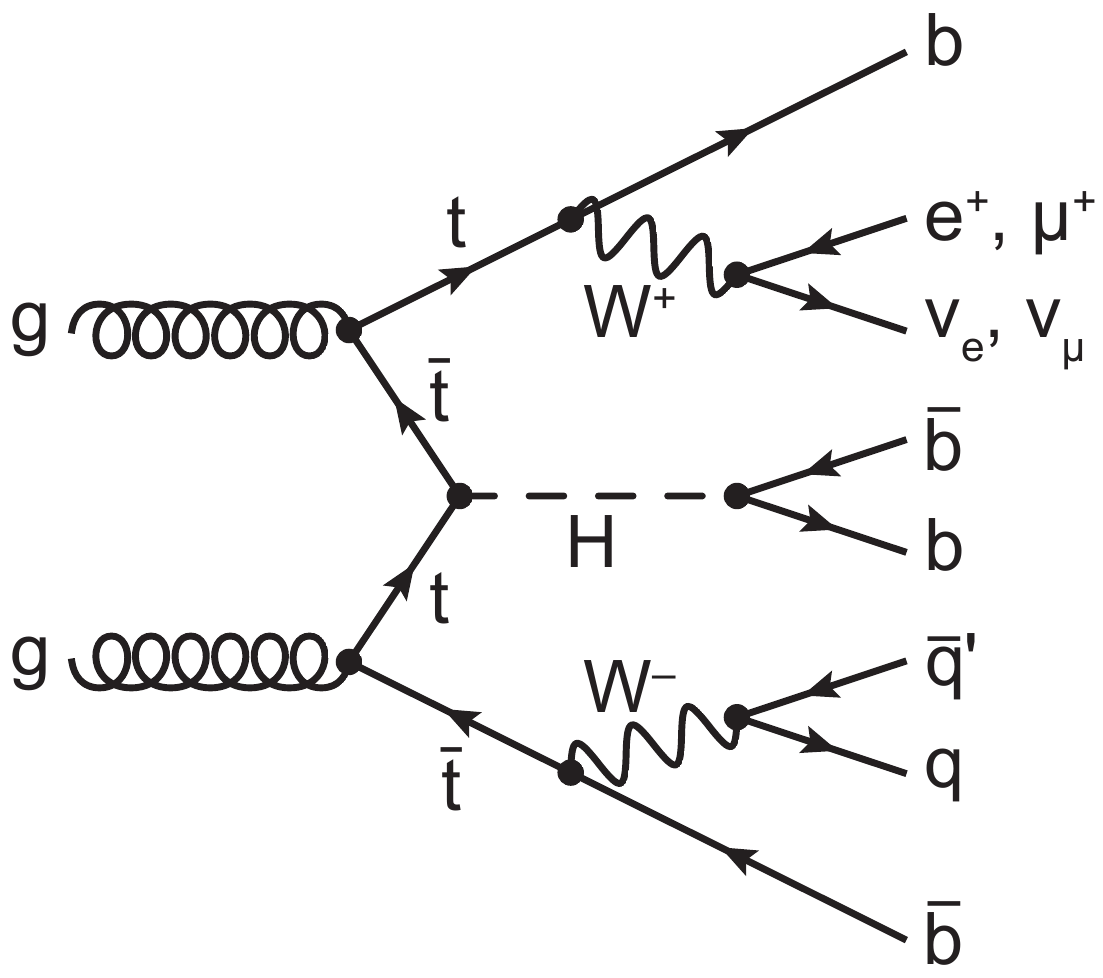} &
   \includegraphics[height=0.2\textheight]{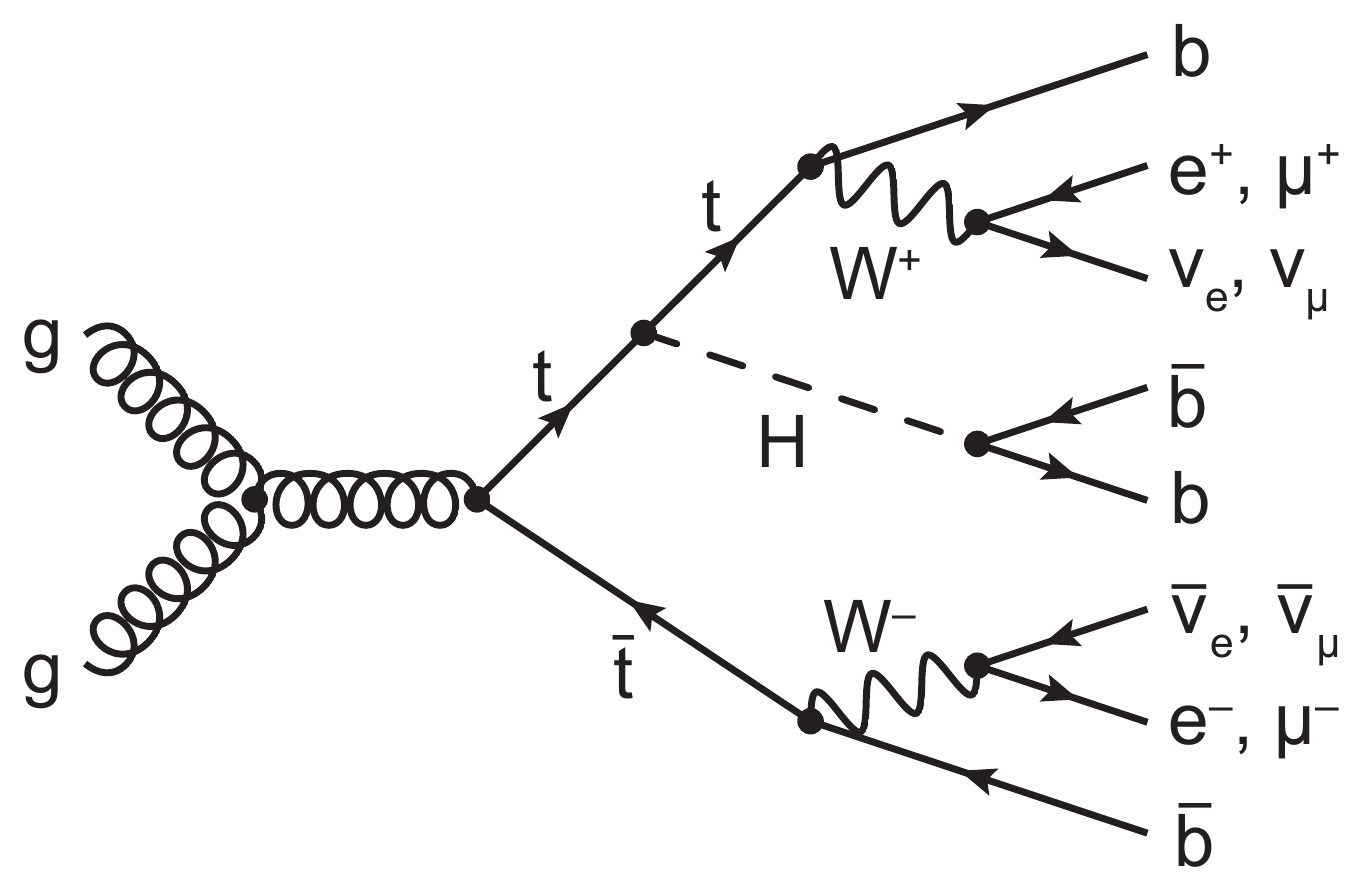}
   \end{tabular}
   \caption{Representative leading-order Feynman diagrams for \ttH
     production, including the subsequent decay of the Higgs boson into a \PQb quark-antiquark pair, and the decay of the top quark-antiquark pair into final states with either one (single-lepton channel, left) or two (dilepton channel, right) electrons or muons.}
   \label{fig:ttH_feynman_diagram}
\end{figure}

In the SM, the Higgs boson couples to fermions with a Yukawa-type
interaction, with a coupling strength proportional to the fermion
mass. Probing the coupling of the Higgs boson to the heaviest known
fermion, the top quark, is therefore very important for testing the SM and
for constraining various models of physics beyond the SM (BSM), some of which predict a different coupling strength than the SM.
Indirect constraints on the coupling between the top quark and the Higgs boson are available from processes
including virtual top quark loops, for example Higgs boson production through
gluon-gluon fusion~\cite{Aad:2013wqa,Khachatryan:2014jba}, as well as from production of four top quarks~\cite{Sirunyan:2017roi}.  On the other
hand, the associated production of a Higgs boson and a top
quark-antiquark pair (\ttH production)
as illustrated by the Feynman diagrams in
Fig.~\ref{fig:ttH_feynman_diagram} is a direct probe of the Higgs boson coupling to fermions with weak isospin $+1/2$. The Higgs boson decay into \bbbar, also shown in
Fig.~\ref{fig:ttH_feynman_diagram}, is experimentally attractive as a final state
because it features the largest branching fraction of $0.58\pm 0.02$
for a 125\GeV Higgs boson~\cite{deFlorian:2016spz}.

Several BSM physics scenarios predict a significantly enhanced production rate of events with \ttH final states, while not modifying the branching fractions of Higgs boson decays by a measurable amount~\cite{Burdman:2002ns, Han:2003wu, Perelstein:2003wd, Cheng:2005as, Cheng:1999bg, Carena:2006bn, Contino:2006qr, Burdman:2007sx, Hill:1991at, Carmona:2012jk}.
In this
context, a measurement of the \ttH production cross section has the
potential to distinguish the SM Higgs mechanism of generating fermion masses from alternative
ones.

Various dedicated searches for \ttH production have been conducted
during Run~1 of the LHC. The CMS Collaboration searches employed proton-proton (\protonproton)
collision data corresponding to an integrated luminosity of 5\fbinv
at a centre-of-mass energy of $\sqrt{s}=7\TeV$ and 19.5\fbinv at
$\sqrt{s}=8\TeV$. These searches have been performed by studying Higgs boson decays to \PQb quarks, photons, and leptons using multivariate
analysis (MVA) techniques, showing a mild excess of the observed \ttH
cross section relative to the SM expectation of
$\mu = \sigma/\sigma_\text{SM} = 2.8\pm
1.0$~\cite{Khachatryan:2014qaa}.
A similar excess of $\mu=2.1^{+1.4}_{-1.2}$ was observed in a search
for \ttH production in multilepton final states by the ATLAS Collaboration using data at $\sqrt{s}=8\TeV$, corresponding to an integrated luminosity of 20.3\fbinv~\cite{Aad:2015iha}. The searches in the \Hbb decay channel were performed with several analysis techniques~\cite{Khachatryan:2014qaa,Khachatryan:2015ila,Aad:2015gra}, yielding a most stringent observed (expected) upper limit on $\mu$ of 3.4\,(2.2) at the 95\% confidence level (\CL).

The increased centre-of-mass energy of
$\sqrt{s}=13\TeV$ results in a \ttH production cross section 3.9 times
larger than at $\sqrt{s}=8\TeV$ based on next-to-leading-order (NLO) calculations; while the cross section for the most
important background, \ttbar production, is increased by a factor
of 3.3~\cite{Dittmaier:2011ti}, resulting in a more favourable
signal-to-background ratio.
The CMS Collaboration has performed searches in the all-jets~\cite{Sirunyan:2018ygk} and multilepton~\cite{Sirunyan:2018shy} final states with 35.9\fbinv of data, achieving evidence for \ttH production with an observed (expected) significance of 3.2 (2.8) standard deviations in the latter case.
Recently, the ATLAS Collaboration reported observed (expected) evidence for \ttH production with a significance of 4.2 (3.8) standard deviations, based on an integrated luminosity of 36.1\fbinv and combining several Higgs boson decay channels~\cite{Aaboud:2017jvq}; in the \Hbb channel alone, an observed (expected) upper limit on $\mu$ of 2.0 (1.2) at 95\% \CL and a best fit value of $\mu = 0.84^{+0.64}_{-0.61}$ were obtained~\cite{Aaboud:2017rss}.

In this paper, a search for \ttH production in the \Hbb final state is presented that has been
performed using \lumivalue of data recorded with the CMS detector at
$\sqrt{s} = 13\TeV$ in 2016.
In the SM, the top quark is expected to decay into a \PW\ boson and a \cPqb\ quark almost exclusively.
Hence different \ttbar decay modes can be identified according to the subsequent decays of the \PW\ bosons.
The event selection is based on the decay topology of \ttH events in which the Higgs boson decays into \bbbar and the \ttbar decay involves at least one lepton, resulting in  either $\ell \Pgn\, \cPq\cPaq^{\prime}\, \bbbar$ (single-lepton) or $\ell^{+} \Pgn\, \ell^{-} \Pagn\, \bbbar$ (dilepton) \ttbar final states, where $\ell = \Pe,\Pgm$ arising either from the prompt decay of a \PW\ boson or from leptonic $\tau$ decays.
Analysis methods
established in Run~1~\cite{Khachatryan:2014qaa,Khachatryan:2015ila} have been significantly improved, and novel
methods have been added.
In particular, two multivariate techniques---namely boosted decision trees (BDTs) and the matrix element method (MEM)~\cite{Hastie:ESL,Pushpalatha:MVA,TMVA,Kondo:1988yd,Abazov:2004cs}---that utilise event information differently in order to discriminate signal from background events have been employed in combination.
Since the two methods aim at separating signal from different background
processes, their combined usage helps to obtain a better sensitivity.
In addition, a new multivariate technique based on deep neural networks (DNNs) has been employed to separate signal from background events.
The best fit value of the signal strength modifier $\mu$ is obtained from a combined profile likelihood fit of the classifier output distributions to the data, correlating processes and their uncertainties where appropriate.

This document is structured as follows.
The CMS detector is described in Section~\ref{sec:detector}.
In Section~\ref{sec:samples}, the simulated signal and background samples are described. The basic
selection of analysis objects and events is discussed in
Section~\ref{sec:reconstruction}. The general analysis strategy and
background estimation methods are introduced in
Section~\ref{sec:classification}. The effect of systematic uncertainties
is studied in Section~\ref{sec:systematics}. Results of the analysis
are presented in
Section~\ref{sec:results}, followed by a summary in
Section~\ref{sec:summary}.

\section{The CMS detector}
\label{sec:detector}
The central feature of the CMS apparatus is a superconducting solenoid of 6\unit{m} internal diameter, providing a magnetic field of 3.8\unit{T}. Within the solenoid volume are a silicon pixel and strip tracker, a lead tungstate crystal electromagnetic calorimeter, and a brass and scintillator hadron calorimeter, each composed of a barrel and two endcap sections. Forward calorimeters extend the pseudorapidity ($\eta$) coverage provided by the barrel and endcap detectors. Muons are detected in gas-ionisation chambers embedded in the steel magnetic flux-return yoke outside the solenoid. A more detailed description of the CMS detector, together with a definition of the coordinate system used and the relevant kinematic variables, can be found in Ref.~\cite{Chatrchyan:2008zzk}. Events of interest are selected using a two-tiered trigger system~\cite{Khachatryan:2016bia}. The first level, composed of custom hardware processors, uses information from the calorimeters and muon detectors to select events, while the second level selects events by running a version of the full event reconstruction software optimised for fast processing on a farm of computer processors.

\section{Simulation of signal and background}
\label{sec:samples}
Several Monte Carlo event generators, interfaced with a detailed detector
simulation, are used to model experimental effects, such as
reconstruction and selection efficiencies, as well as detector
resolutions. The CMS detector response is simulated using
\GEANTfour~(v.9.4)~\cite{bib:geant}.

For the simulation of the \ttH signal sample, the NLO event generator \POWHEG~(v.2)~\cite{powheg:general:1,powheg:general:2,powheg:general:3,powheg:tth:1} is used.
Standard model backgrounds are simulated using \POWHEG~(v.2), \PYTHIA~(v.8.200)~\cite{Sjostrand:2007gs}, or \MGvATNLO (v.2.2.2)~\cite{Alwall:2014hca}, depending on the process.
The value of the Higgs boson mass is assumed to be 125\GeV, while the top quark mass value is set to 172.5\GeV.
The proton structure is described by the parton distribution functions (PDF) NNPDF3.0~\cite{Ball:2014uwa}.

The main background contribution originates from \ttbar
production, the production of \PW\ and \PZ/$\gamma^{*}$ bosons with
additional jets (referred to as \Wjets and \Zjets, or commonly as \Vjets), single top quark production (\cPqt\PW\ and $t$-channel production), diboson (\PW\PW, \PW\PZ, and $\PZ\PZ$) processes, and \ttbar production in
association with a \PW\ or \PZ\ boson (referred to as \ttW and \ttZ, or commonly as \ttV).
Both the \ttbar and the single top quark processes in the $t$- and \cPqt\PW-channels are simulated with \POWHEG~\cite{powheg:singlet:st,powheg:singlet:tW}.
The $s$-channel single top quark processes, as well as \Vjets and \ttV processes are simulated at NLO with \MGvATNLO, where for the \Vjets processes the matching of matrix-element (ME) jets to parton showers (PS) is performed using the \textsc{FxFx}~\cite{Frederix:2012ps} prescription.
The \PYTHIA event generator is used to simulate diboson events.

Parton showering and hadronisation are simulated with \PYTHIA~(v.~8.200) for all signal and background processes. The \PYTHIA CUETP8M2T4~\cite{bib:CMS:2016kle}
 tune is used to characterise the underlying event in
the \ttH signal and \ttbar and single top quark background processes, while
the CUETP8M1~\cite{bib:CUETP8tune} tune is used for all other
background processes.

For comparison with the observed distributions, the events in the
simulated samples are normalised to the same integrated luminosity of the data sample, according to their predicted cross sections. These are
taken from theoretical calculations at next-to-next-to-leading order
(NNLO, for \Vjets production), approximate NNLO (single top quark \cPqt\PW\
channel~\cite{bib:twchan}), and NLO (single top quark $t$- and $s$-channels~\cite{Aliev:2010zk,Kant:2014oha}, \ttV production~\cite{Maltoni:2015ena}, and diboson production~\cite{bib:mcfm:diboson}).
The \ttH cross section of $507^{+35}_{-50}\,\text{fb}$ and Higgs boson branching fractions used in the analysis also correspond to NLO accuracy~\cite{deFlorian:2016spz}.
The \ttbar simulated
sample is normalised to the full NNLO calculation with resummation to
next-to-next-to-leading-logarithmic accuracy~\cite{bib:xs1,bib:xs2,bib:xs3,bib:xs4,bib:xs7,Czakon:2013goa,Czakon:2011xx},
assuming a top quark mass value of 172.5\GeV and using the NNPDF3.0
PDF set. This sample is further separated into the following processes
based on the flavour of additional jets that do not originate from the
top quark decays in the event: \ttbb, defined at generator level as
the events in which two additional \cPqb\ jets are generated within the
acceptance requirements (see Section~\ref{sec:reconstruction}), each of which
originates from one or more \PB{} hadrons; \ttb, for which only
one additional \cPqb\ jet within the acceptance originates from a single \PB{} hadron; \tttwob, which
corresponds to events with two additional \PB{} hadrons that are close
enough in direction to produce a single \cPqb\ jet; \ttcc, for which events
have at least one additional \cPqc\ jet within the acceptance and no additional \cPqb\ jets;
\ttbar + light flavour jets (\ttlf), which corresponds to events that do
not belong to any of the above processes.
The \ttbb, \ttb, \tttwob, and \ttcc processes are collectively referred to as \tthf in the following.
This categorisation is
important because the subsamples originate from different
physics processes and have different systematic uncertainties.

Effects from additional \protonproton interactions in the same bunch
crossings (pileup) are modelled by adding simulated minimum-bias events
(generated with \PYTHIA~v.8.212, tune~CUETP8M1) to all simulated processes. The pileup
multiplicity distribution in simulation is reweighted to reflect the
luminosity profile of the observed \protonproton collisions.
Correction factors described in Section~\ref{sec:reconstruction} are applied to the simulation where necessary to improve the description of the data.

\section{Object and event reconstruction}
\label{sec:reconstruction}
The event selection is optimised to identify events from the production of a Higgs boson in association with \ttbar events, where the Higgs boson decays into \bbbar.
Two \ttbar decay modes are considered: the single-lepton mode
($\ttbar \to \ell \Pgn\, \cPq\cPaq^{\prime}\, \bbbar$), where one \PW{}
boson decays into a charged lepton and a neutrino, and the dilepton
mode ($\ttbar \to \ell^{+} \Pgn\, \ell^{-} \Pagn\, \bbbar$), where
both \PW\ bosons decay into a charged lepton and a neutrino. These
signatures imply the presence of isolated leptons ($\ell =
\Pe,\,\Pgm$), missing transverse momentum due
to the neutrinos from \PW\ boson decays, and highly energetic jets
originating from the final-state quarks.
Jets originating from the hadronisation of \cPqb\ quarks are identified through \cPqb\ tagging techniques~\cite{Sirunyan:2017ezt}.

Online, events in the single-lepton channel were selected by
single-lepton triggers which require the presence of one electron
(muon) with a transverse momentum (\pt) threshold of $\pt > 27 (24)\GeV$. Events in the dilepton channel were selected
either by the single-lepton trigger (retaining events with an additional lepton)
or by dilepton triggers that require the presence of two
electrons or muons. The same-flavour dilepton triggers required two
electrons with $\pt > 23$ and 12\GeV, or two muons with
$\pt > 17$ and 8\GeV, respectively. The different-flavour
dilepton triggers required either a muon with $\pt > 23\GeV$ and an
electron with $\pt > 12\GeV$, or an electron with $\pt > 23\GeV$ and a
muon with $\pt > 8\GeV$.

Events are reconstructed using a
particle-flow (PF) technique~\cite{bib:PF}, which
combines information from all subdetectors to enhance the reconstruction
performance by identifying individual particle candidates in
\protonproton~collisions.
An interaction vertex~\cite{Chatrchyan:2014fea} is required within 24\cm of the detector centre along the beam line direction, and within 2\cm of the beam line in the transverse plane. 
Among all such vertices, the reconstructed vertex with the largest value of summed physics-object $\pt^{2}$ is taken to be the primary $\Pp\Pp$ interaction vertex. The physics objects are the jets, clustered using a jet finding algorithm~\cite{Cacciari:2008gp,Cacciari:2011ma} with the tracks assigned to the vertex as inputs, and the associated missing transverse momentum, taken as the negative vector sum of the \pt of those jets.
All other interaction vertices are considered as pileup vertices.
Charged tracks identified as hadrons from pileup vertices are omitted in the subsequent event reconstruction.

The electron and muon candidates are required to be sufficiently
isolated from nearby jet activity as follows. For each electron (muon)
candidate, a cone of $\Delta R = 0.3\,(0.4)$ is
constructed around the direction of the track at the event vertex,
where $\Delta R$ is defined as
$\sqrt{\smash[b]{(\Delta \eta)^2 + (\Delta \phi)^2}}$, and
$\Delta\eta$ and $\Delta \phi$ are the distances in the pseudorapidity and
azimuthal angle. Excluding the contribution from the lepton candidate,
the scalar \pt sum of all particle candidates inside the cone
consistent with arising from the chosen primary event vertex is
calculated. The neutral component from pileup interactions is subtracted event-by-event, based on the average transverse energy deposited by neutral particles in the event in the case of electrons, and half the transverse momentum carried by charged-particles identified to come from pileup vertices in the case of muons.
A relative isolation discriminant $I_\text{rel}$ is defined as the ratio of this sum to the \pt of the lepton candidate.
Electron candidates are selected if they have values of
$I_\text{rel}<0.06$, while muons are selected if they fulfil the requirement
$I_\text{rel}<0.15$ in the single-lepton channel and $I_\text{rel}<0.25$
in the dilepton channel. In addition, electrons from identified photon
conversions are rejected~\cite{Khachatryan:2015hwa}. To further increase the purity of muons
originating from the primary interaction and to suppress misidentified
muons or muons from decay-in-flight processes, additional quality
criteria, such as a minimal number of hits associated with the muon
track, are required in both the silicon tracker and the muon
system~\cite{Sirunyan:2018fpa}.

For the single-lepton channel, events are selected containing exactly
one energetic, isolated lepton (\Pe\ or \Pgm), which is required to
have $\pt > 30 (26)\GeV$ in the case of the electron (muon), and $\abs{\eta} < 2.1$. Electron candidates in the transition region between the barrel and endcap calorimeters, $1.4442< \abs{\eta}< 1.5560$, are excluded.
The flavour of the lepton must match the flavour of the trigger that accepted the event (\eg if an electron is identified, the single-electron trigger must have accepted the event).
For the dilepton channel, events are
required to have a pair of oppositely charged energetic leptons
(\ee, \mue, \mumu). The
lepton with the highest \pt out of the pair is required to have $\pt > 25 \GeV$, and the other lepton $\pt > 15 \GeV$; both leptons are
required to fulfil the requirement $\abs{\eta} < 2.4$, excluding electrons in the transition region. The flavours
of the lepton pair must match the flavour of the trigger that accepted the event.
The events are unambiguously classified as \ee, \mue, or \mumu,
depending on the type of the selected lepton pair, and there is no
overlap with the other channels under study. The invariant mass of the
selected lepton pair, $m_{\ell\ell}$, is required to be larger
than 20\GeV to suppress events from heavy-flavour resonance decays and
low-mass Drell--Yan processes. In the same-flavour channels, events are
also rejected if $76< m_{\ell\ell} < 106\GeV$, thereby
suppressing further contribution from \Zjets events.
In both the single- and dilepton channel, events with additional isolated
leptons with $\pt > 15 \GeV$ and $\abs{\eta} < 2.4$ are
excluded from further analysis.

The missing transverse momentum vector \ptvecmiss is defined as the
projection of the negative
vector sum of the momenta of all reconstructed PF objects in an
event on the plane perpendicular to the beams. Its magnitude is referred to as \ptmiss. Events are required to fulfil $\ptmiss > 20\GeV$ in the single-lepton and $\ptmiss > 40\GeV$ in the dilepton same-flavour channels to further suppress background contribution.

Jets are reconstructed from the PF particle candidates using the
anti-\kt clustering algorithm~\cite{Cacciari:2008gp} with a distance
parameter of 0.4, as implemented in \FASTJET~\cite{Cacciari:2011ma}.
Charged hadrons that are associated to pileup vertices are discarded from the clustering.
The jet energy is corrected for the remaining neutral-hadron pileup component
in a manner similar to that used to find the energy within the
lepton isolation cone~\cite{bib:PUSubtraction}.
Jet energy corrections are also applied as a
function of jet \pt and $\eta$~\cite{JESPUB:JME-13-004} to data and
simulation.
All reconstructed jets in the single-lepton channel and the two jets leading in \pt in the dilepton channel are required to satisfy $\abs{\eta}< 2.4$ and $\pt> 30\GeV$.
Other jets in the dilepton channel are selected if $\pt> 20\GeV$.
Events are selected if they contain at least four jets in the single-lepton channel or at least two jets in the dilepton channel.

Jets originating from the hadronisation of \cPqb\ quarks are identified
using a combined secondary vertex algorithm (\btagger)~\cite{Sirunyan:2017ezt},
which provides a \cPqb\ tagging discriminant by combining identified
secondary vertices and track-based lifetime information. A
discriminant value is chosen such that the
probability of tagging jets originating from light-flavour quarks
($\cPqu$, $\cPqd$, or $\cPqs$) or gluons is about 1\%, and the corresponding efficiency for tagging
jets from \cPqb\ (\cPqc) quarks is ${\approx}65\%$ (10\%).
The shape of the \btagger discriminant distribution in simulation is corrected by scale factors to better describe the data.
This correction is derived separately for light-flavour and \cPqb\ jets with a tag-and-probe approach.
Control samples enriched in events with a \PZ\ boson and exactly two jets where a \cPqb\ jet veto is applied are used to obtain the correction for light-flavour jets.
The correction for \cPqb\ jets is estimated using a sample enriched in \ttbar events with no additional jets~\cite{Sirunyan:2017ezt}.
For \cPqc\ jets, the data-to-simulation scale factor is set to unity with an uncertainty twice the one of the correction for \cPqb\ jets.
Events are required to have at least two (one) \cPqb-tagged jets in the single-lepton (dilepton) channels.

\begin{figure}[!htb]
  \centering
  \begin{tabular}{cc}
    \includegraphics[width=0.47\textwidth]{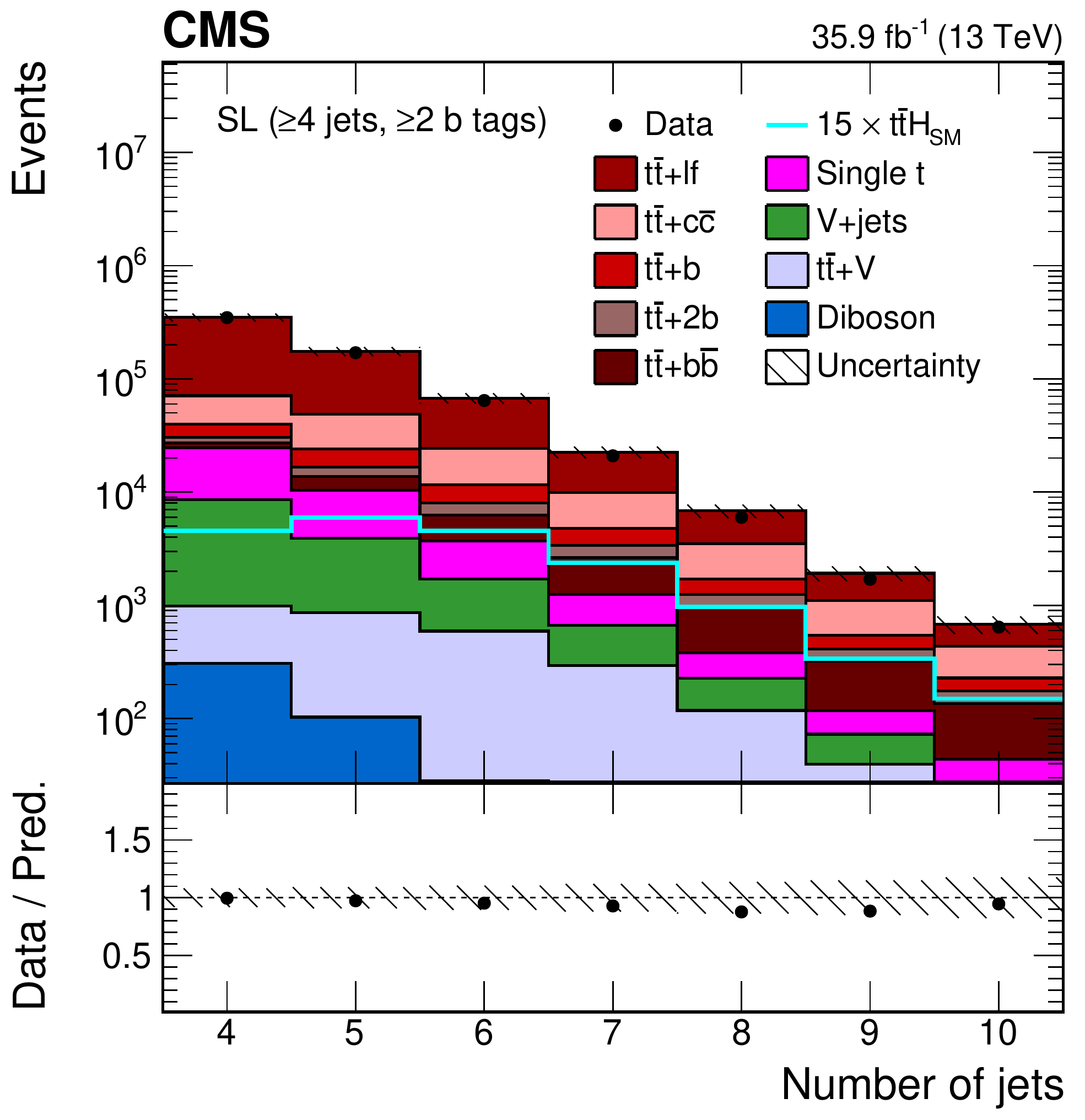} &
    \includegraphics[width=0.47\textwidth]{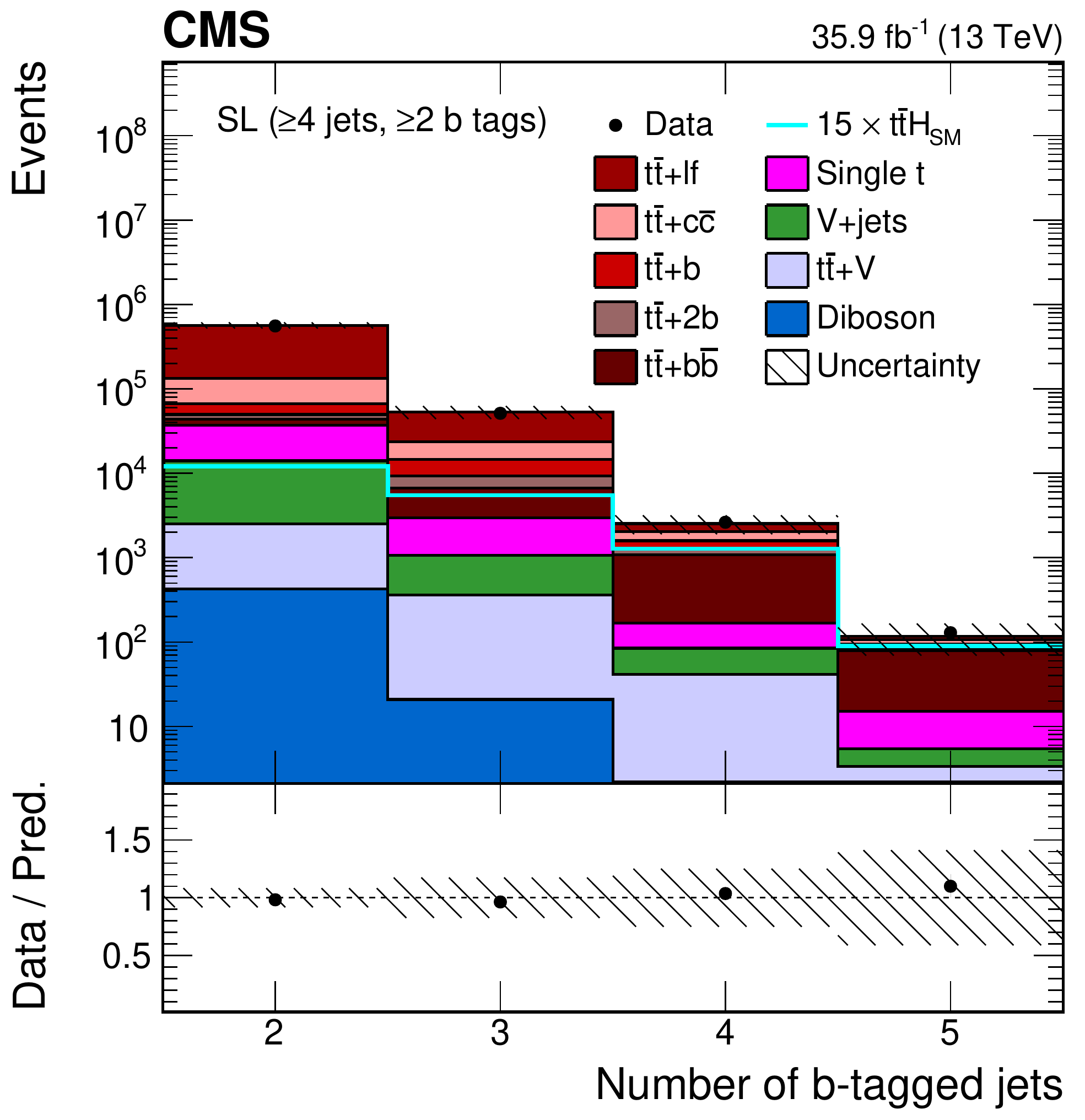} \\
  \end{tabular}
  \caption{
    Jet (\cmsLeft) and \cPqb-tagged jet (\cmsRight) multiplicity in the single-lepton (SL) channel after the baseline selection.
    The expected background contributions (filled histograms) are stacked, and the expected signal
    distribution (line), which includes \Hbb and all other Higgs boson decay modes, is superimposed. Each contribution
    is normalised to an integrated luminosity of \lumivalue, and the
    signal distribution is additionally scaled by a factor of 15
    for better visibility.
    The hatched uncertainty bands correspond to the total statistical and systematic uncertainties (excluding uncertainties that affect only the normalisation of the distribution) added in quadrature.
    The distributions observed in data (markers) are overlayed.
    The last bin includes overflow events.
    The lower plots show the ratio of the data to the background prediction.
  }
  \label{fig:baselineselection:slnjets}
\end{figure}

\begin{figure}[!htb]
  \centering
  \begin{tabular}{cc}
    \includegraphics[width=0.47\textwidth]{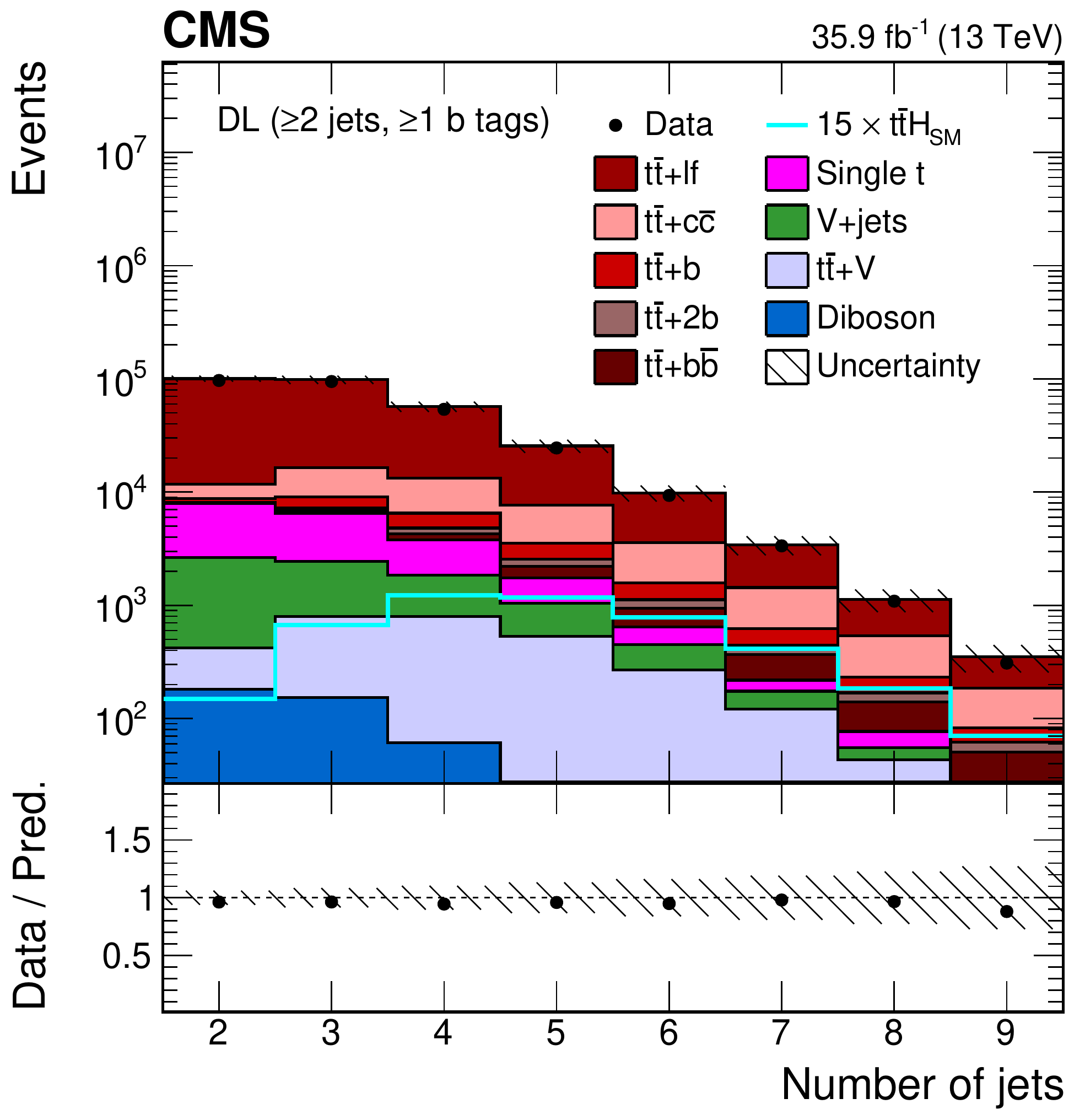} &
    \includegraphics[width=0.47\textwidth]{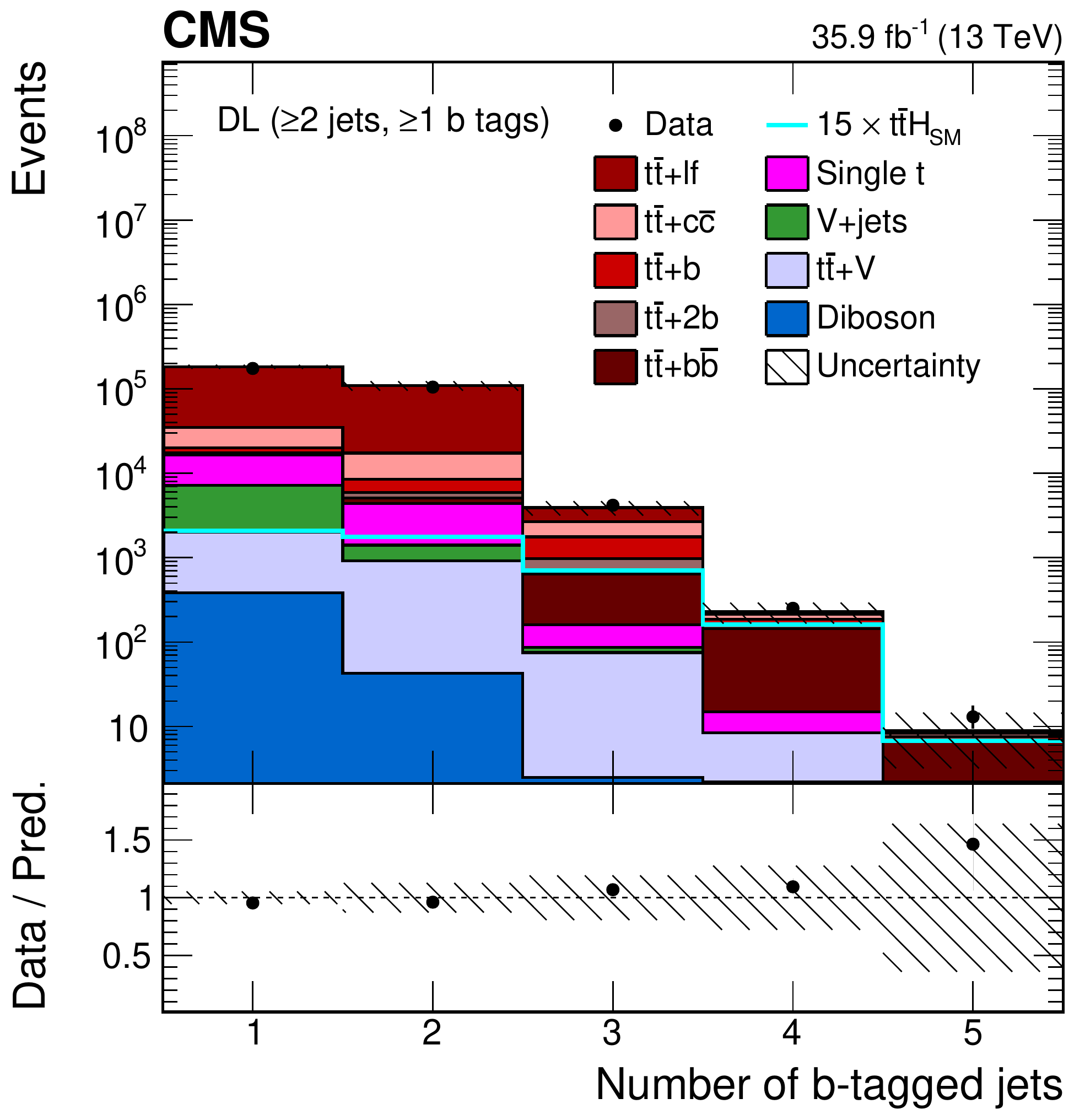} \\
  \end{tabular}
  \caption{
    Jet (\cmsLeft) and \cPqb-tagged jet (\cmsRight) multiplicity in the dilepton (DL) channel after the baseline selection.
    The expected background contributions (filled histograms) are stacked, and the expected signal
    distribution (line), which includes \Hbb and all other Higgs boson decay modes, is superimposed. Each contribution
    is normalised to an integrated luminosity of \lumivalue, and the
    signal distribution is additionally scaled by a factor of 15
    for better visibility.
    The hatched uncertainty bands correspond to the total statistical and systematic uncertainties (excluding uncertainties that affect only the normalisation of the distribution) added in quadrature.
    The distributions observed in data (markers) are overlayed.
    The last bin includes overflow events.
    The lower plots show the ratio of the data to the background prediction.
  }
  \label{fig:baselineselection:dlnjets}
\end{figure}

Event yields observed in data and predicted by the simulation after this selection (referred to as baseline selection in the following) are listed in Table~\ref{tab:baselineselection:yields} for the single-lepton and dilepton channels.
The corresponding jet and \cPqb-tagged jet multiplicity distributions are shown in Figs.~\ref{fig:baselineselection:slnjets} and~\ref{fig:baselineselection:dlnjets}, respectively. The \ttH signal includes \Hbb and all other Higgs boson decay modes.
Background contributions from QCD multijet production, estimated using a low-\ptmiss control region in data, have been found to be negligible in this analysis.

\begin{table}[!htb]
  \centering
  \topcaption{
    Event yields observed in data and predicted by the simulation after the baseline selection requirements in the single-lepton (SL) and dilepton (DL) channels.
    The \ttH signal includes \Hbb and all other Higgs boson decay modes.
    The quoted uncertainties are statistical only.
  }
  \label{tab:baselineselection:yields}
  \begin{tabular}{l r@{\,}c@{\,}l r@{\,}c@{\,}l}
    \hline
    Process     & \multicolumn{3}{c}{SL channel} & \multicolumn{3}{c}{DL channel} \\
    \hline
    \ttlf       &         463\;658 &$\pm$& 174 &         241\;032 &$\pm$&  99 \\
    \ttcc       & \hphantom{5pt}76\;012 &$\pm$&  70 & \hphantom{5pt} 24\;550 &$\pm$&  32 \\
    \ttb        &          22\;416 &$\pm$&  38 &           5\;979 &$\pm$&  16 \\
    \tttwob     &           9\;052 &$\pm$&  24 &           1\;785 &$\pm$&   9 \\
    \ttbb       &          10\;897 &$\pm$&  27 &           1\;840 &$\pm$&   9 \\
    \singlet    &          25\;215 &$\pm$& 166 &          12\;206 &$\pm$& 125 \\
    \Vjets      &          12\;309 &$\pm$&  58 &           5\;684 &$\pm$& 209 \\
    \ttV        &         2\;457 &$\pm$&  12 &           2\;570 &$\pm$&  23 \\
    \diboson    &            449 &$\pm$&  14 &            430 &$\pm$&  15 \\[\cmsTabSkip]
    Total bkg.  &         622\;466 &$\pm$& 263 &         296\;077 &$\pm$& 266 \\[\cmsTabSkip]
    \ttH        &           1\;232 &$\pm$&   2 &          314.0 &$\pm$& 0.9 \\[\cmsTabSkip]
    Data        & \multicolumn{3}{c}{610\;556} & \multicolumn{3}{c}{283\;942} \\
    \hline
  \end{tabular}
\end{table}

\section{Analysis strategy and event classification}
\label{sec:classification}
In both the single-lepton and
dilepton channels, events with at least four jets of which at least three are
\cPqb-tagged are selected among those passing the baseline selection
described in Section~\ref{sec:reconstruction}. These events are
then further divided into categories with varying signal purity and
different background composition. In each category, combinations of several
multivariate discriminants are optimised to
separate signal from background. The signal is extracted in a
simultaneous template fit of the discriminant output obtained from the simulation to the data
across all the categories, correlating processes and their
uncertainties where appropriate. In this way, the different background
composition in the different categories helps to constrain the
uncertainties of the different processes and increases the overall
sensitivity of the search.

\begin{figure}[!htbp]
  \centering
  \includegraphics[width=0.7\textwidth]{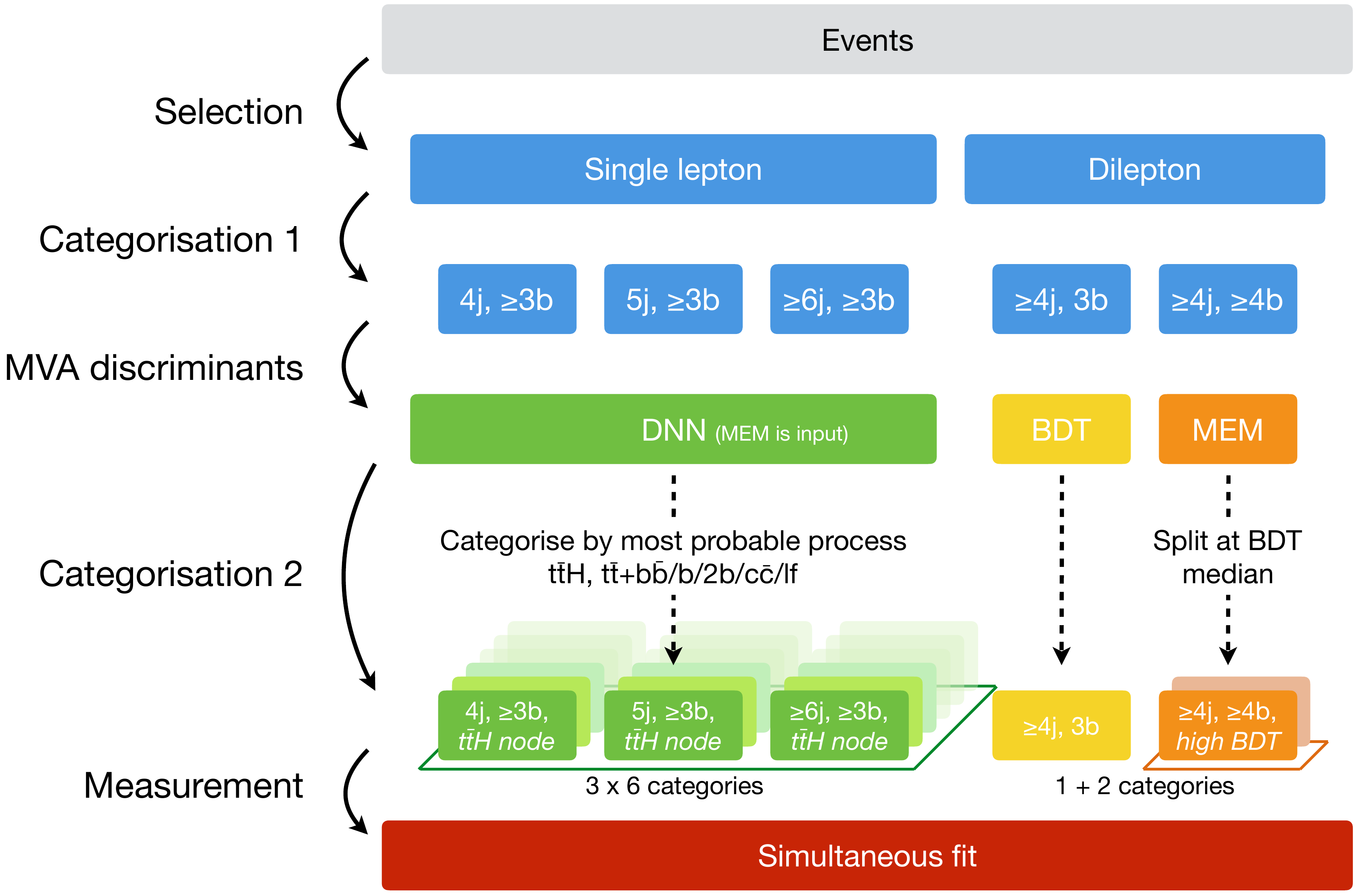}
  \caption{Illustration of the analysis strategy.}
  \label{fig:analysisflow}
\end{figure}

Several methods that classify events as signal- or background-like were explored to achieve optimal sensitivity:
DNNs and BDTs, combined with a MEM. In the DNN approach, the jet multiplicity and the DNN classification output, described below, are used for the event categorisation (``jet-process categories''). In the BDT approach, events are divided into categories based on their jet and \cPqb-tagged jet multiplicity (``jet-tag categories''). The approach that provided the best expected sensitivity in each channel, evaluated on fits to simulated data, was chosen for obtaining the final result from data.
Therefore, in the single-lepton channel the DNN approach is used, while in the dilepton channel a BDT+MEM classification is chosen. The methods and the corresponding categorisation are illustrated in Fig.~\ref{fig:analysisflow} and described in the following.

In the single-lepton channel, events are separated depending on the jet multiplicity into three categories with \ljFourThreeIncl, \ljFiveThreeIncl, and \ljSixThreeIncl.
Dedicated mul\-ti-classification DNNs~\cite{Goodfellow-et-al-2016} are trained in each jet multiplicity category to separate signal and each of the five \ttjets background processes \ttbb, \tttwob, \ttb, \ttcc, or \ttlf.

The DNN training is performed using simulated \ttH and \ttjets events as signal and background, respectively.
The overall set of events is split into a training set (30\%), an independent set (20\%) for validation and optimisation of the DNN configuration (hyper parameters), such as the number of nodes per layer, and a set that is reserved for the fit to the data (50\%).
The hyper parameters and input variables are detailed in Appendix~\ref{sec:appendix:MVAs}.

The training is conducted in two stages.
In the first stage, a DNN is trained to predict which of the reconstructed physics objects originate from the expected underlying hard process, such as for example the \cPqb\ quark jet from the decay of a top quark.
In the second stage, the initial network is extended by adding hidden layers, which take as input the variables and the output values of the first stage, and the resulting network is trained to predict the physics process of an event.
The values obtained in the output nodes of the second stage are normalised to unity using a ``softmax'' function~\cite{Goodfellow-et-al-2016}, and, as a result, can be interpreted as probabilities describing the likelihood of the event being a \ttH signal or one of the five \ttjets background processes.
Events are divided into subcategories of the most
probable process according to this DNN classification. Thus, there are
in total 18 jet-process categories in the single-lepton channel.
In each of the jet-process categories, the DNN classifier output
distribution of the node that matches the process category is used as
the final discriminant.

The DNNs utilise input variables related to kinematic properties of individual objects, event shape, and the jet \btagger \cPqb\ tagging discriminant, and additionally the MEM discriminant output, described in the following.

The MEM discriminant is constructed as the ratio of the probability density values for the signal (\ttH) and background (\ttbb) hypotheses, following the algorithm described in Ref.~\cite{Khachatryan:2015ila}. Each event is assigned a probability density value computed from the four-momenta of the reconstructed particles, which is based on the leading order scattering amplitudes for the \ttH and \ttbb processes and integrated over the particle-level quantities that are either unknown or poorly measured.
The probability density functions are constructed at leading order, assuming gluon-gluon fusion production both for signal and background processes as it represents the majority of the event rate.
In each event, the four jets that are most likely to originate from \cPqb\ quarks are considered explicitly as candidates for the \cPqb\ quarks from the decay of the Higgs boson and the top quarks.
All permutations of jets, regardless of their \cPqb\ tagging discriminant, are considered when associating the \cPqb-quark-like jets to the top quark or Higgs boson decays in the matrix element.
The four \cPqb-like jets are selected using a likelihood ratio criterion as follows.
The likelihoods are computed under either the hypothesis that four jets or that two jets in the event originate from \cPqb\ quarks, based on the expected \cPqb\ tagging discriminant probability densities from simulation.
The used ratio is computed as the four-\cPqb-jets likelihood, normalised to the sum of the four- and the two-\cPqb-jets likelihoods. When computing the MEM in the single-lepton channel, up to four additional light jets, ordered in \pt, are permuted over as candidates for the light quarks from the hadronic decay of the \PW\ boson.

In the dilepton channel, events are separated into two jet-tag
categories with \dlFourThree and \dlFourFour. In each jet-tag
category, a dedicated BDT is trained to separate signal from background processes. The BDTs utilise input variables related to kinematic properties of individual objects, event
shape, and the jet \btagger \cPqb\ tagging discriminant, similar as the DNNs, but no MEM information.
The training is
performed using simulated \ttH and \ttjets events as signal and
background, respectively, which are weighted to achieve equal yields
of signal and background events.
In order to avoid a biased performance estimate,
the events are separated in half for training and validation.
The specific BDT boosting method used is the stochastic
gradient boost~\cite{Friedman2002367,Hastie:ESL}, available as part of the TMVA package~\cite{TMVA}.
The choice of the BDT architecture and the input variables was optimised with a procedure based on the particle swarm algorithm~\cite{ParticleSwarm,ElMorabit:48864}, selecting the configuration and set of variables that yields the highest discrimination power.
They are detailed in Appendix~\ref{sec:appendix:MVAs}.

In the \dlFourThree category, the BDT output distribution is used as the
final discriminant. The \dlFourFour category is further divided
into two subcategories, one with small values of the BDT output (background-like) and one with large output values (signal-like).
The division is taken at the median of
the BDT output distribution for simulated signal events. In each
subcategory, the MEM discriminant output is used as the final
discriminant.
The high BDT output subcategory is expected to be enhanced with signal events and residual \ttbb background
events, and the MEM discriminant achieves by construction particularly
powerful additional separation against the \ttbb background
contributions. The choice of the median contributes to a robust
result by ensuring a sufficient number of events in each
subcategory. Including the low \PQb tag multiplicity and the low BDT
output subcategories into the fit constrains the background
contributions and systematic uncertainties for each of the different
event topologies. Thus, there are in total three categories in
the dilepton channel.

In summary, in the single-lepton channel events are
subdivided into 18 jet-process categories and the DNN output
distribution of the most probable process is used as the final
discriminant. In the dilepton channel events are subdivided into three
jet-tag categories and either the BDT or MEM output distribution is
used as the final discriminant.

\section{Systematic uncertainties}
\label{sec:systematics}
In Table~\ref{tab:systsummary}, all sources of systematic
uncertainties considered in the analysis are listed. They affect
either the rate of the signal or background processes, or the
discriminant shape, or both. In the last case, the rate and shape
effects are treated as entirely correlated and are varied
simultaneously. The uncertainties are taken into account via nuisance
parameters in the final fit procedure described in
Section~\ref{sec:results}, where the effects from the same source are treated as fully correlated among the different categories.
The impact of the uncertainties on the final result is discussed in Section~\ref{sec:results}.

\begin{table}[hbtp] \small
  \centering
  \topcaption{Systematic uncertainties considered in the analysis, their corresponding type (affecting rate or shape of the distributions), and additional remarks.}
  \label{tab:systsummary}
  \begin{tabular}{lcp{0.55\textwidth}}
    \hline
    Source                        & Type     & Remarks \\
    \hline
    Integrated luminosity                    & rate     & Signal and all backgrounds \\
    Lepton identification/isolation                 & shape    & Signal and all backgrounds \\
    Trigger efficiency            & shape    & Signal and all backgrounds \\
    Pileup                        & shape    & Signal and all backgrounds \\
    Jet energy scale              & shape    & Signal and all backgrounds \\
    Jet energy resolution         & shape    & Signal and all backgrounds \\
    \cPqb\ tag hf fraction             & shape    & Signal and all backgrounds \\
    \cPqb\ tag hf stats (linear)       & shape    & Signal and all backgrounds \\
    \cPqb\ tag hf stats (quadratic)    & shape    & Signal and all backgrounds \\
    \cPqb\ tag lf fraction             & shape    & Signal and all backgrounds \\
    \cPqb\ tag lf stats (linear)       & shape    & Signal and all backgrounds \\
    \cPqb\ tag lf stats (quadratic)    & shape    & Signal and all backgrounds \\
    \cPqb\ tag charm (linear)          & shape    & Signal and all backgrounds \\
    \cPqb\ tag charm (quadratic)       & shape    & Signal and all backgrounds \\[\cmsTabSkip]
    Renorm./fact. scales (\ttH)              & rate     & Scale uncertainty of NLO \ttH prediction \\
    Renorm./fact. scales (\ttbar)            & rate     & Scale uncertainty of NNLO \ttbar prediction \\
    Renorm./fact. scales (\tthf)             & rate     & Additional 50\% rate uncertainty of \tthf predictions \\
    Renorm./fact. scales (\cPqt)                 & rate     & Scale uncertainty of NLO single \cPqt\ prediction \\
    Renorm./fact. scales (V)                 & rate     & Scale uncertainty of NNLO \PW\ and \PZ\ prediction \\
    Renorm./fact. scales (VV)                & rate     & Scale uncertainty of NLO diboson prediction \\[\cmsTabSkip]
    PDF ($\Pg\Pg$)                      & rate     & PDF uncertainty for $\Pg\Pg$ initiated processes except \ttH \\
    PDF ($\Pg\Pg$ \ttH)                 & rate     & PDF uncertainty for \ttH \\
    PDF ($\Pq\Paq$)       & rate     & PDF uncertainty of $\Pq\Paq$ initiated processes (\ttW,\PW,\PZ) \\
    PDF ($\Pq\Pg$)                      & rate     & PDF uncertainty of $\Pq\Pg$ initiated processes (single t) \\[\cmsTabSkip]
    $\mu_{\mathrm{R}}$ scale (\ttbar)         & shape    & Renormalisation scale uncertainty of the \ttbar ME generator (\POWHEG), same for additional jet flavours   \\
    $\mu_{\mathrm{F}}$ scale (\ttbar)         & shape    & Factorisation scale uncertainty of the \ttbar ME generator (\POWHEG), same for additional jet flavours   \\
    PS scale: ISR (\ttbar)       & rate     & Initial state radiation uncertainty of the PS (for \ttbar events), jet multiplicity dependent rate uncertainty, independent for additional jet flavours   \\
    PS scale: FSR (\ttbar)       & rate     & Final state radiation uncertainty (for \ttbar events), jet multiplicity dependent rate uncertainty, independent for additional jet flavours   \\
    ME-PS matching (\ttbar)       & rate     & NLO ME to PS matching, {\it hdamp}~\cite{bib:CMS:2016kle} (for \ttbar events), jet multiplicity dependent rate uncertainty, independent for additional jet flavours  \\
    Underlying event  (\ttbar)   & rate     & Underlying event (for \ttbar events), jet multiplicity dependent rate uncertainty, independent for additional jet flavours\\
    NNPDF3.0NLO (\ttH, \ttbar) & shape & Based on the NNPDF replicas, same for \ttH and additional jet flavours \\[\cmsTabSkip]
    Bin-by-bin event count            & shape     & Statistical uncertainty of the signal and background prediction due to the limited sample size \\
    \hline
  \end{tabular}
\end{table}

The uncertainty in the integrated luminosity estimate is 2.5\%~\cite{CMS-PAS-LUM-17-001}.
The trigger efficiency in the single-lepton channel and the electron and muon identification efficiency uncertainties are estimated by comparing variations in measured efficiency between data and simulation using a high-purity sample of \PZ\ boson decays. In the dilepton channel, the trigger efficiency is measured in data with a method based on triggers that are uncorrelated with those used in the analysis, in particular based on \ptmiss requirements. These uncertainties are found to be small, typically below 1--2\%.
Effects of the uncertainty in the distribution of the number of pileup interactions are evaluated by varying the total inelastic cross section used to predict the number of pileup interactions in the simulated events by $\pm4.6\%$ from its nominal value~\cite{Aaboud:2016mmw}. The uncertainty due to the limited knowledge of the jet energy scale (resolution) is determined by variations of the energy scale (resolution) correction of all jets in the signal and background predictions by one standard deviation. In the case of the jet energy scale uncertainty, these variations are divided into 26 sources, which include uncertainties owing to the extrapolation between samples of different jet-flavour composition and the presence of pileup collisions in the derivation of the corrections~\cite{JESPUB:JME-13-004}. The effect of each source is evaluated individually.
The uncertainty of the \btagger
\cPqb\ tagging scale factors is evaluated by applying alternative scale
factors based on varying the following systematic effects~\cite{Sirunyan:2017ezt} by one
standard deviation, separately for the different jet flavours: the
contamination of background processes in the control samples, the jet
energy scale uncertainty---which is correlated with the overall jet
energy scale uncertainty---and the statistical uncertainty in the
scale factor evaluation. The impact of the statistical uncertainty is
parameterised as the sum of two contributions:
one term with linear dependence on the \cPqb\ tagging discriminant value, allowing an overall tilt of the discriminant distribution, and another term with quadratic dependence, allowing an overall shift of the discriminant distribution.

Theoretical uncertainties of the cross sections used to predict the
rates of various processes are propagated to the yield estimates.  All
rates are estimated using cross sections with at least NLO accuracy,
which have uncertainties arising primarily from PDFs and the choice of
factorisation and renormalisation scales (both in the ME and the PS).
The cross section uncertainties are each separated into their PDF and scale components (renorm./fact. scales)
and are correlated where appropriate between processes.
For example, the PDF uncertainties for background processes originating primarily from gluon-gluon initial states are treated as 100\% correlated.
The PDF uncertainty of the \ttH signal production is treated separately from the background processes.

The \ttbb process, and to lesser extent the \tttwob, \ttb, and \ttcc production, represent important sources of irreducible background.
Neither previous measurements of \tthf production~\cite{CMS:2014yxa,Aad:2015yja,Khachatryan:2015mva,Sirunyan:2017snr} nor higher-order theoretical calculations can currently constrain the normalisation of these contributions to better than 35\% accuracy~\cite{Bevilacqua:2017cru,Jezo:2018yaf}.
The shape of the final discriminant distributions as well as important input variable distributions of the sum of the \ttbb, \tttwob, and \ttb processes obtained with the nominal \ttbar simulation were compared to those obtained from a 4-flavour scheme \SHERPA~(v.2.2.2)~\cite{Gleisberg:2008ta} \ttbb simulation combined with \textsc{OpenLoops}~(v.1.3.1)~\cite{Cascioli:2011va}. The shapes agree within the statistical precision.
Therefore, an additional 50\% rate uncertainty is assigned to each of the \tthf processes to account also for differences in the phase space with respect to Ref.~\cite{Sirunyan:2017snr}.
Moreover, the robustness of the fit model was verified using simulated toy data, which were sampled from the templates of the fit model.
The background templates were modified in the following ways to sample the toy data:
increasing the normalisation of the \ttbb background template by 30\% in accordance with the results in Ref.~\cite{Sirunyan:2017snr} or replacing the sum of the templates of the \ttbb, \tttwob, and \ttb processes obtained with the nominal \ttbar simulation by those obtained from the 4-flavour scheme \SHERPA plus \textsc{OpenLoops} mentioned above.
In each case, a fit of the nominal model to the toy data is performed as described in Section~\ref{sec:results}, including the full set of systematic uncertainties.
The injected signal is recovered within a few percent, well within the uncertainties assigned to these processes.

The uncertainty arising from the missing higher-order terms in the simulation with \POWHEG of the \ttjets process at the ME level is assessed by varying the renormalisation and factorisation scales in the simulation up and down by factors of two with respect to the nominal values, using event weights obtained directly from the generator. At the PS level, the corresponding uncertainty is estimated by varying the parameters controlling the amount of initial- and final-state radiation independently by factors of 0.5 and 2~\cite{Skands:2014pea}. These sources of uncertainties are treated as uncorrelated. The uncertainty originating from the scheme used to match the ME level calculation to the PS simulation is derived by comparing the reference \ttjets simulation with two samples with varied {\it hdamp} parameter~\cite{bib:CMS:2016kle}, which controls the ME and PS matching and effectively regulates the high-\pt radiation.
The effect on the final discriminators owing to uncertainties in the underlying event tune of the \ttjets event generator are estimated using simulations with varied parameters with respect to those used to derive the CUETP8M2T4 tune in the default setup. The event count in the additional samples required to estimate the modelling uncertainties was small and induced changes to the discriminant distributions comparable in size to the statistical fluctuations of the additional samples.
For this reason, the uncertainties were estimated conservatively as the changes in the rates of the different \ttbar subprocesses independently for different jet multiplicities.
If the statistical uncertainty owing to the size of the simulated samples was larger than the rate change, the former was assigned as uncertainty.
The derived rate uncertainties were then correlated between jet multiplicities to account for migration effects and are treated as uncorrelated among the \ttbar subprocesses.
Possible shape variations of the final discriminant distributions due to the PDF uncertainty have been estimated by evaluating the PDF replicas provided with the NNPDF set~\cite{Ball:2014uwa}.
The impact of the mismodelling of the top quark \pt spectrum in the \ttbar simulation~\cite{Sirunyan:2018wem} was found to be negligible.

The impact of statistical fluctuations in the signal and background prediction due to the limited number of simulated events is accounted for using the Barlow--Beeston approach described in Refs.~\cite{Barlow:1993dm,Conway:2011in}.

\section{Results}
\label{sec:results}
\begin{figure}[hbtp]
  \centering
  \begin{tabular}{c@{\hskip 0.05\textwidth}c}
    \includegraphics[width=0.46\textwidth]{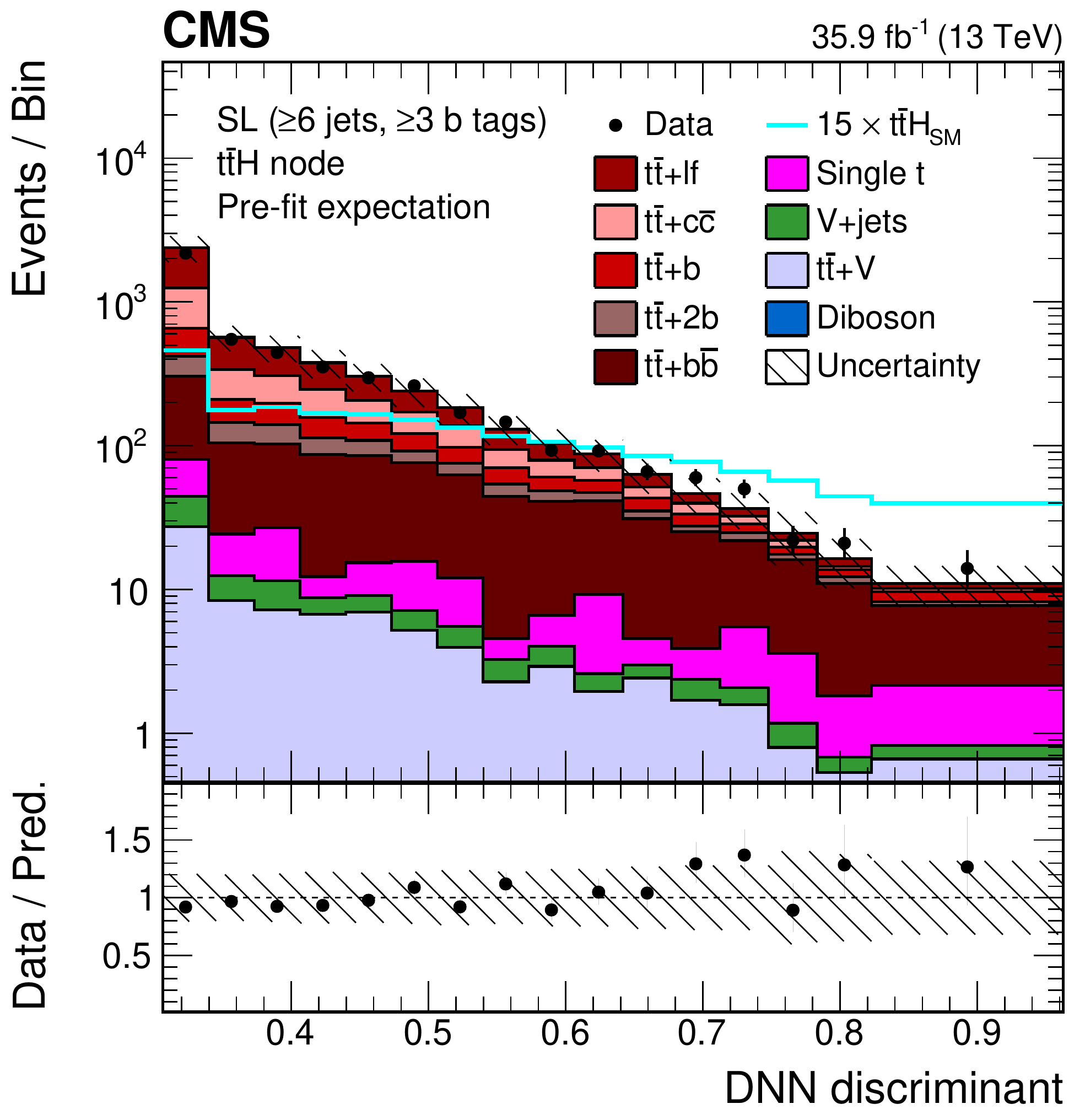} &
    \includegraphics[width=0.46\textwidth]{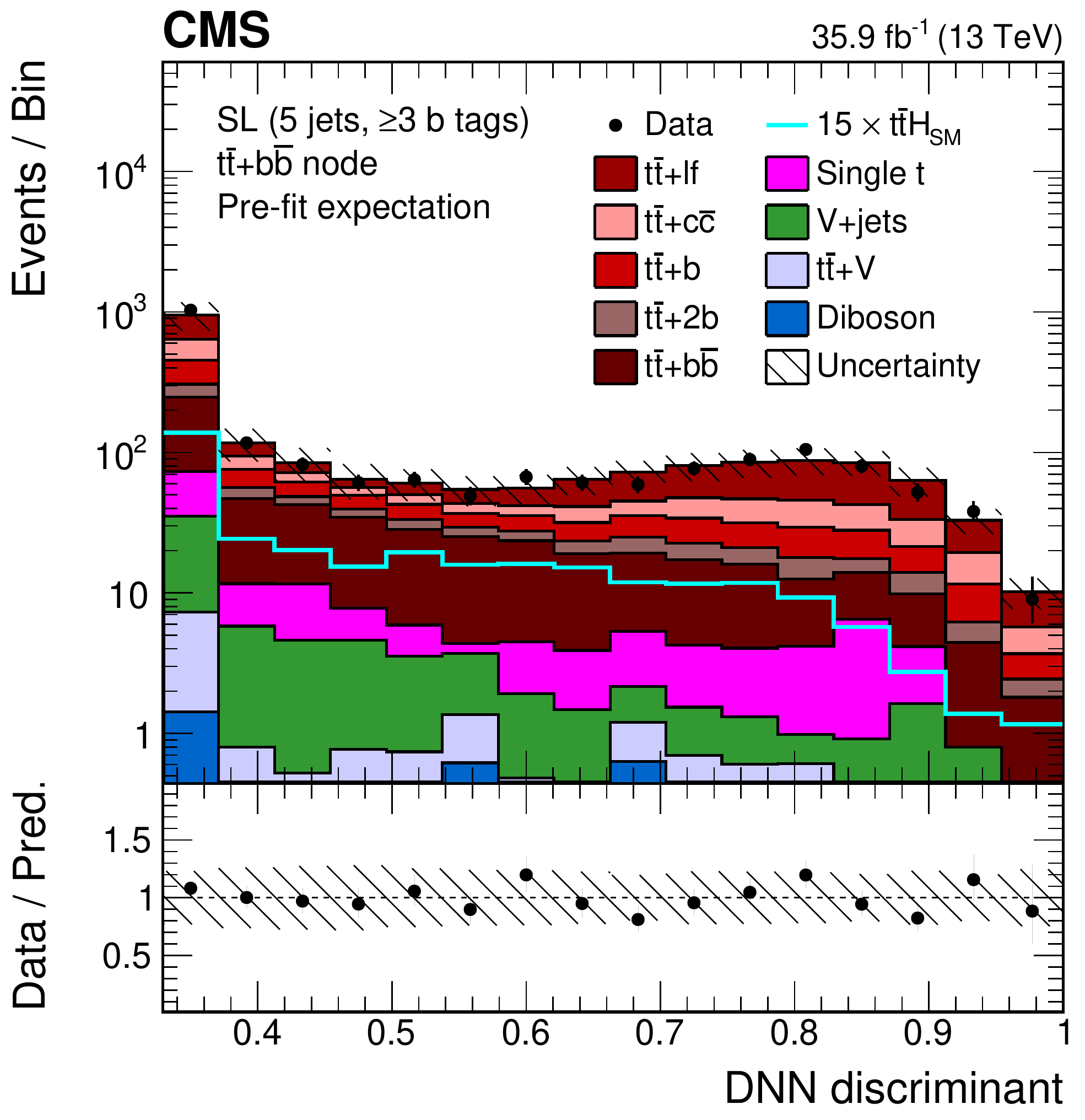}\\
    \includegraphics[width=0.46\textwidth]{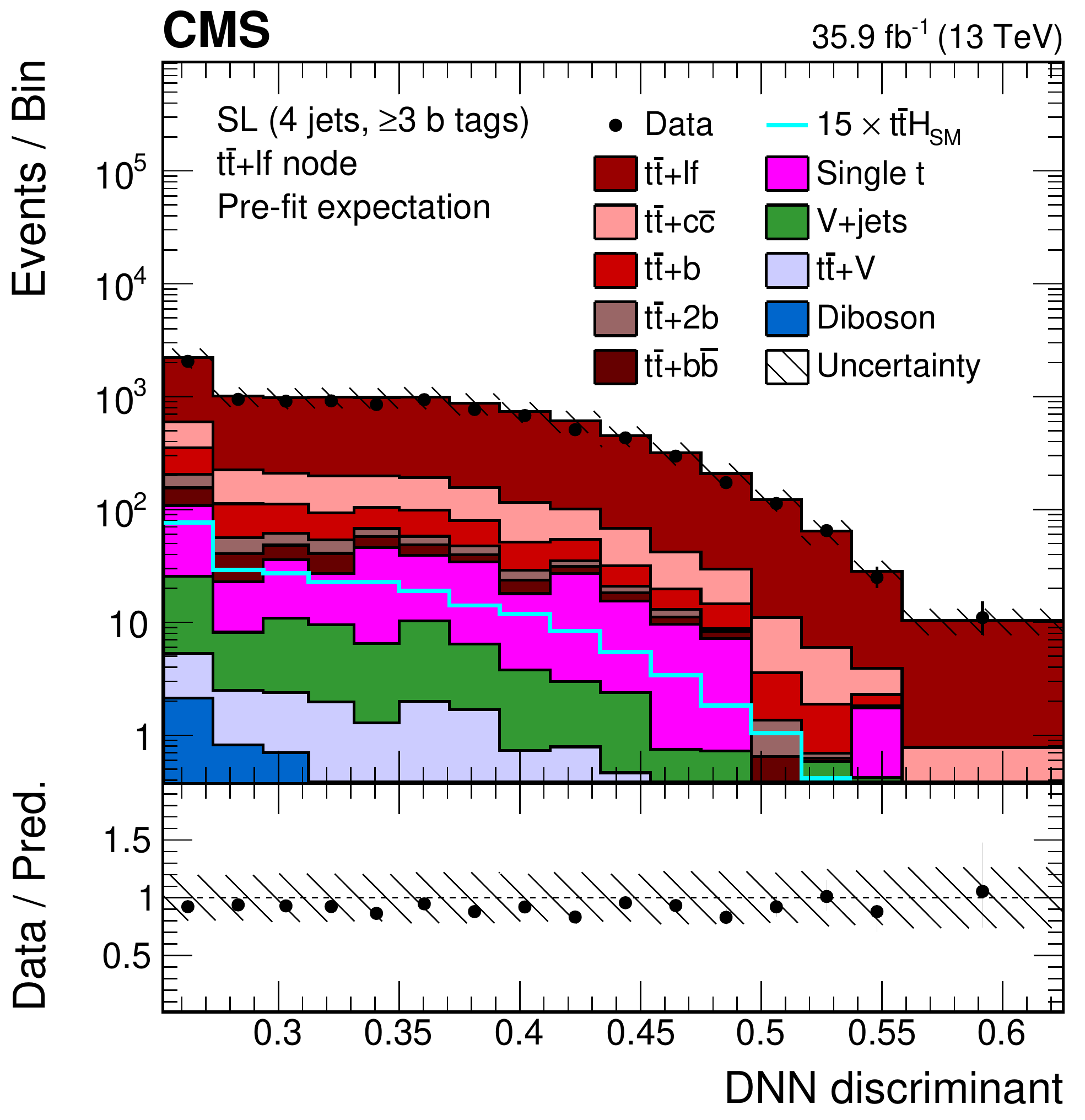} &
    \includegraphics[width=0.46\textwidth]{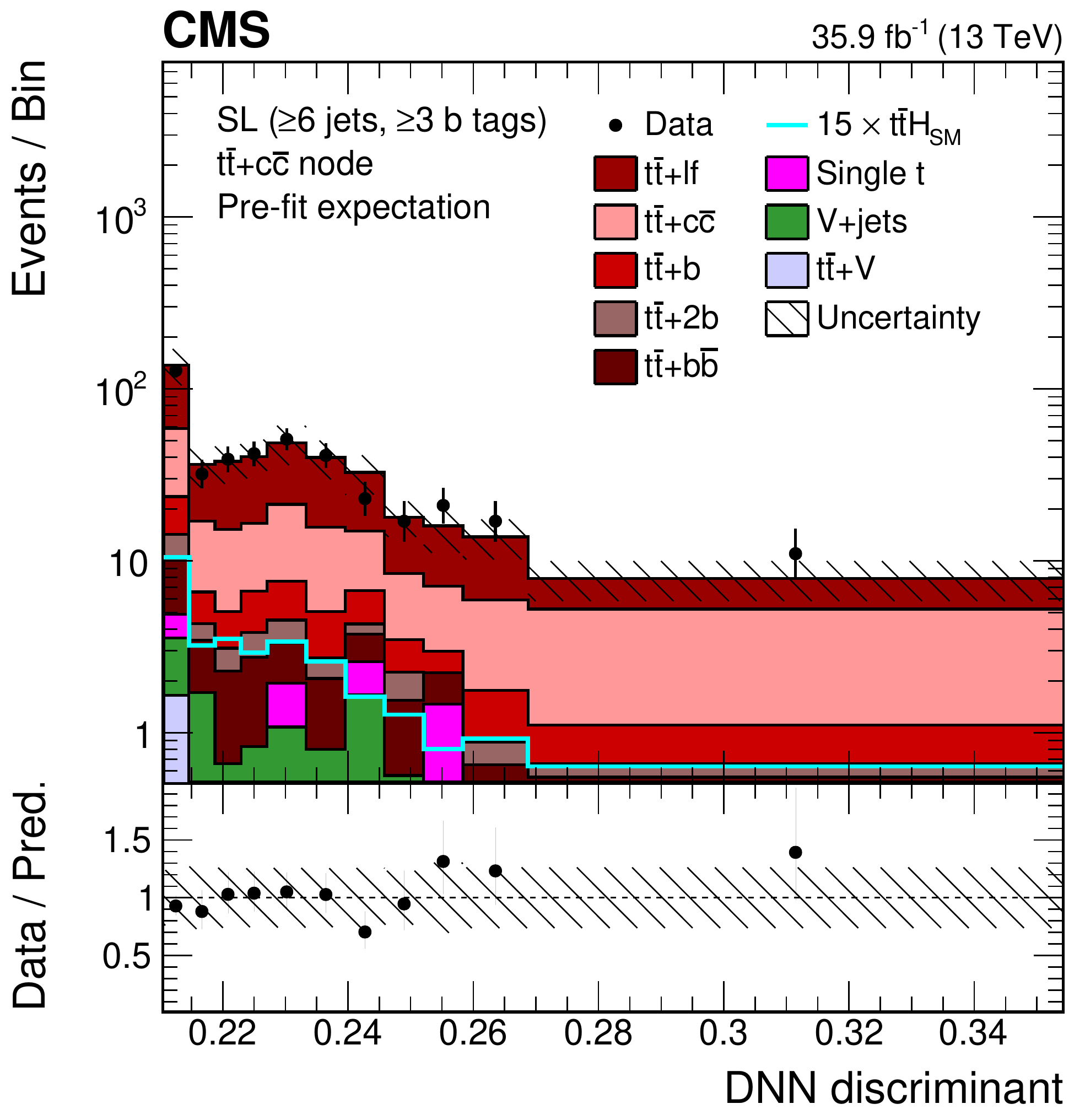}\\
  \end{tabular}
  \caption{Final discriminant shapes in the single-lepton (SL) channel before the fit to data:
    DNN discriminant in the jet-process categories with $\geq$6 jets-\ttH (upper \cmsLeft); 5 jets-\ttbb (upper \cmsRight); 4 jets-\ttlf (lower \cmsLeft); and $\geq$6 jets-\ttcc (lower \cmsRight).
    The expected background contributions (filled histograms) are stacked, and the expected signal
    distribution (line), which includes \Hbb and all other Higgs boson decay modes, is superimposed.
    Each contribution
    is normalised to an integrated luminosity of \lumivalue, and the
    signal distribution is additionally scaled by a factor of 15
    for better visibility.
    The hatched uncertainty bands include the total uncertainty of the fit model.
    The distributions observed in data (markers) are overlayed.
    The first and the last bins include underflow and overflow events, respectively.
    The lower plots show the ratio of the data to the background prediction.
  }
  \label{fig:ljdiscriminants_1}
\end{figure}

\begin{figure}[hbtp]
  \centering
  \includegraphics[width=0.46\textwidth]{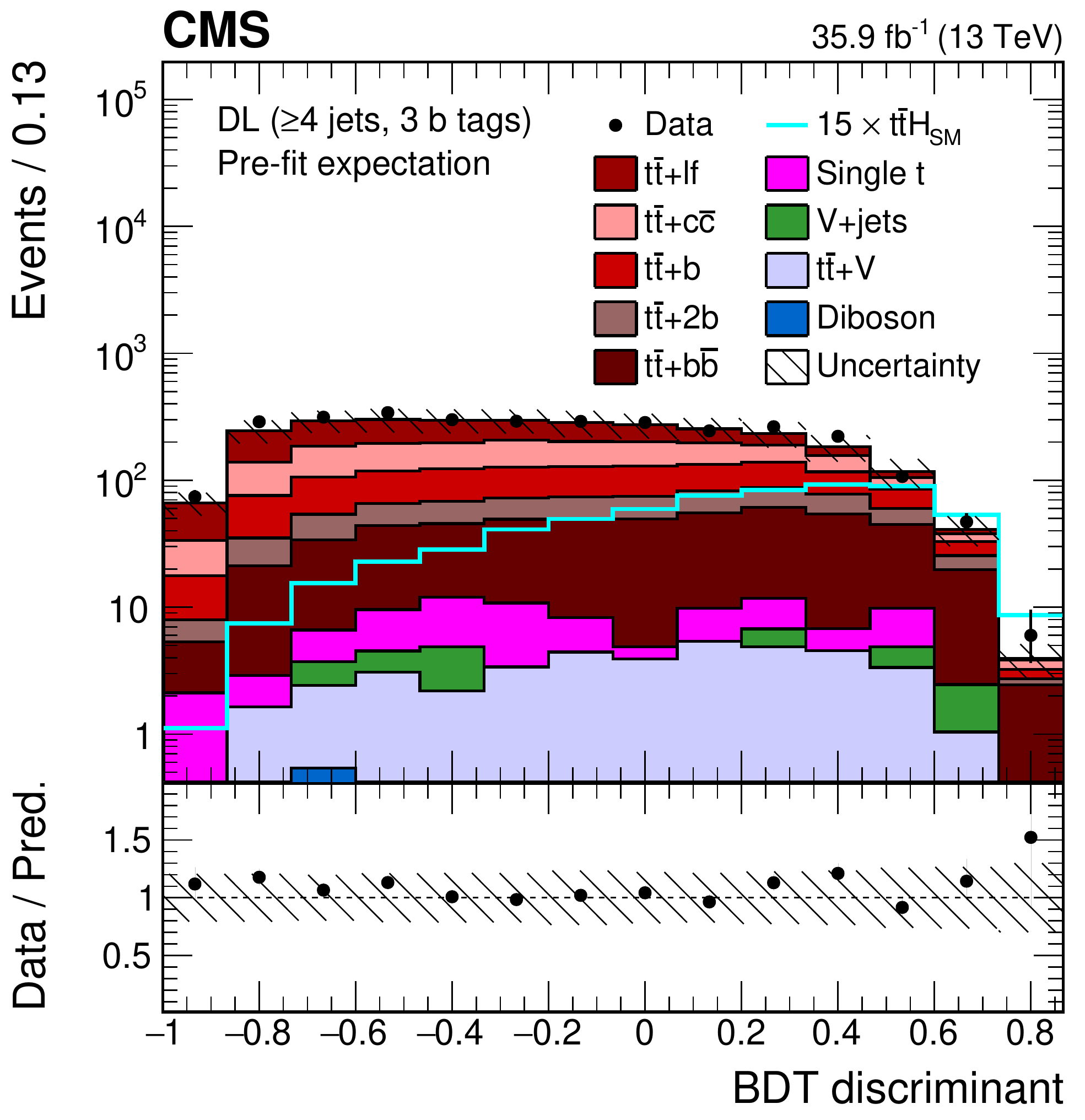} \\
  \begin{tabular}{c@{\hskip 0.05\textwidth}c}
    \includegraphics[width=0.46\textwidth]{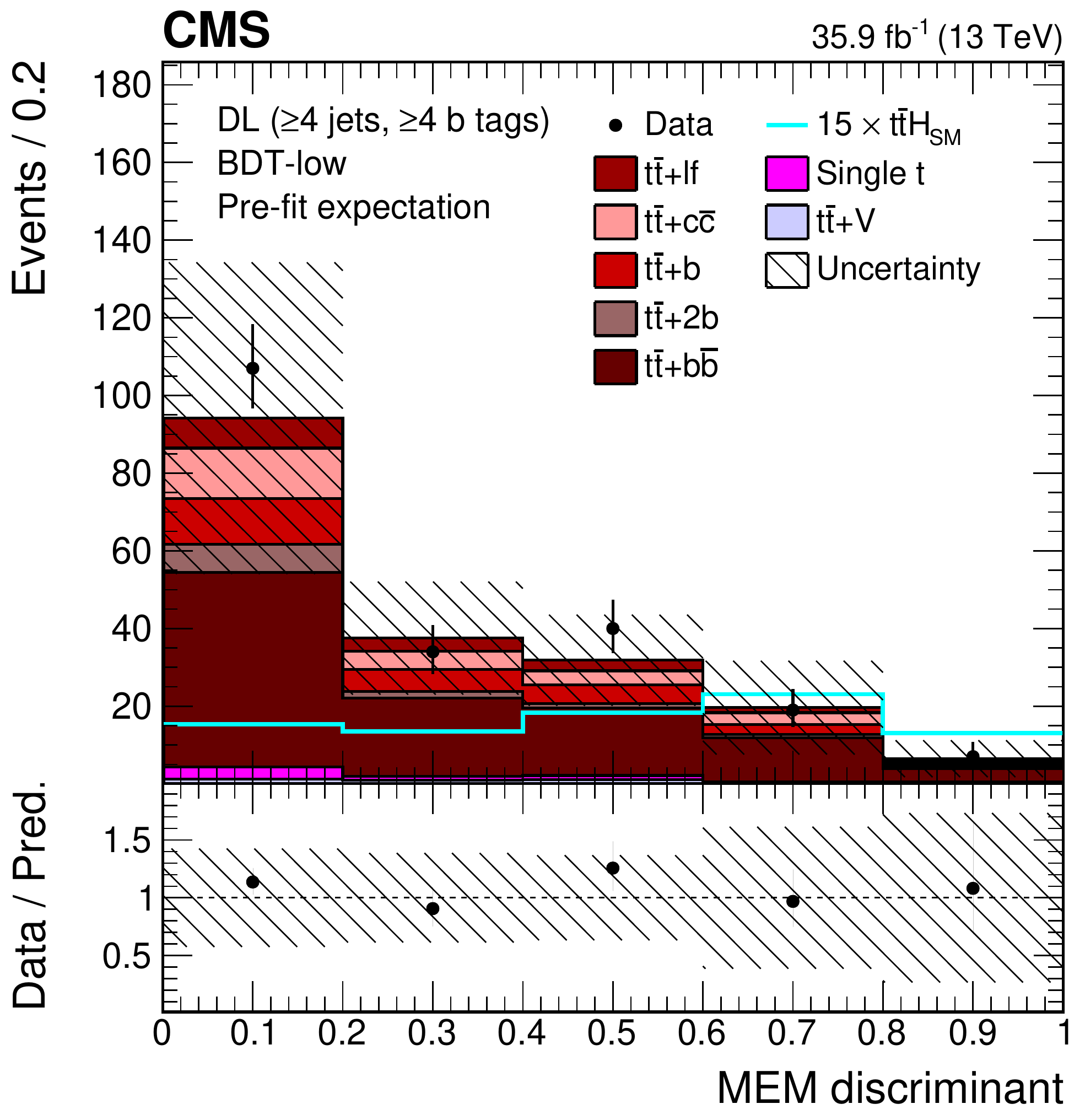} &
    \includegraphics[width=0.46\textwidth]{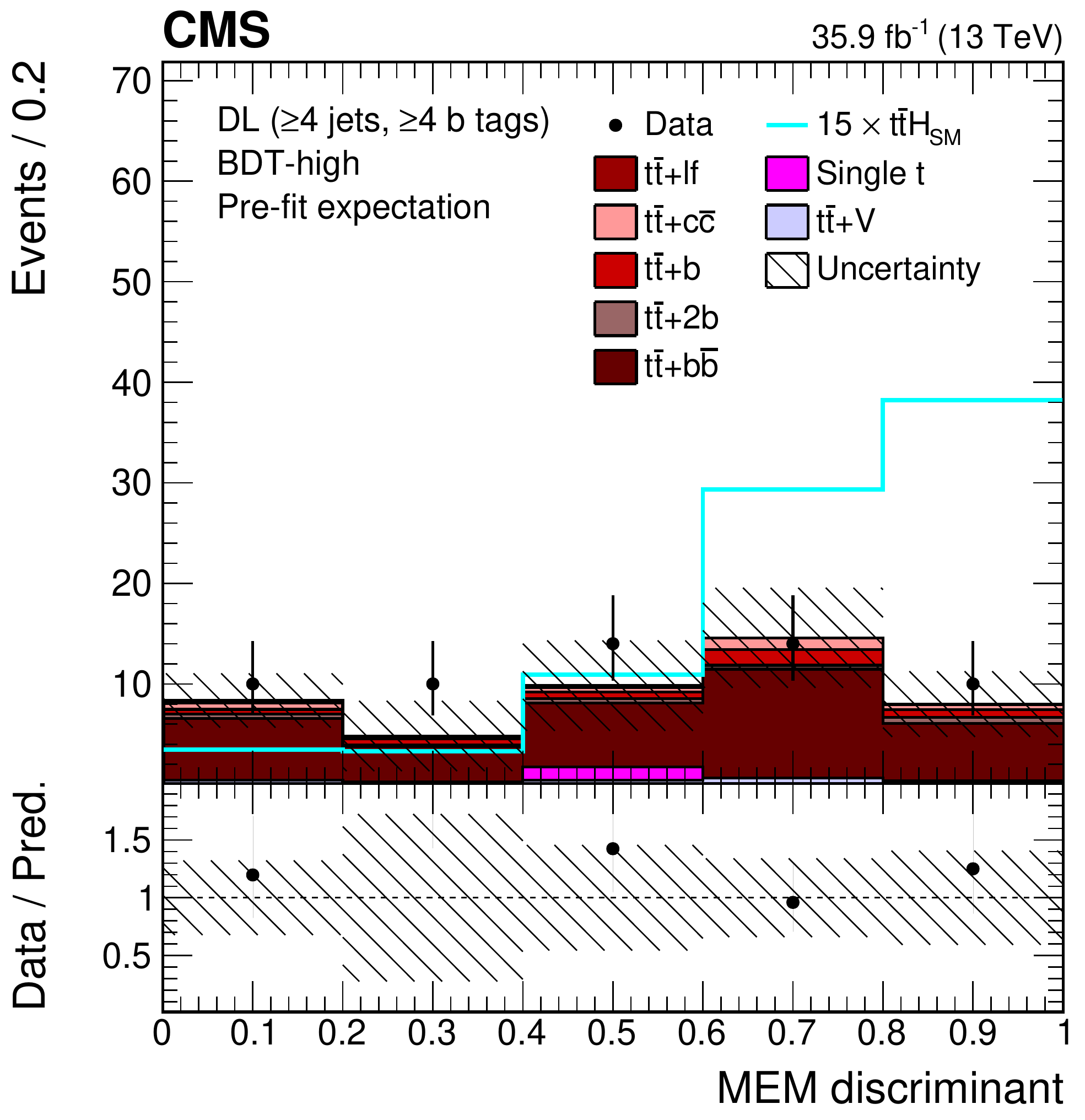} \\
  \end{tabular}
  \caption{
    Final discriminant shapes in the dilepton (DL) channel before the fit to data:
    BDT discriminant in the analysis category with \dlFourThree (upper row) and MEM discriminant in the analysis categories with \dlFourFour (lower row) with low (\cmsLeft) and high (\cmsRight) BDT output.
    The expected background contributions (filled histograms) are stacked, and the expected signal
    distribution (line), which includes \Hbb and all other Higgs boson decay modes, is superimposed.
    Each contribution
    is normalised to an integrated luminosity of \lumivalue, and the
    signal distribution is additionally scaled by a factor of 15
    for better visibility.
    The hatched uncertainty bands include the total uncertainty of the fit model.
    The distributions observed in data (markers) are overlayed.
    The first and the last bins include underflow and overflow events, respectively.
    The lower plots show the ratio of the data to the background prediction.
  }
  \label{fig:dildiscriminants_1}
\end{figure}

\begin{figure}[hbtp]
  \centering
  \begin{tabular}{c@{\hskip 0.05\textwidth}c}
    \includegraphics[width=0.46\textwidth]{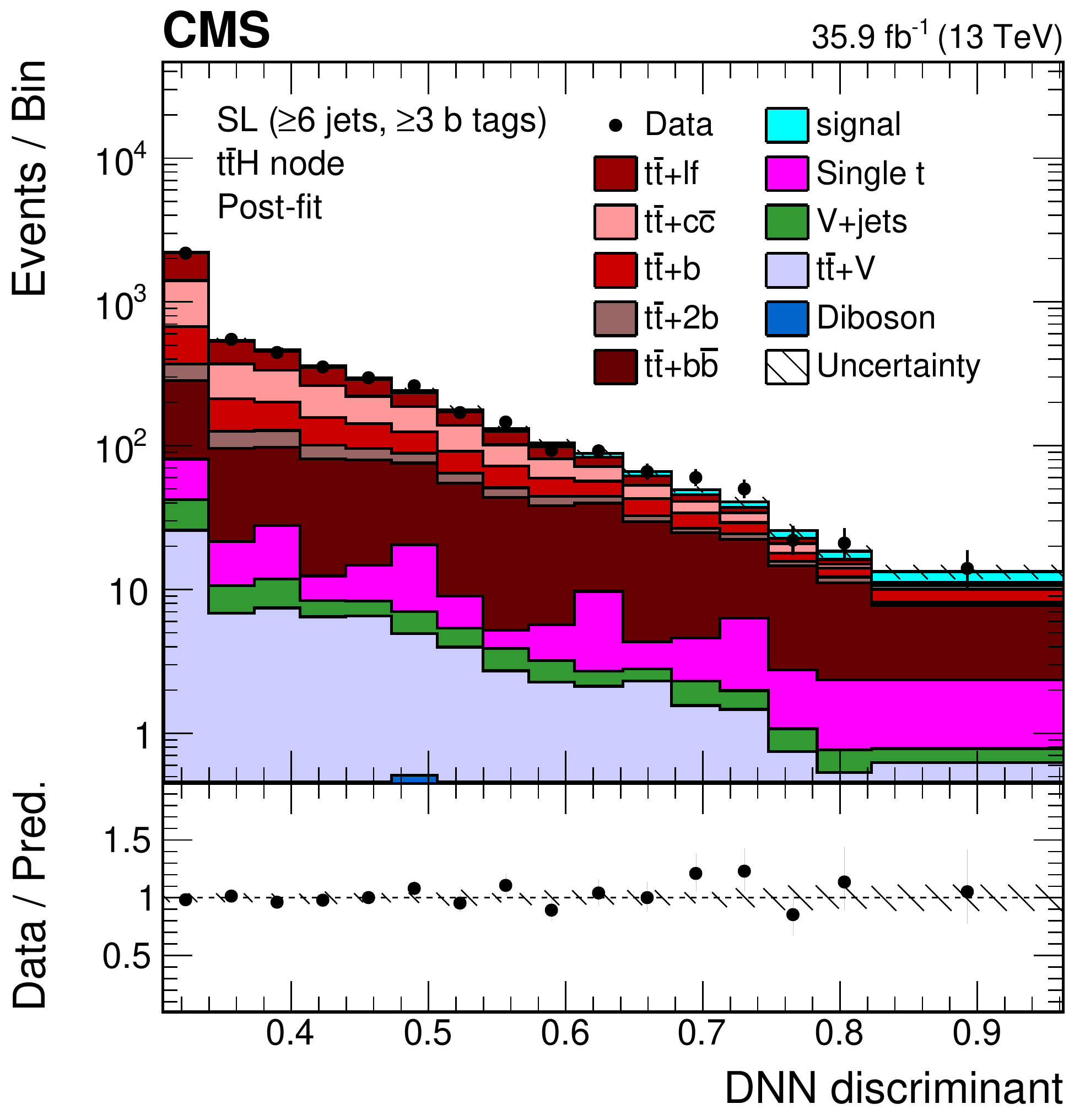} &
    \includegraphics[width=0.46\textwidth]{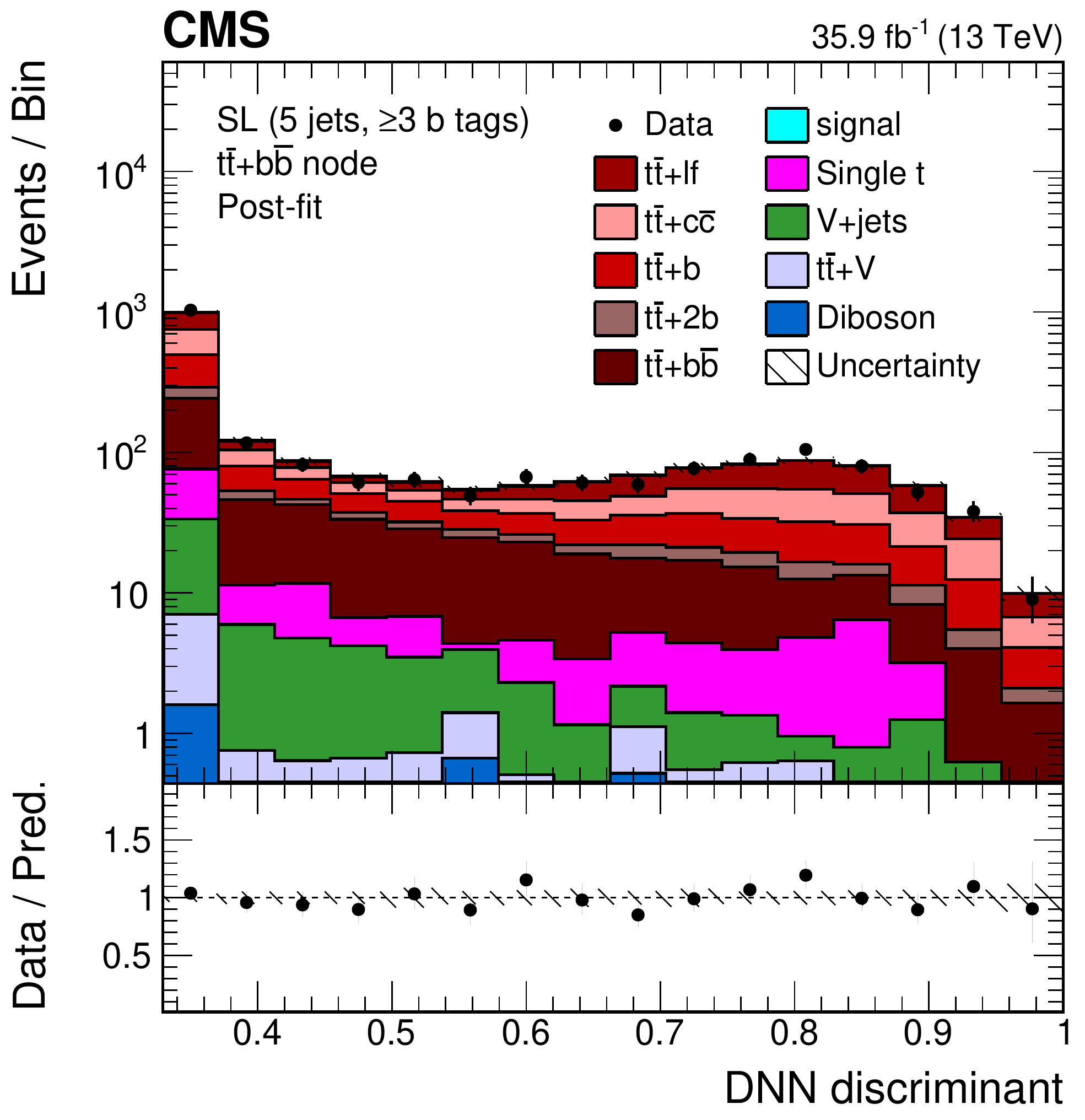}\\
    \includegraphics[width=0.46\textwidth]{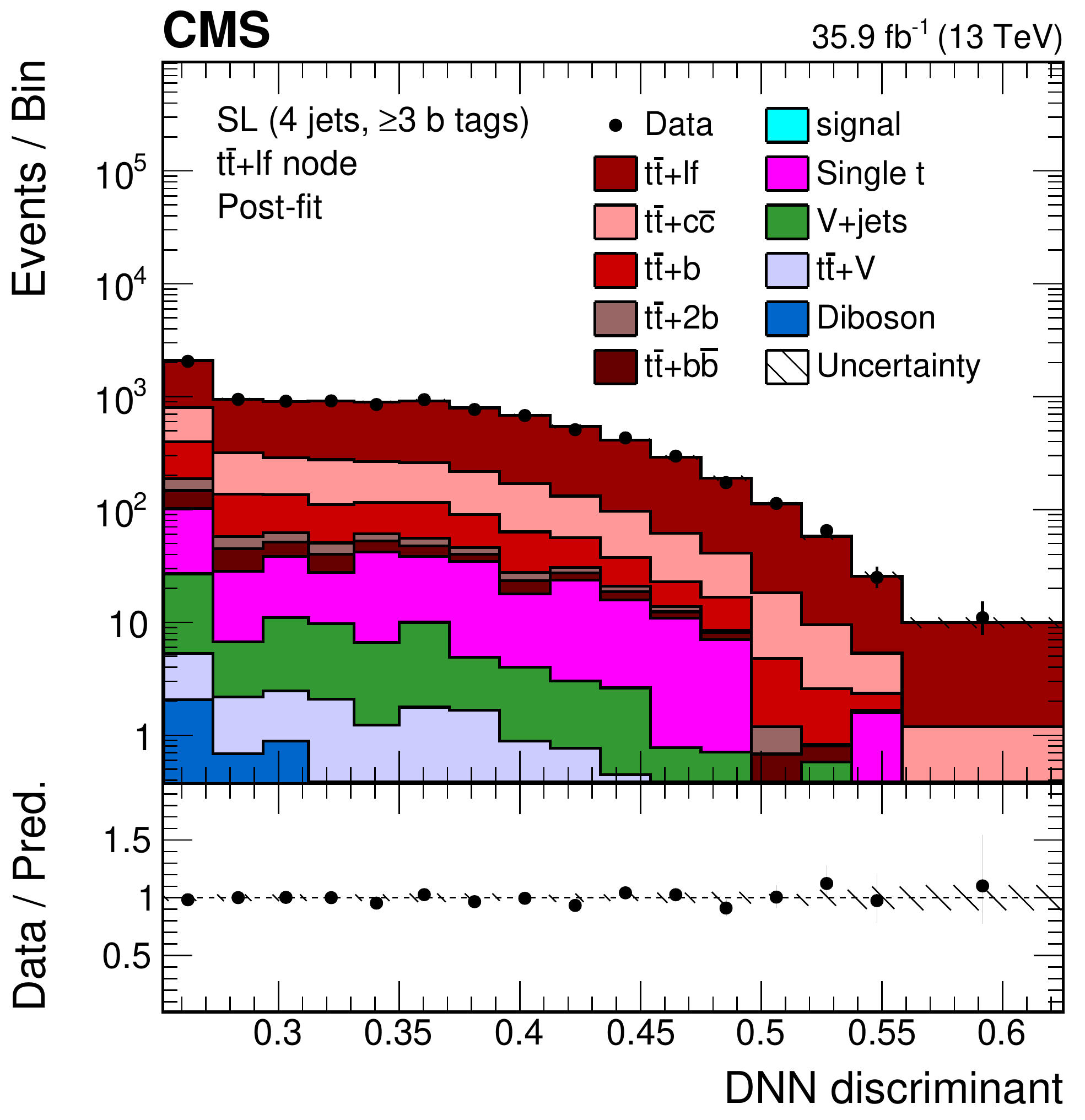} &
    \includegraphics[width=0.46\textwidth]{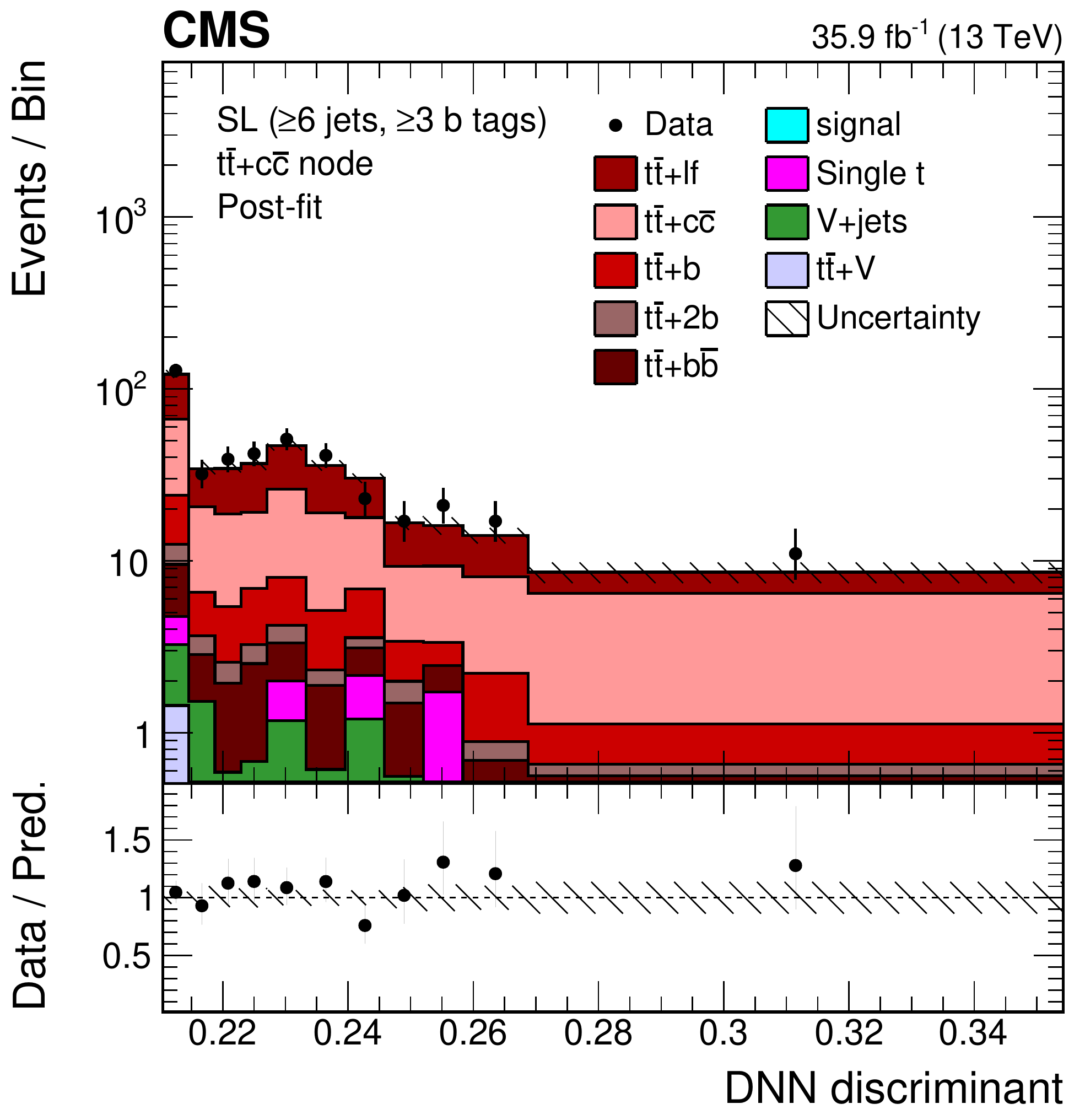}\\
  \end{tabular}
  \caption{
    Final discriminant shapes in the single-lepton (SL) channel after the fit to data:
    DNN discriminant in the jet-process categories with $\geq$6 jets-\ttH (upper \cmsLeft); 5 jets-\ttbb (upper \cmsRight); 4 jets-\ttlf (lower \cmsLeft); and $\geq$6 jets-\ttcc (lower \cmsRight).
    The hatched uncertainty bands include the total uncertainty after the fit to data.
    The distributions observed in data (markers) are overlayed.
    The first and the last bins include underflow and overflow events, respectively.
    The lower plots show the ratio of the data to the post-fit background plus signal distribution.
  }
  \label{fig:ljdiscriminantsPostFit_1}
\end{figure}

\begin{figure}[hbtp]
  \centering
  \includegraphics[width=0.46\textwidth]{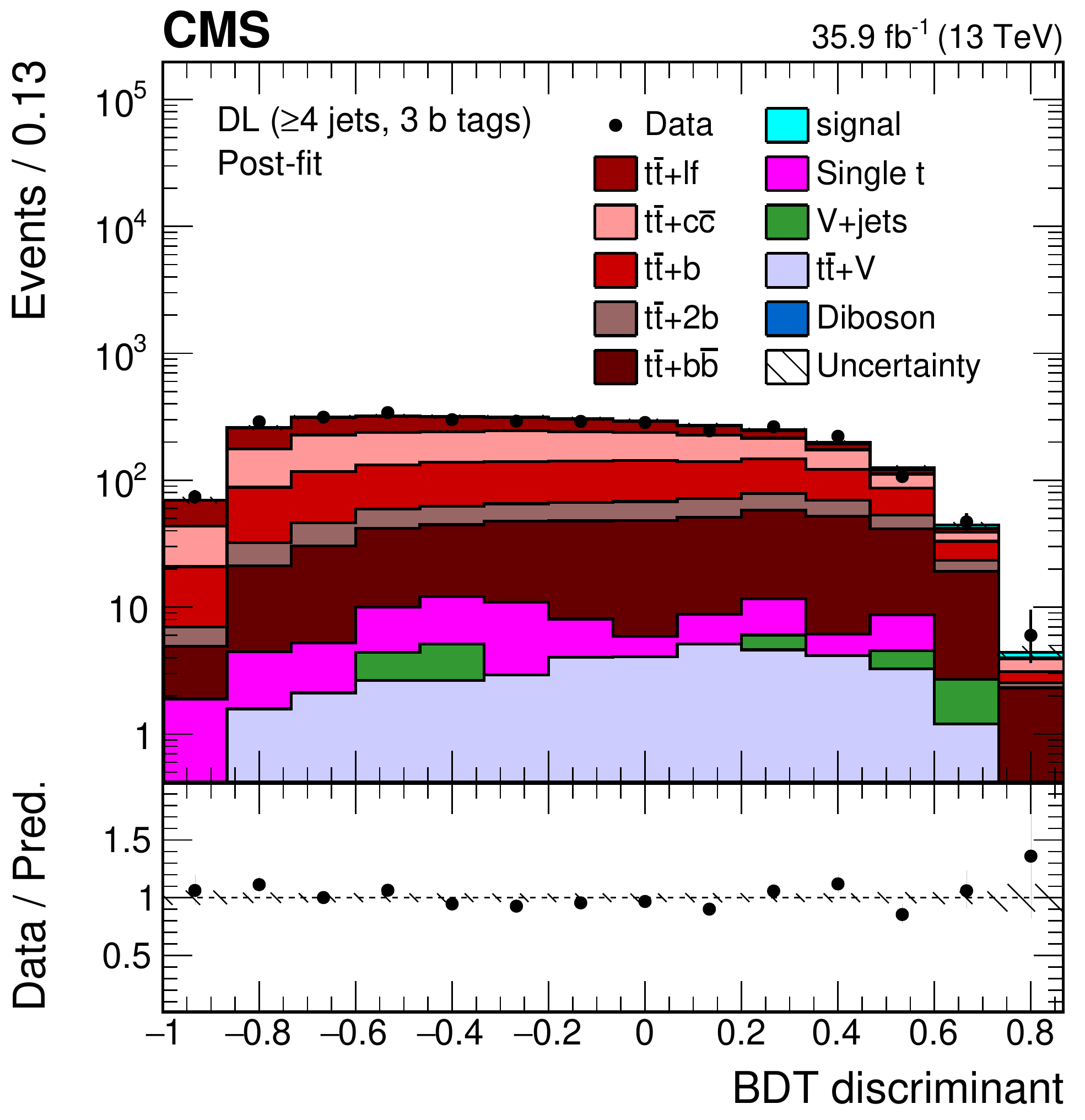} \\
  \begin{tabular}{c@{\hskip 0.05\textwidth}c}
    \includegraphics[width=0.46\textwidth]{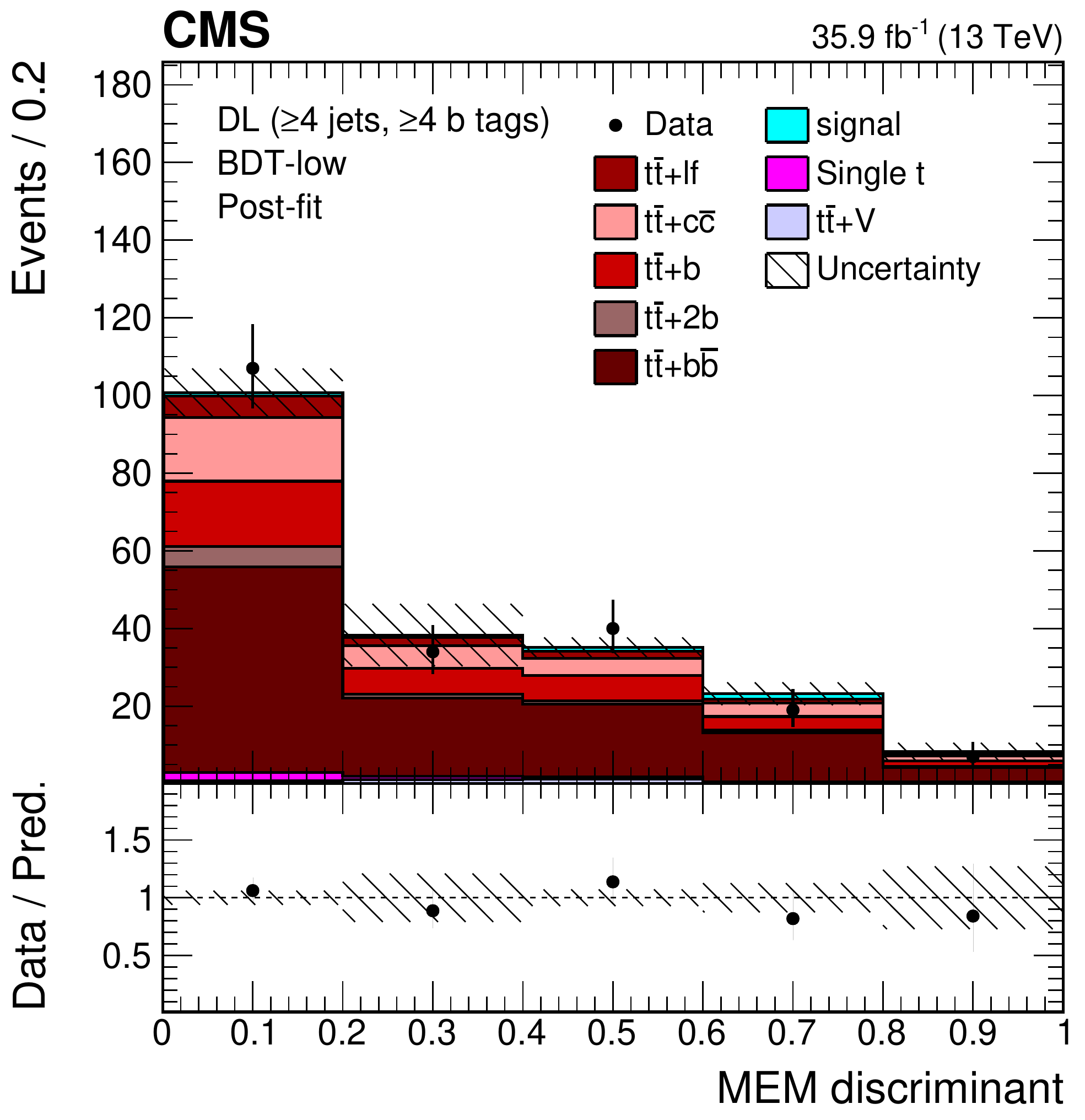} &
    \includegraphics[width=0.46\textwidth]{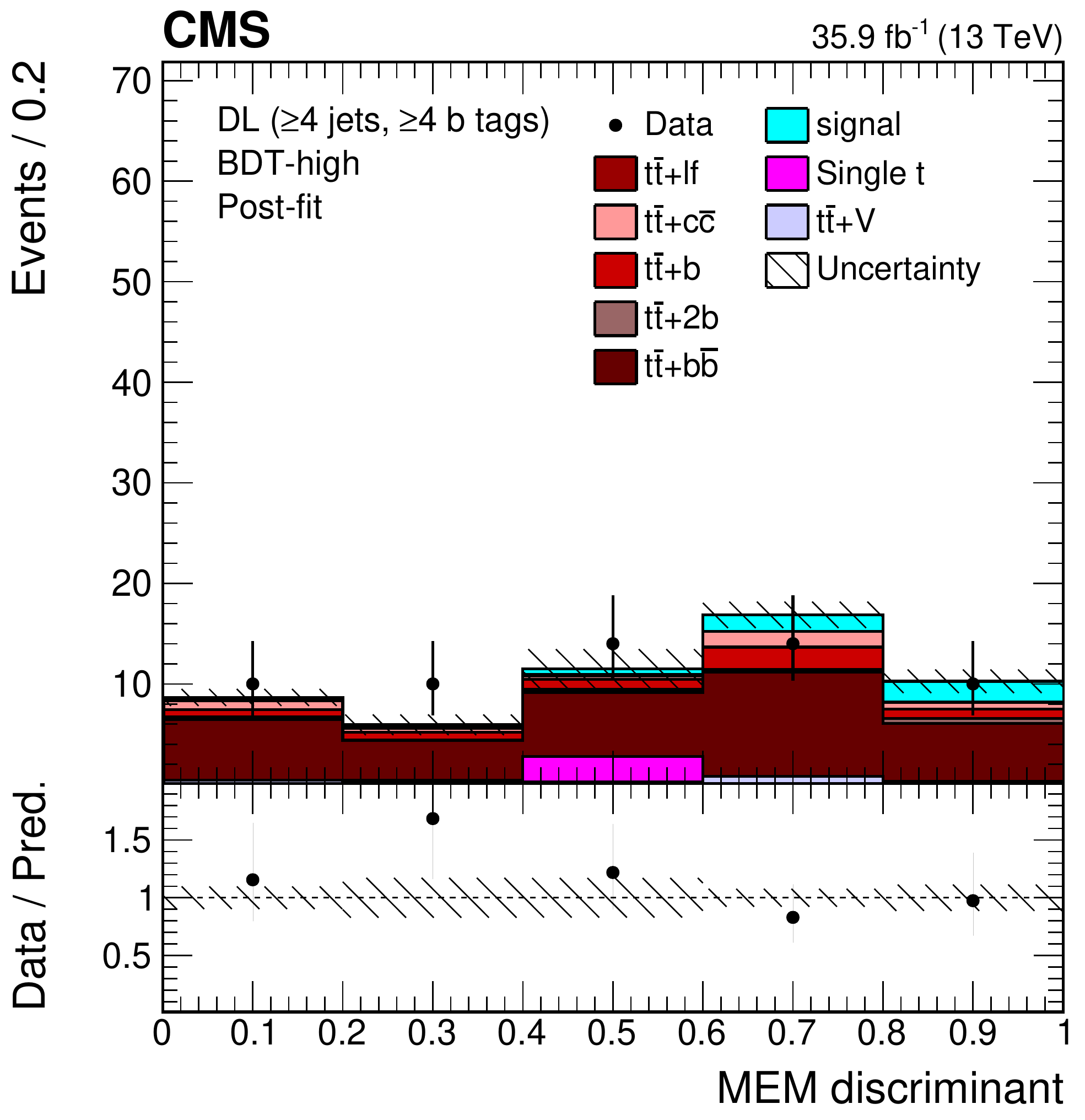} \\
  \end{tabular}
  \caption{
    Final discriminant shapes in the dilepton (DL) channel after the fit to data:
    BDT discriminant in the analysis category with \dlFourThree (upper row) and MEM discriminant in the analysis categories with \dlFourFour (lower row) with low (\cmsLeft) and high (\cmsRight) BDT output.
    The hatched uncertainty bands include the total uncertainty after the fit to data.
    The distributions observed in data (markers) are overlayed.
    The first and the last bins include underflow and overflow events, respectively.
    The lower plots show the ratio of the data to the post-fit background plus signal distribution.
  }
  \label{fig:dildiscriminantsPostFit_1}
\end{figure}

The numbers of events selected in the jet-process categories of the single-lepton channel and in the jet-tag categories of the dilepton channel, before and after the fit of the signal strength modifier and the nuisance parameters, are listed in Tables~\ref{tab:dnn:yields-jet-process-categories:sl4}--\ref{tab:bdtmem:yields-jet-tag-categories:dl}.
The final discriminants in some example categories in the single-lepton channel and the three dilepton categories before and after the fit to data are displayed in Figs.~\ref{fig:ljdiscriminants_1}--\ref{fig:dildiscriminants_1} and Figs.~\ref{fig:ljdiscriminantsPostFit_1}--\ref{fig:dildiscriminantsPostFit_1}, respectively.
All final discriminants in the single-lepton channel before and after the fit to data are displayed in Appendices~\ref{sec:appendix:prefitshapes} and~\ref{sec:appendix:postfitshapes}.

\begin{table}[!hbtp]
  \centering
  \topcaption{
    Observed and expected event yields per jet-process category (node) in the single-lepton channel with 4 jets and at least 3 \PQb tags, prior to the fit to data (after the fit to data).
    The quoted uncertainties denote the total statistical and systematic components.
  }
  \label{tab:dnn:yields-jet-process-categories:sl4}
  \cmsTable{
    \begin{tabular}{l r@{\;}r r@{\;}r r@{\;}r r@{\;}r r@{\;}r r@{\;}r }
      \hline
      & \multicolumn{ 12 }{c}{pre-fit (post-fit) yields}
      \\
      Process  & \multicolumn{2}{c}{ \ttH  node}  & \multicolumn{2}{c}{ \ttbb  node}  & \multicolumn{2}{c}{ \tttwob  node}  & \multicolumn{2}{c}{ \ttb  node}  & \multicolumn{2}{c}{ \ttcc  node}  & \multicolumn{2}{c}{ \ttlf  node}  \\
      \hline
      \ttlf      & $  1249 $ & ($   962 $) & $   727 $ & ($   572 $) & $  1401 $ & ($  1090 $) & $  1035 $ & ($   823 $) & $  2909 $ & ($  2296 $) & $  8463 $ & ($  6829 $)  \\
      \ttcc      & $   298 $ & ($   458 $) & $   232 $ & ($   359 $) & $   428 $ & ($   678 $) & $   251 $ & ($   400 $) & $   686 $ & ($  1068 $) & $  1022 $ & ($  1652 $)  \\
      \ttb       & $   253 $ & ($   356 $) & $   215 $ & ($   311 $) & $   370 $ & ($   530 $) & $   326 $ & ($   484 $) & $   308 $ & ($   437 $) & $   469 $ & ($   683 $)  \\
      \tttwob    & $   124 $ & ($    96 $) & $    77 $ & ($    62 $) & $   317 $ & ($   254 $) & $    90 $ & ($    73 $) & $   100 $ & ($    79 $) & $   134 $ & ($   108 $)  \\
      \ttbb      & $   139 $ & ($   137 $) & $   191 $ & ($   192 $) & $   149 $ & ($   140 $) & $   105 $ & ($   103 $) & $   119 $ & ($   114 $) & $   133 $ & ($   128 $)  \\
      \singlet   & $    96 $ & ($    96 $) & $   117 $ & ($   109 $) & $   167 $ & ($   162 $) & $    93 $ & ($    96 $) & $   231 $ & ($   232 $) & $   304 $ & ($   307 $)  \\
      \Vjets     & $    37 $ & ($    37 $) & $    76 $ & ($    74 $) & $    48 $ & ($    46 $) & $    27 $ & ($    27 $) & $    97 $ & ($    89 $) & $    69 $ & ($    69 $)  \\
      \ttV       & $    13 $ & ($    13 $) & $     6 $ & ($     6 $) & $    12 $ & ($    11 $) & $     6 $ & ($     6 $) & $    10 $ & ($    10 $) & $    16 $ & ($    16 $)  \\
      \diboson   & $     4 $ & ($     4 $) & $     5 $ & ($     5 $) & $   0.9 $ & ($   0.8 $) & $   0.6 $ & ($   0.7 $) & $     2 $ & ($     2 $) & $     4 $ & ($     4 $)  \\[\cmsTabSkip]
      Total\;bkg. & $  2213 $ & ($  2158 $) & $  1645 $ & ($  1688 $) & $  2892 $ & ($  2911 $) & $  1935 $ & ($  2012 $) & $  4462 $ & ($  4328 $) & $ 10614 $ & ($  9795 $)  \\
      $\pm$ tot unc. & $\pm 508 $ & ($\pm  58 $) & $\pm 415 $ & ($\pm  53 $) & $\pm 588 $ & ($\pm  89 $) & $\pm 402 $ & ($\pm  67 $) & $\pm 1051 $ & ($\pm 120 $) & $\pm 2359 $ & ($\pm 270 $)  \\[\cmsTabSkip]
      \ttH       & $    27 $ & ($    21 $) & $     9 $ & ($     7 $) & $    16 $ & ($    12 $) & $     7 $ & ($     5 $) & $     9 $ & ($     7 $) & $    16 $ & ($    13 $)  \\
      $\pm$ tot unc. & $\pm   4 $ & ($\pm   3 $) & $\pm   1 $ & ($\pm   1 $) & $\pm   2 $ & ($\pm   2 $) & $\pm   1 $ & ($\pm   1 $) & $\pm   1 $ & ($\pm   1 $) & $\pm   2 $ & ($\pm   2 $)  \\[\cmsTabSkip]
      Data       & \multicolumn{2}{c}{$  2125 $} & \multicolumn{2}{c}{$  1793 $} & \multicolumn{2}{c}{$  2896 $} & \multicolumn{2}{c}{$  2027 $} & \multicolumn{2}{c}{$  4366 $} & \multicolumn{2}{c}{$  9693 $}  \\
      \hline
    \end{tabular}
}
\end{table}

\begin{table}[!hbtp]
  \centering
  \topcaption{
    Observed and expected event yields per jet-process category (node) in the single-lepton channel with 5 jets and at least 3 \cPqb\ tags, prior to the fit to data (after the fit to data).
    The quoted uncertainties denote the total statistical and systematic uncertainty.
  }
  \label{tab:dnn:yields-jet-process-categories:sl5}
  \cmsTable{
    \begin{tabular}{l r@{\;}r r@{\;}r r@{\;}r r@{\;}r r@{\;}r r@{\;}r }
      \hline
      & \multicolumn{ 12 }{c}{pre-fit (post-fit) yields}
      \\
      Process  & \multicolumn{2}{c}{ \ttH  node}  & \multicolumn{2}{c}{ \ttbb  node}  & \multicolumn{2}{c}{ \tttwob  node}  & \multicolumn{2}{c}{ \ttb  node}  & \multicolumn{2}{c}{ \ttcc  node}  & \multicolumn{2}{c}{ \ttlf  node}  \\
      \hline
      \ttlf      & $   785 $ & ($   570 $) & $   647 $ & ($   467 $) & $   830 $ & ($   604 $) & $   683 $ & ($   525 $) & $  1148 $ & ($   848 $) & $  4903 $ & ($  3697 $)  \\
      \ttcc      & $   336 $ & ($   455 $) & $   341 $ & ($   469 $) & $   445 $ & ($   633 $) & $   264 $ & ($   382 $) & $   552 $ & ($   756 $) & $  1207 $ & ($  1726 $)  \\
      \ttb       & $   257 $ & ($   351 $) & $   290 $ & ($   399 $) & $   355 $ & ($   494 $) & $   321 $ & ($   477 $) & $   219 $ & ($   301 $) & $   494 $ & ($   692 $)  \\
      \tttwob    & $   136 $ & ($   104 $) & $   128 $ & ($    99 $) & $   324 $ & ($   253 $) & $    89 $ & ($    73 $) & $    85 $ & ($    65 $) & $   184 $ & ($   143 $)  \\
      \ttbb      & $   266 $ & ($   251 $) & $   410 $ & ($   397 $) & $   224 $ & ($   207 $) & $   150 $ & ($   143 $) & $   144 $ & ($   132 $) & $   228 $ & ($   212 $)  \\
      \singlet   & $    62 $ & ($    63 $) & $    82 $ & ($    84 $) & $    98 $ & ($    96 $) & $    45 $ & ($    58 $) & $   114 $ & ($   113 $) & $   189 $ & ($   193 $)  \\
      \Vjets     & $    25 $ & ($    23 $) & $    54 $ & ($    53 $) & $    34 $ & ($    31 $) & $    11 $ & ($    12 $) & $    46 $ & ($    41 $) & $    54 $ & ($    51 $)  \\
      \ttV       & $    20 $ & ($    20 $) & $    14 $ & ($    13 $) & $    17 $ & ($    16 $) & $     7 $ & ($     7 $) & $    11 $ & ($    10 $) & $    25 $ & ($    24 $)  \\
      \diboson   & $     1 $ & ($     1 $) & $     3 $ & ($     3 $) & $   0.4 $ & ($   0.4 $) &      \NA  & (\NA) & $   0.6 $ & ($   0.4 $) & $     3 $ & ($     3 $)  \\[\cmsTabSkip]
      Total\;bkg. & $  1889 $ & ($  1838 $) & $  1969 $ & ($  1985 $) & $  2326 $ & ($  2332 $) & $  1570 $ & ($  1676 $) & $  2320 $ & ($  2268 $) & $  7287 $ & ($  6742 $)  \\
      $\pm$ tot unc. & $\pm 459 $ & ($\pm  57 $) & $\pm 485 $ & ($\pm  70 $) & $\pm 489 $ & ($\pm  71 $) & $\pm 334 $ & ($\pm  47 $) & $\pm 597 $ & ($\pm  79 $) & $\pm 1655 $ & ($\pm 219 $)  \\[\cmsTabSkip]
      \ttH       & $    53 $ & ($    41 $) & $    21 $ & ($    17 $) & $    20 $ & ($    15 $) & $     8 $ & ($     6 $) & $    11 $ & ($     8 $) & $    28 $ & ($    22 $)  \\
      $\pm$ tot unc. & $\pm   7 $ & ($\pm   6 $) & $\pm   3 $ & ($\pm   3 $) & $\pm   2 $ & ($\pm   2 $) & $\pm   1 $ & ($\pm   1 $) & $\pm   1 $ & ($\pm   1 $) & $\pm   3 $ & ($\pm   3 $)  \\[\cmsTabSkip]
      Data       & \multicolumn{2}{c}{$  1848 $} & \multicolumn{2}{c}{$  2040 $} & \multicolumn{2}{c}{$  2299 $} & \multicolumn{2}{c}{$  1690 $} & \multicolumn{2}{c}{$  2302 $} & \multicolumn{2}{c}{$  6918 $}  \\
      \hline
    \end{tabular}
  }
\end{table}

\begin{table}[!hbtp]
  \centering
  \topcaption{
    Observed and expected event yields per jet-process category (node) in the single-lepton channel with at least 6 jets and at least 3 \cPqb\ tags, prior to the fit to data (after the fit to data).
    The quoted uncertainties denote the total statistical and systematic uncertainty.
  }
  \label{tab:dnn:yields-jet-process-categories:sl6}
  \cmsTable{
    \begin{tabular}{l r@{\;}r r@{\;}r r@{\;}r r@{\;}r r@{\;}r r@{\;}r }
      \hline
      & \multicolumn{ 12 }{c}{pre-fit (post-fit) yields}
      \\
      Process  & \multicolumn{2}{c}{ \ttH  node}  & \multicolumn{2}{c}{ \ttbb  node}  & \multicolumn{2}{c}{ \tttwob  node}  & \multicolumn{2}{c}{ \ttb  node}  & \multicolumn{2}{c}{ \ttcc  node}  & \multicolumn{2}{c}{ \ttlf  node}  \\
      \hline
      \ttlf      & $  1982 $ & ($  1381 $) & $  1280 $ & ($   897 $) & $   852 $ & ($   595 $) & $   916 $ & ($   661 $) & $   243 $ & ($   172 $) & $    50 $ & ($    36 $)  \\
      \ttcc      & $  1150 $ & ($  1415 $) & $   998 $ & ($  1230 $) & $   636 $ & ($   805 $) & $   444 $ & ($   567 $) & $   115 $ & ($   147 $) & $    16 $ & ($    19 $)  \\
      \ttb       & $   549 $ & ($   705 $) & $   575 $ & ($   746 $) & $   314 $ & ($   409 $) & $   253 $ & ($   338 $) & $    28 $ & ($    35 $) & $     4 $ & ($     5 $)  \\
      \tttwob    & $   306 $ & ($   233 $) & $   282 $ & ($   215 $) & $   372 $ & ($   293 $) & $    78 $ & ($    62 $) & $    10 $ & ($     8 $) & $     1 $ & ($   0.8 $)  \\
      \ttbb      & $   834 $ & ($   769 $) & $  1156 $ & ($  1082 $) & $   299 $ & ($   266 $) & $   145 $ & ($   129 $) & $    17 $ & ($    15 $) & $     3 $ & ($     2 $)  \\
      \singlet   & $   110 $ & ($   116 $) & $   146 $ & ($   145 $) & $    92 $ & ($    82 $) & $    53 $ & ($    53 $) & $     4 $ & ($     4 $) & $     3 $ & ($     3 $)  \\
      \Vjets     & $    38 $ & ($    37 $) & $    78 $ & ($    76 $) & $    34 $ & ($    30 $) & $    10 $ & ($     9 $) & $     7 $ & ($     6 $) & $   0.6 $ & ($   0.6 $)  \\
      \ttV       & $    80 $ & ($    75 $) & $    58 $ & ($    54 $) & $    31 $ & ($    28 $) & $    11 $ & ($    11 $) & $     4 $ & ($     4 $) & $   0.4 $ & ($   0.4 $)  \\
      \diboson   & $   0.9 $ & ($   0.9 $) & $   0.5 $ & ($   0.5 $) & $   0.4 $ & ($   0.4 $) & $   0.4 $ & ($   0.4 $) &      \NA  &       (\NA) &      \NA  &       (\NA)  \\[\cmsTabSkip]
      Total\;bkg. & $  5049 $ & ($  4733 $) & $  4575 $ & ($  4447 $) & $  2629 $ & ($  2509 $) & $  1911 $ & ($  1831 $) & $   429 $ & ($   392 $) & $    77 $ & ($    67 $)  \\
      $\pm$ tot unc. & $\pm 1216 $ & ($\pm 186 $) & $\pm 1156 $ & ($\pm 142 $) & $\pm 603 $ & ($\pm  80 $) & $\pm 422 $ & ($\pm  65 $) & $\pm 107 $ & ($\pm  14 $) & $\pm  18 $ & ($\pm   3 $)  \\[\cmsTabSkip]
      \ttH       & $   142 $ & ($   108 $) & $    53 $ & ($    40 $) & $    24 $ & ($    18 $) & $    10 $ & ($     7 $)             & $    2.1 $ &  ($  1.5 $)  &    $ 0.30 $ & ($ 0.23 $) \\
      $\pm$ tot unc. & $\pm  19 $ & ($\pm  15 $) & $\pm   8 $ & ($\pm   6 $) & $\pm   3 $ & ($\pm   2 $) & $\pm   1 $ & ($\pm   1 $) & $\pm 0.2 $ & ($\pm 0.2 $) & $\pm 0.03 $ & ($\pm 0.03 $)  \\[\cmsTabSkip]
      Data       & \multicolumn{2}{c}{$  4822 $} & \multicolumn{2}{c}{$  4400 $} & \multicolumn{2}{c}{$  2484 $} & \multicolumn{2}{c}{$  1852 $} & \multicolumn{2}{c}{$   422 $} & \multicolumn{2}{c}{$    76 $}  \\
      \hline
    \end{tabular}
  }
\end{table}

\begin{table}[!hbtp]
  \centering
  \topcaption{
    Observed and expected event yields per jet-tag category in the dilepton channel, prior to the fit to data (after the fit to data).
    The quoted uncertainties denote the total statistical and systematic uncertainty.
  }
  \label{tab:bdtmem:yields-jet-tag-categories:dl}
    \begin{tabular}{l r@{\;}r r@{\;}r r@{\;}r }
      \hline
      & \multicolumn{ 6 }{c}{pre-fit (post-fit) yields}
      \\
      Process  & \multicolumn{2}{c}{$\geq4$ jets, 3 \cPqb\ tags}  & \multicolumn{4}{c}{$\geq4$ jets, $\geq4$ \cPqb\ tags}  \\
      & & & \multicolumn{2}{c}{BDT-low}  & \multicolumn{2}{c}{BDT-high}  \\
      \hline
      \ttlf      & \hspace{20pt}$   845 $ & ($   637 $) & $    16 $ & ($    11 $) & $   0.7 $ & ($   0.5 $)  \\
      \ttcc      & $   712 $ & ($   966 $) & $    25 $ & ($    31 $) & $     3 $ & ($     4 $)  \\
      \ttb       & $   546 $ & ($   747 $) & $    26 $ & ($    35 $) & $     4 $ & ($     6 $)  \\
      \tttwob    & $   252 $ & ($   196 $) & $    11 $ & ($     8 $) & $     2 $ & ($     1 $)  \\
      \ttbb      & $   439 $ & ($   415 $) & $   103 $ & ($   109 $) & $    33 $ & ($    32 $)  \\
      \singlet   & $    47 $ & ($    51 $) & $     5 $ & ($     3 $) & $     1 $ & ($     2 $)  \\
      \Vjets     & $    10 $ & ($     8 $) &     \NA   &       (\NA) &     \NA   &       (\NA)  \\
      \ttV       & $    40 $ & ($    38 $) & $     4 $ & ($     4 $) & $     2 $ & ($     2 $)  \\
      \diboson   & $   0.9 $ & ($   0.7 $) &     \NA   &       (\NA) &     \NA   &       (\NA)  \\[\cmsTabSkip]
      Total\;bkg. & $  2893 $ & ($  3058 $) & $   190 $ & ($   201 $) & $    46 $ & ($    48 $)  \\
      $\pm$ tot unc. & $\pm 705 $ & ($\pm  98 $) & $\pm  67 $ & ($\pm  10 $) & $\pm  17 $ & ($\pm   3 $)  \\[\cmsTabSkip]
      \ttH       & $    42 $ & ($    32 $) & $     6 $ & ($     5 $) & $     6 $ & ($     5 $)  \\
      $\pm$ tot unc. & $\pm   6 $ & ($\pm   5 $) & $\pm   1 $ & ($\pm   1 $) & $\pm   1 $ & ($\pm   1 $)  \\[\cmsTabSkip]
      Data       & \multicolumn{2}{c}{\hspace{20pt}$  3077 $} & \multicolumn{2}{c}{$   207 $} & \multicolumn{2}{c}{$    58 $}  \\
      \hline
    \end{tabular}
\end{table}

\begin{figure}[hbtp]
  \centering
  \includegraphics[width=0.75\textwidth]{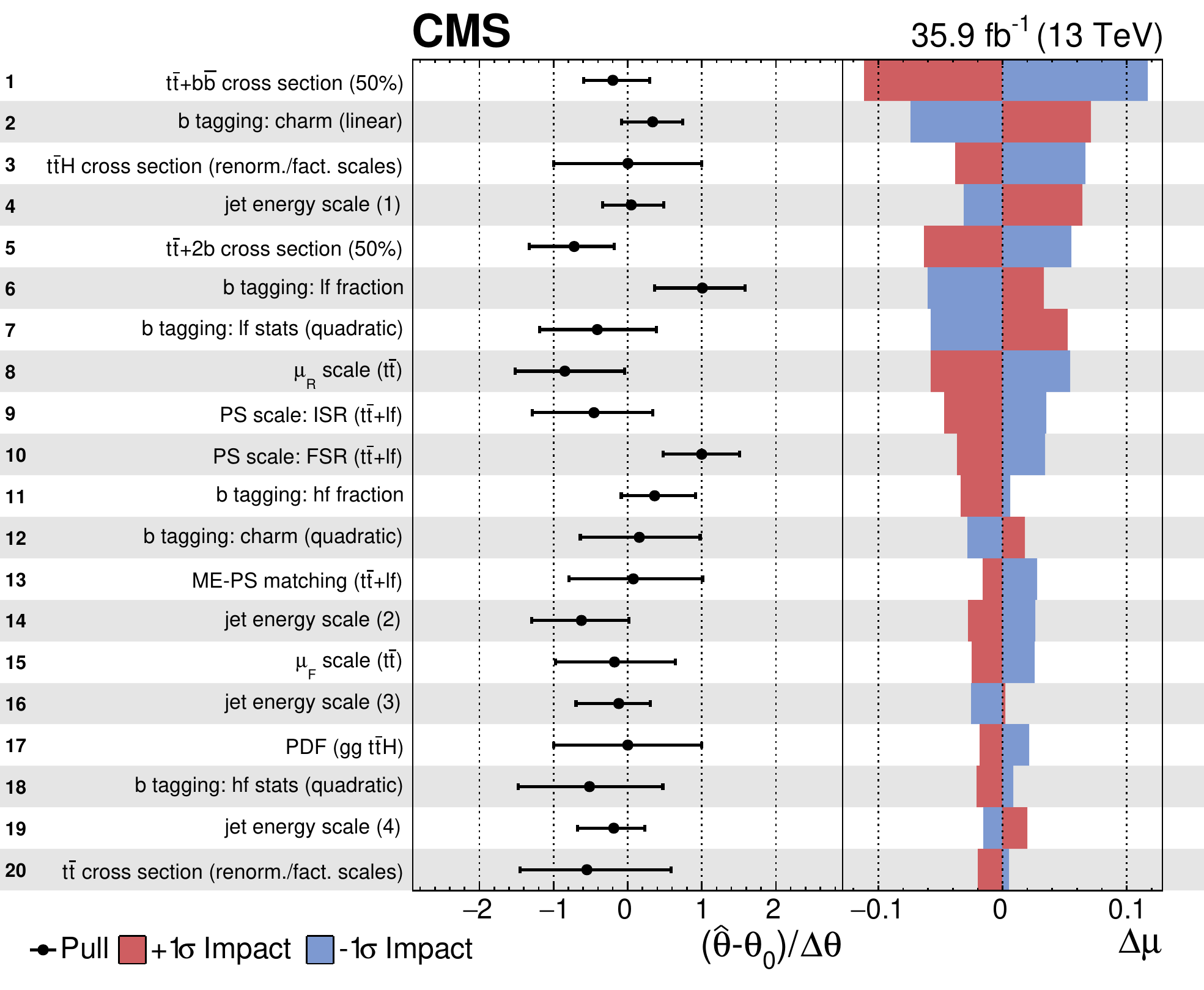}
  \caption{
    Post-fit pull and impact on the signal strength $\mu$ of the nuisance parameters included in the fit, ordered by their impact.
    Only the 20 highest ranked parameters are shown, not including nuisance parameters describing the uncertainty due to the size of the simulated samples.
    The four highest-ranked nuisance parameters related to the jet energy scale uncertainty sources are shown as indicated in parentheses.
    The pulls of the nuisance parameters (black markers) are computed relative to their pre-fit values $\theta_{0}$ and uncertainties $\Delta\theta$.
    The impact $\Delta\mu$ is computed as the difference of the nominal best fit value of $\mu$ and the best fit value obtained when fixing the nuisance parameter under scrutiny to its best fit value $\hat{\theta}$ plus/minus its post-fit uncertainty (coloured areas).
  }
  \label{fig:impacts}
\end{figure}

The signal strength modifier $\mu = \sigma/\sigma_\mathrm{SM}$ of the \ttH production cross section is determined in a simultaneous binned maximum likelihood fit to the data across all analysis categories.
The fit procedure takes into account systematic uncertainties that modify the shape and normalisation of the final discriminant distributions, as described in
Section~\ref{sec:systematics}.
The best fit values of the nuisance parameters are within 1 standard deviation of the prior uncertainty for more than 95\% of the total number of nuisance parameters.
The best fit values of the 20 parameters ranked highest in impact are presented in Fig.~\ref{fig:impacts}.
As expected, the fit constrains the nuisance parameters related to the conservatively assigned 50\% prior uncertainties on the \tthf cross section to 40--60\% of the prior.
A few other nuisance parameters that are related to jet energy scale and \cPqb\ tagging uncertainties are constrained up to a factor of 50\%.
These constraints are not due to conservatively assigned prior uncertainties but are attributed to the fact that events are selected according to different, large multiplicities of jets and \cPqb-tagged jets, thus increasing the sensitivity of the analysis to changes of the jet energy scale and \cPqb\ tagging efficiency, \eg by their effect on the event yield per analysis category.
Furthermore, the impact on $\mu$ of the most relevant sources of uncertainty is shown in Fig.~\ref{fig:impacts}, which is computed as the difference of the nominal best fit value of $\mu$ and the best fit value obtained when fixing the nuisance parameter under scrutiny to its best fit value plus/minus its post-fit uncertainty.
In particular, the 20 parameters with the highest impact are shown, excluding nuisance parameters describing the statistical uncertainties due to the size of the simulated samples.
The nuisance parameters with the highest impact are related to the uncertainty in the \tthf and signal cross sections, as well as in the \cPqb\ tagging scale factors.

\begin{figure}[hbtp]
  \centering
  \begin{tabular}{@{}c@{\hskip 0.05\textwidth}c@{}}
    \includegraphics[width=0.45\textwidth]{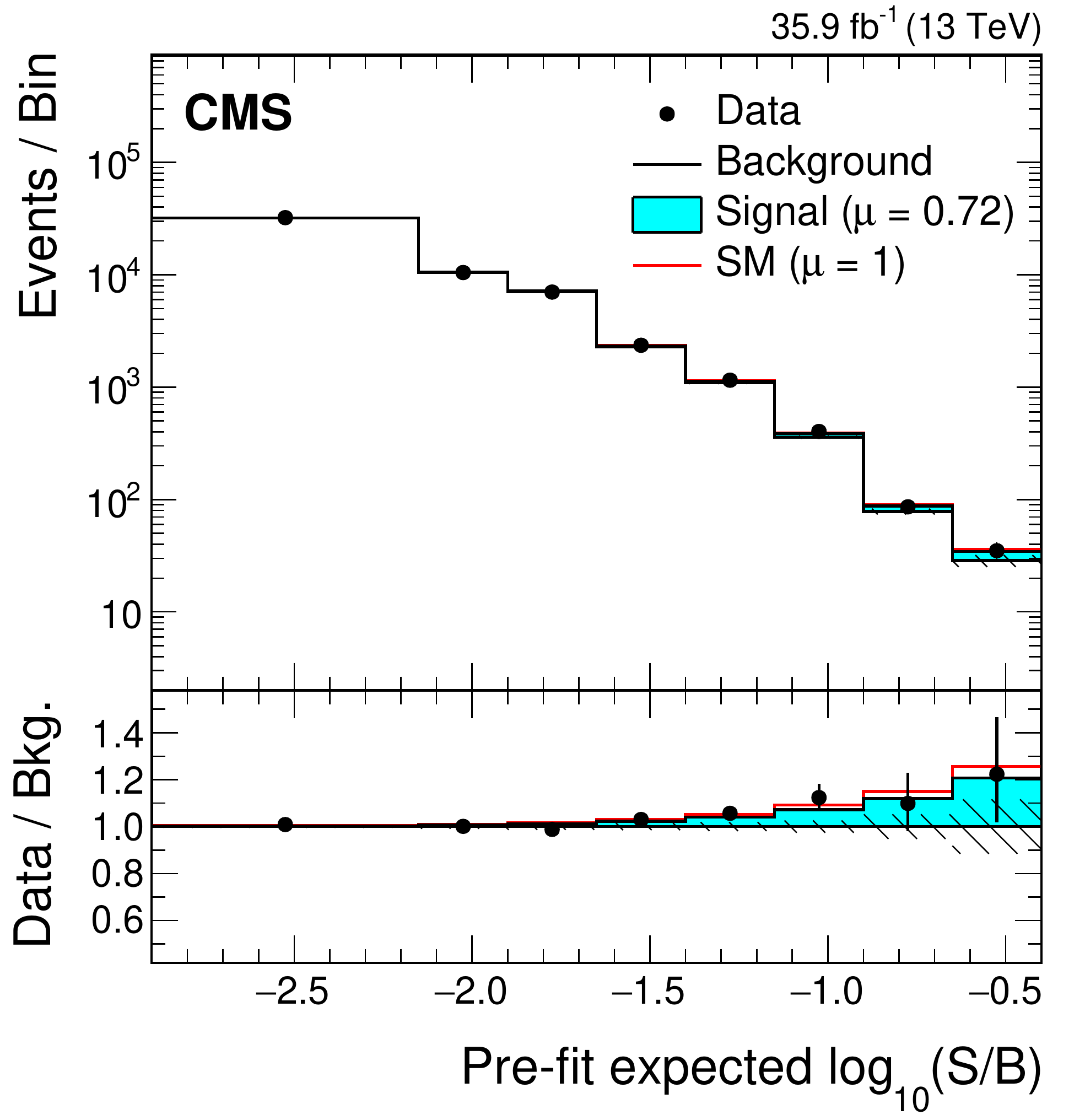} &
    \includegraphics[width=0.45\textwidth]{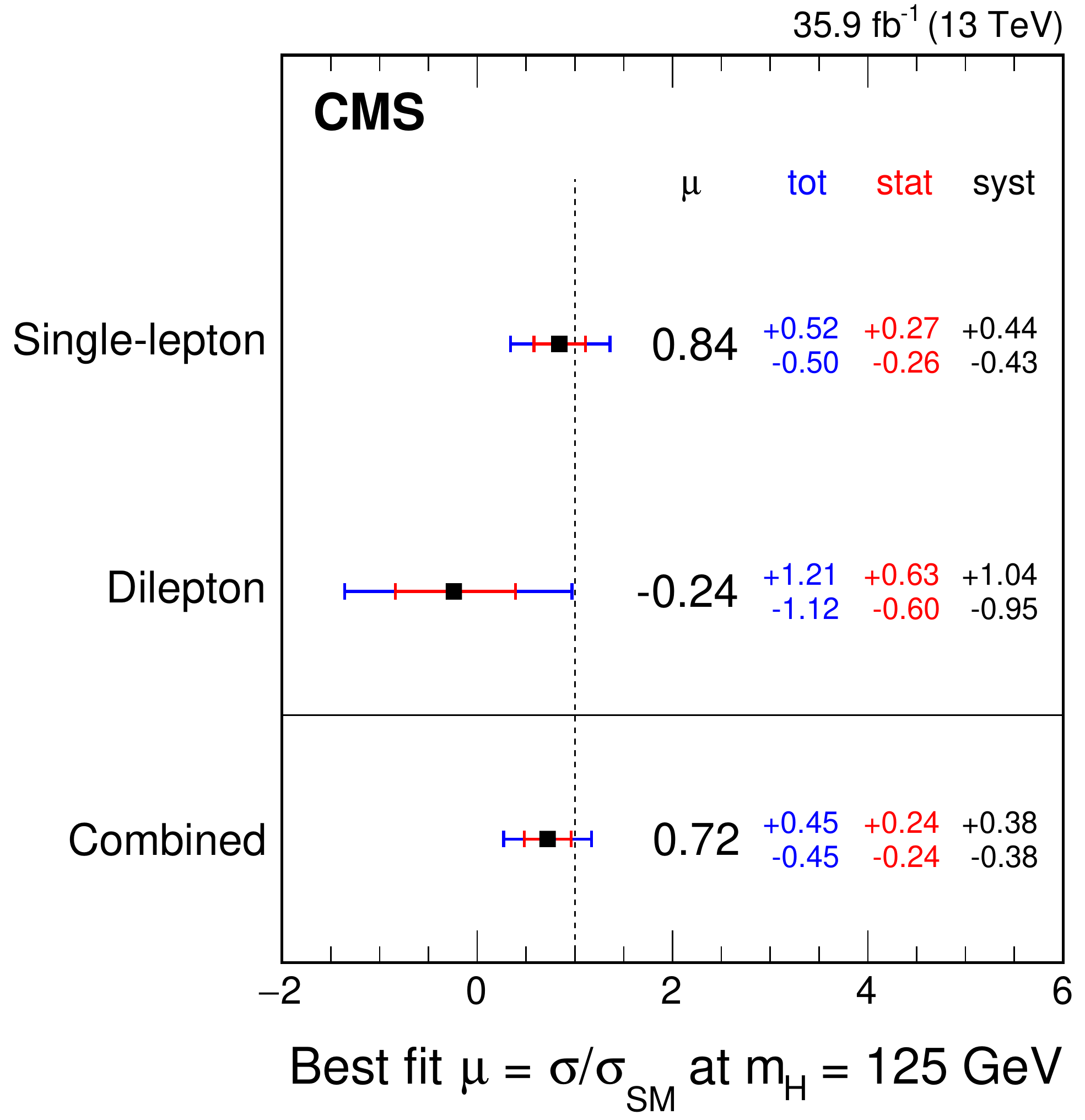} \\
  \end{tabular}
  \caption{
    Bins of the final discriminants as used in the fit (\cmsLeft), reordered by the pre-fit expected signal-to-background ratio (S/B).
    Each of the shown bins includes multiple bins of the final discriminants with similar S/B.
    The fitted signal (cyan) is compared to the expectation for the SM Higgs boson $\mu = 1$ (red).
    Best fit values of the signal strength modifiers $\mu$ (\cmsRight) with their 68\% expected confidence intervals (outer error bar), also split into their statistical (inner error bar) and systematic components.
  }
  \label{fig:bestfit}
\end{figure}

The obtained best fit value of $\mu$ is $0.72 \pm 0.24\stat \pm 0.38\syst$ with a total uncertainty of $\pm0.45$.
This corresponds to an observed (expected) significance of 1.6 (2.2) standard deviations above the background-only hypothesis.
The observed and predicted event yields in all the bins of the final discriminants, ordered by the pre-fit expected signal-to-background ratio (S/B) are shown in Fig.~\ref{fig:bestfit} (\cmsLeft).
The best fit values in each analysis channel separately and in the combination are listed in
Table~\ref{tab:limits} and displayed in Fig.~\ref{fig:bestfit} (\cmsRight).

\begin{table}[hbtp]
  \centering
  \topcaption{Best fit value of the signal strength modifier $\mu$ and the
    observed and median expected 95\% \CL upper limits in the single-lepton and the dilepton
    channels as well as the combined results.
    The one standard deviation confidence intervals of the expected limit
    and the best fit value are also quoted, split into the statistical and systematic components
    in the latter case.
  }
  \label{tab:limits}
  \renewcommand{\arraystretch}{1.8}
  \begin{tabular}{lccr}
    \hline
    Channel     & \multicolumn{2}{c}{95\% \CL upper limit} & \multicolumn{1}{c}{Best-fit $\mu$} \\[-8pt]
                & \multicolumn{1}{c}{observed} & \multicolumn{1}{c}{expected} & \multicolumn{1}{c}{$\pm\text{tot}\,(\pm\text{stat}\,\,\pm\text{syst})$} \\
    \hline
    Single-lepton & $1.75 $ & $ 1.03^{+0.44}_{-0.29} $&  $0.84^{+0.52}_{-0.50}\; \left(^{+0.27}_{-0.26}\,\,^{+0.44}_{-0.43}\right)$ \\
    Dilepton    & $2.34 $ & $ 2.48^{+1.17}_{-0.76} $& $-0.24^{+1.21}_{-1.12}\; \left(^{+0.63}_{-0.60}\,\,^{+1.04}_{-0.95}\right)$ \\[\cmsTabSkip]
    Combined    & $1.51 $ & $ 0.92^{+0.39}_{-0.26} $&  $0.72^{+0.45}_{-0.45}\; \left(^{+0.24}_{-0.24}\,\,^{+0.38}_{-0.38}\right)$ \\
    \hline
  \end{tabular}
  \renewcommand{\arraystretch}{1.0}
\end{table}

The contributions of the statistical and various systematic uncertainties to the uncertainty in $\mu$ are listed in Table~\ref{tab:uncertaintysplit}.
The statistical uncertainty is evaluated by fixing all nuisance parameters to their post-fit values.
The impact of the systematic uncertainties is evaluated by repeating the fit fixing only the nuisance parameters related to the uncertainty under scrutiny to their post-fit values and subtracting the obtained uncertainty in quadrature from the total uncertainty of the fit where no parameters are fixed.
The total uncertainty of the full fit (0.45) is different from the quadratic sum of the listed contributions because of correlations between the nuisance parameters.
\begin{table}[!htbp]
  \centering
  \topcaption{
    Contributions of different sources of uncertainties to the result for the fit to the data (observed) and to the expectation from simulation (expected).
    The quoted uncertainties $\Delta\mu$ in $\mu$ are obtained by fixing the listed sources of uncertainties to their post-fit values in the fit and subtracting the obtained result in quadrature from the result of the full fit.
    The statistical uncertainty is evaluated by fixing all nuisance parameters to their post-fit values.
    The quadratic sum of the contributions is different from the total uncertainty because of correlations between the nuisance parameters.
  }
  \label{tab:uncertaintysplit}
  \renewcommand{\arraystretch}{1.5}
  \begin{tabular}{lcc}
    \hline
    Uncertainty source & $\pm\Delta\mu$ (observed) & $\pm\Delta\mu$ (expected) \\
    \hline
    Total experimental & $+0.15$/$-0.16$ & $+0.19$/$-0.17$\\
    \hspace{12pt}\cPqb\ tagging         & $+0.11$/$-0.14$ & $+0.12$/$-0.11$\\
    \hspace{12pt}jet energy scale and resolution & $+0.06$/$-0.07$ & $+0.13$/$-0.11$\\[\cmsTabSkip]
    Total theory & $+0.28$/$-0.29$ & $+0.32$/$-0.29$\\
    \hspace{12pt}\tthf cross section and parton shower  & $+0.24$/$-0.28$ & $+0.28$/$-0.28$\\[\cmsTabSkip]
    Size of the simulated samples  & $+0.14$/$-0.15$ & $+0.16$/$-0.16$ \\[\cmsTabSkip]
    Total systematic  & $+0.38$/$-0.38$ & $+0.45$/$-0.42$\\
    Statistical       & $+0.24$/$-0.24$ & $+0.27$/$-0.27$\\[\cmsTabSkip]
    Total             & $+0.45$/$-0.45$ & $+0.53$/$-0.49$\\
    \hline
  \end{tabular}
  \renewcommand{\arraystretch}{1.0}
\end{table}

The total uncertainty of 0.45 is dominated by contributions from systematic effects, while the statistical component is 0.24.
The largest contributions originate from the theoretical uncertainties amounting to $+0.28$/$-0.29$, where the \tthf modelling uncertainties have a major contribution.
Experimental uncertainties amount to $+0.15$/$-0.16$, dominated by the \cPqb\ tagging related uncertainties.
Systematic uncertainties due to the size of the various simulated samples used to model the background and signal templates are at the same order and amount to $+0.14$/$-0.15$.

\begin{figure}[hbtp]
  \centering
  \includegraphics[width=0.9\textwidth]{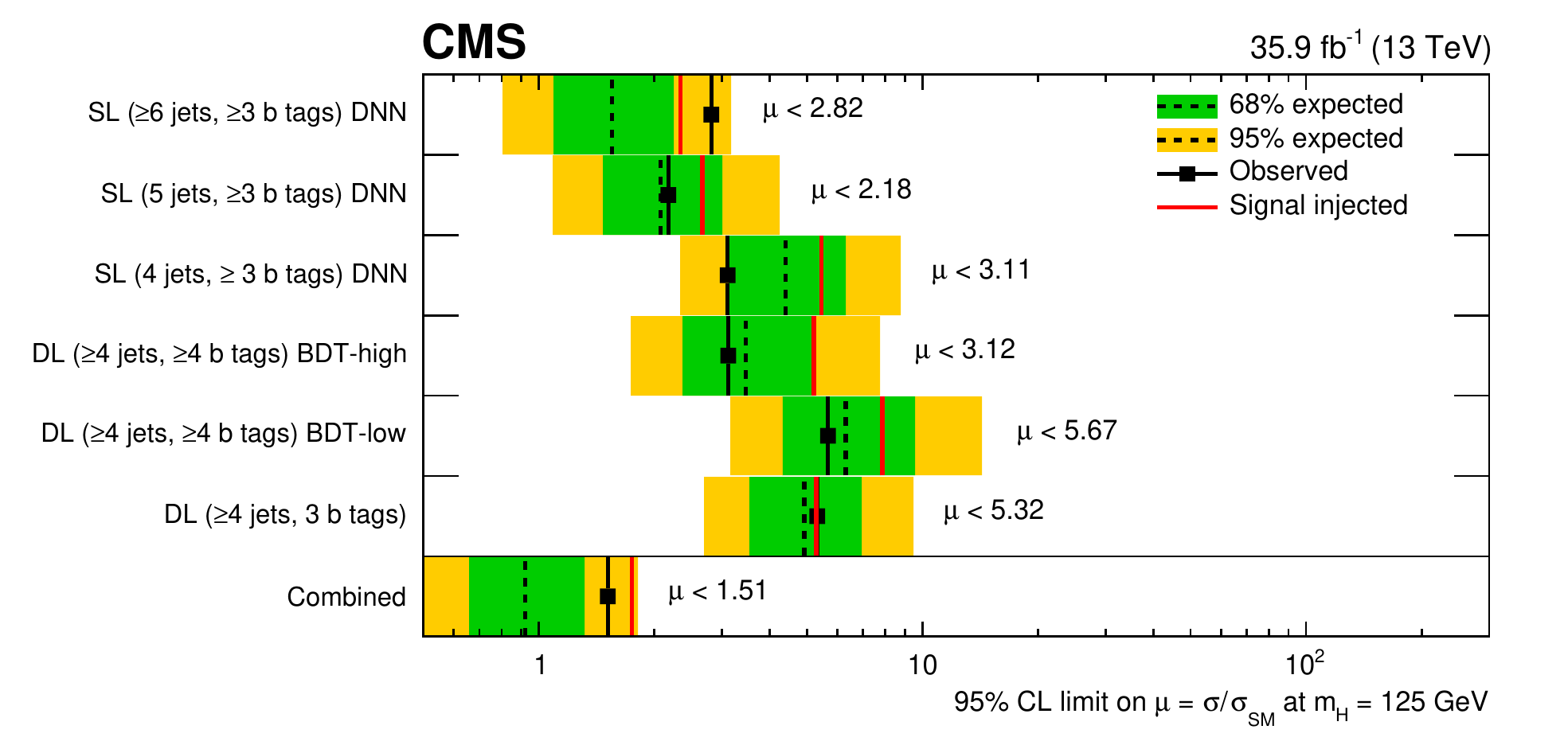}
  \caption{
    Median expected (dashed line) and observed (markers) 95\% \CL upper limits on $\mu$.
    The inner (green) band and the outer (yellow) band indicate the regions containing 68 and 95\%, respectively, of the distribution of limits expected under the background-only hypothesis.
    Also shown is the limit that is expected in case a SM \ttH signal ($\mu=1$) is present in the data (solid red line).
  }
  \label{fig:limitschannels}
\end{figure}

An upper limit on $\mu$ under the background-only hypothesis is also determined, using a modified frequentist $\text{CL}_{\text{S}}$ procedure~\cite{Junk:1999kv,Read:2002hq} with the asymptotic method~\cite{Cowan:2010js}.
When combining all categories and channels, an observed
(expected) upper limit at 95\% \CL on $\mu$ of 1.5 (0.9) is obtained.
The observed and expected upper limits in each channel and in the combination are listed in
Table~\ref{tab:limits} and visualised in Fig.~\ref{fig:limitschannels}.

In addition, the statistical analysis has been performed using the jet-process categorisation and DNN output in both channels and their combination, as well as using the jet-tag categorisation and the BDT or MEM in both channels. The results obtained in each channel and the combination are compatible within 1.7 standard deviations or better, evaluated using a jackknife procedure~\cite{Jackknife}. This serves as an important cross check and validation of the complex analysis methods.

\section{Summary}
\label{sec:summary}
A search for the associated production of a Higgs boson and a top
quark-antiquark pair (\ttH) is performed using \protonproton collision data recorded
with the CMS detector at a centre-of-mass energy of 13\TeV in
2016, corresponding to an integrated luminosity of \lumivalue. Candidate events are selected in final states compatible with
the Higgs boson decaying into a \PQb quark-antiquark pair and the single-lepton and dilepton decay
channels of the \ttbar system. Selected events are split into mutually
exclusive categories according to their \ttbar decay channel and jet
content.
In each category a powerful discriminant is constructed to separate the
\ttH signal from the dominant \ttjets background, based on several multivariate analysis techniques (boosted decision trees, matrix element method, and deep neural networks).
An observed (expected) upper limit on the
\ttH production cross section $\mu$ relative to the SM expectations of
$1.5\,(0.9)$ at 95\% confidence level is obtained.
The best fit value of $\mu$ is $0.72 \pm 0.24\stat \pm 0.38\syst$.
These results correspond to an observed (expected) significance of 1.6 (2.2) standard deviations above the background-only hypothesis.

\begin{acknowledgments}
\hyphenation{Bundes-ministerium Forschungs-gemeinschaft Forschungs-zentren Rachada-pisek} We congratulate our colleagues in the CERN accelerator departments for the excellent performance of the LHC and thank the technical and administrative staffs at CERN and at other CMS institutes for their contributions to the success of the CMS effort. In addition, we gratefully acknowledge the computing centres and personnel of the Worldwide LHC Computing Grid for delivering so effectively the computing infrastructure essential to our analyses. Finally, we acknowledge the enduring support for the construction and operation of the LHC and the CMS detector provided by the following funding agencies: the Austrian Federal Ministry of Science, Research and Economy and the Austrian Science Fund; the Belgian Fonds de la Recherche Scientifique, and Fonds voor Wetenschappelijk Onderzoek; the Brazilian Funding Agencies (CNPq, CAPES, FAPERJ, and FAPESP); the Bulgarian Ministry of Education and Science; CERN; the Chinese Academy of Sciences, Ministry of Science and Technology, and National Natural Science Foundation of China; the Colombian Funding Agency (COLCIENCIAS); the Croatian Ministry of Science, Education and Sport, and the Croatian Science Foundation; the Research Promotion Foundation, Cyprus; the Secretariat for Higher Education, Science, Technology and Innovation, Ecuador; the Ministry of Education and Research, Estonian Research Council via IUT23-4 and IUT23-6 and European Regional Development Fund, Estonia; the Academy of Finland, Finnish Ministry of Education and Culture, and Helsinki Institute of Physics; the Institut National de Physique Nucl\'eaire et de Physique des Particules~/~CNRS, and Commissariat \`a l'\'Energie Atomique et aux \'Energies Alternatives~/~CEA, France; the Bundesministerium f\"ur Bildung und Forschung, Deutsche Forschungsgemeinschaft, and Helmholtz-Gemeinschaft Deutscher Forschungszentren, Germany; the General Secretariat for Research and Technology, Greece; the National Research, Development and Innovation Fund, Hungary; the Department of Atomic Energy and the Department of Science and Technology, India; the Institute for Studies in Theoretical Physics and Mathematics, Iran; the Science Foundation, Ireland; the Istituto Nazionale di Fisica Nucleare, Italy; the Ministry of Science, ICT and Future Planning, and National Research Foundation (NRF), Republic of Korea; the Lithuanian Academy of Sciences; the Ministry of Education, and University of Malaya (Malaysia); the Mexican Funding Agencies (BUAP, CINVESTAV, CONACYT, LNS, SEP, and UASLP-FAI); the Ministry of Business, Innovation and Employment, New Zealand; the Pakistan Atomic Energy Commission; the Ministry of Science and Higher Education and the National Science Centre, Poland; the Funda\c{c}\~ao para a Ci\^encia e a Tecnologia, Portugal; JINR, Dubna; the Ministry of Education and Science of the Russian Federation, the Federal Agency of Atomic Energy of the Russian Federation, Russian Academy of Sciences and the Russian Foundation for Basic Research; the Ministry of Education, Science and Technological Development of Serbia; the Secretar\'{\i}a de Estado de Investigaci\'on, Desarrollo e Innovaci\'on, Programa Consolider-Ingenio 2010, Plan Estatal de Investigaci\'on Cient\'{\i}fica y T\'ecnica y de Innovaci\'on 2013-2016, Plan de Ciencia, Tecnolog\'{i}a e Innovaci\'on 2013-2017 del Principado de Asturias and Fondo Europeo de Desarrollo Regional, Spain; the Swiss Funding Agencies (ETH Board, ETH Zurich, PSI, SNF, UniZH, Canton Zurich, and SER); the Ministry of Science and Technology, Taipei; the Thailand Center of Excellence in Physics, the Institute for the Promotion of Teaching Science and Technology of Thailand, Special Task Force for Activating Research and the National Science and Technology Development Agency of Thailand; the Scientific and Technical Research Council of Turkey, and Turkish Atomic Energy Authority; the National Academy of Sciences of Ukraine, and State Fund for Fundamental Researches, Ukraine; the Science and Technology Facilities Council, UK; the US Department of Energy, and the US National Science Foundation.

Individuals have received support from the Marie-Curie programme and the European Research Council and Horizon 2020 Grant, contract No. 675440 (European Union); the Leventis Foundation; the A. P. Sloan Foundation; the Alexander von Humboldt Foundation; the Belgian Federal Science Policy Office; the Fonds pour la Formation \`a la Recherche dans l'Industrie et dans l'Agriculture (FRIA-Belgium); the Agentschap voor Innovatie door Wetenschap en Technologie (IWT-Belgium); the F.R.S.-FNRS and FWO (Belgium) under the ``Excellence of Science - EOS" - be.h project n. 30820817; the Ministry of Education, Youth and Sports (MEYS) of the Czech Republic; the Lend\"ulet (``Momentum") Programme and the J\'anos Bolyai Research Scholarship of the Hungarian Academy of Sciences, the New National Excellence Program \'UNKP, the NKFIA research grants 123842, 123959, 124845, 124850 and 125105 (Hungary); the Council of Scientific and Industrial Research, India; the HOMING PLUS programme of the Foundation for Polish Science, cofinanced from European Union, Regional Development Fund, the Mobility Plus programme of the Ministry of Science and Higher Education, the National Science Center (Poland), contracts Harmonia 2014/14/M/ST2/00428, Opus 2014/13/B/ST2/02543, 2014/15/B/ST2/03998, and 2015/19/B/ST2/02861, Sonata-bis 2012/07/E/ST2/01406; the National Priorities Research Program by Qatar National Research Fund; the Programa de Excelencia Mar\'{i}a de Maeztu and the Programa Severo Ochoa del Principado de Asturias; the Thalis and Aristeia programmes cofinanced by EU-ESF and the Greek NSRF; the Rachadapisek Sompot Fund for Postdoctoral Fellowship, Chulalongkorn University and the Chulalongkorn Academic into Its 2nd Century Project Advancement Project (Thailand); the Welch Foundation, contract C-1845; and the Weston Havens Foundation (USA).
\end{acknowledgments}

\bibliography{auto_generated}

\clearpage
\appendix

\section{BDT and DNN input variables and configuration}
\label{sec:appendix:MVAs}
All input variables used in the DNNs and BDTs are listed in Tables~\ref{tab:classifiers:inputs_part1}-\ref{tab:classifiers:inputs_part3}.
\begin{table}[hbtp]
  \centering
  \topcaption{
    Input variables used in the DNNs or BDTs in the different categories of the single-lepton and dilepton channels.
    Variables used in a specific multivariate method and analysis category are denoted by a ``$+$'' and unused variables by a ``$-$''.
    (Continued in Tables~\ref{tab:classifiers:inputs_part2} and \ref{tab:classifiers:inputs_part3}.)
  }
  \label{tab:classifiers:inputs_part1}
  \cmsTable{
    \renewcommand{\arraystretch}{1.6}
    \begin{tabular}{cp{10cm}ccccc}
      \hline
      Variable & Definition & \rotatebox{90}{SL  \ljFourThreeIncl \xspace } &  \rotatebox{90}{SL  \ljFiveThreeIncl \xspace } &  \rotatebox{90}{SL  \ljSixThreeIncl \xspace } &  \rotatebox{90}{DL  \dlFourThree \xspace } &  \rotatebox{90}{DL  \dlFourFour \xspace } \\
      \hline
      \varfirstjetpt        & \pt of the highest-\pt jet        & +             & +             & -                 & -                 &  -    \\
      $\eta(\text{jet 1})$  & $\eta$ of the highest-\pt  jet       & -             & +             & +                 & -                 &  -    \\
      $d(\text{jet 1})$     & \cPqb\ tagging discriminant of the highest-\pt  jet  & +             & +             & +                 & -                 &  -    \\
      \varsecondjetpt       & \pt of the second highest-\pt jet           & -             & +             & -                 & -                 & -     \\
      $\eta(\text{jet 2})$  & $\eta$ of the second highest-\pt jet          & +             & +             & +                 & -                 & -     \\
      $d(\text{jet 2})$     & \cPqb\ tagging discriminant of the second highest-\pt jet     & +             & +             & +                 & -                 & -     \\
      \varthirdjetpt        & \pt of the third highest-\pt jet            & -             & +             & -                 & -                 & -     \\
      $\eta(\text{jet 3})$  & $\eta$ of the third highest-\pt jet           & +             & +             & +                 & -                 & -     \\
      $d(\text{jet 3})$     & \cPqb\ tagging discriminant of the third highest-\pt jet      & +             & +             & +                 & -                 & -     \\
      \varfourthjetpt       & \pt of the fourth highest-\pt jet           & +             & +             & -                 & -                 & -     \\
      $\eta(\text{jet 4})$  & $\eta$ of the fourth highest-\pt jet          & +             & +             & +                 & -                 & -     \\
      $d(\text{jet 4})$     & \cPqb\ tagging discriminant of the fourth highest-\pt jet     & +             & -             & +                 & -                 & -     \\
      $\pt(\text{lep 1})$   & \pt of the highest-\pt lepton               & -             & +             & +                 & -                 & -     \\
      $\eta(\text{lep 1})$  & $\eta$ of the highest-\pt lepton              & +             & -             & +                 & -                 & -     \\
      \varavgbtagdiscjetsForm  &  average \cPqb\ tagging discriminant value of all jets                  & +             & +             & +                 & -                 & -     \\
      \varavgbtagdiscbtagsForm & average \cPqb\ tagging discriminant value of \cPqb-tagged jets                   & +             & +             & +                 & +                 & +     \\
      \varavgbtagdiscuntaggedjetsForm & average \cPqb\ tagging discriminant value of non-\cPqb-tagged jets                   & -             & -             & -                 & +                 & +     \\
      \vardevfromavgdiscbtagsAlt  &   squared difference between the \cPqb\ tagging discriminant value of a \cPqb-tagged jet and the average \cPqb\ tagging discriminant values of all \cPqb-tagged jets, summed over all \cPqb-tagged jets             &   +             & +             & +                 & -                 & -      \\
      $d_{\text{j}}^{\text{max}}$ & maximal \cPqb\ tagging discriminant value of all jets  & +             & +             & +                 & -                 & -     \\
      $d_{\text{b}}^{\text{max}}$ & maximal \cPqb\ tagging discriminant value of \cPqb-tagged jets  & +             & +             & +                 & -                 & -     \\
      $d_{\text{j}}^{\text{min}}$ & minimal \cPqb\ tagging discriminant value of all jets  & +             & +             & +                 & -                 & -     \\
      $d_{\text{j}}^{\text{min}}$ & minimal \cPqb\ tagging discriminant value of \cPqb-tagged jets  & +             & +             & +                 & -                 & -     \\
      $d_{2}$                 & second highest \cPqb\ tagging discriminant value of all jets               & +             & +             & +                 & -                 & -     \\
      \hline
    \end{tabular}
  }
\end{table}

\clearpage

\begin{table}[hbtp]
  \centering
  \topcaption{Continued from Table~\ref{tab:classifiers:inputs_part1} and continued in Table~\ref{tab:classifiers:inputs_part3}.}
  \label{tab:classifiers:inputs_part2}
  \cmsTable{
    \renewcommand{\arraystretch}{1.5}
    \begin{tabular}{cp{10cm}ccccc}
      \hline
      Variable & Definition & \rotatebox{90}{SL  \ljFourThreeIncl \xspace } &  \rotatebox{90}{SL  \ljFiveThreeIncl \xspace } &  \rotatebox{90}{SL  \ljSixThreeIncl \xspace } &  \rotatebox{90}{DL  \dlFourThree \xspace } &  \rotatebox{90}{DL  \dlFourFour \xspace } \\
      \hline

      $N_{\text{b}}(\text{tight})$ & number of \cPqb-tagged jets at a working point with a 0.1\% probability of tagging gluon and light-flavour jets & +             & +             & +                 & -                 & -     \\
      $\text{BLR}$ & likelihood ratio discriminating between 4 \cPqb\ quark jets and 2 \cPqb\ quark jets events                               & +             & +             & +                 & -                 &  -    \\
      $\text{BLR}^{\text{trans}}$ & transformed BLR defined as $\ln[\text{BLR}/(1.0 - \text{BLR})]$                 & +             & +             & +                 & -                 & -     \\

      \varmindrjets & \(\Delta R\) between the two closest jets                         & +             & +             & +                 & -                 & -     \\
      \varmindrtaggedjets & \(\Delta R\) between the two closest \cPqb-tagged jets              & +             & +             & +                 & -                 & -     \\
      \varmaxdrjets & \(\Delta R\) between the two jets furthest apart                             & -             & +             & -                 & -                 & -     \\
      \varmaxdrtaggedjets & \(\Delta R\) between the two \cPqb-tagged jets furthest apart                  & -             & -             & +                 & -                 & -     \\
      \varmaxdetajetsAlt  & \(\Delta \eta\) between the two jets furthest apart in $\eta$                                 & -             & -             & -                 & -                 & +     \\
      \varmaxdetataggedjetsAlt  & \(\Delta \eta\) between the two \cPqb-tagged jets furthest apart in $\eta$                                  & -             & -             & -                 & +                 & +     \\
      \varEvtDetaTaggedJetsAverageAlt & average \(\Delta \eta\) between \cPqb-tagged jets          & -             & -             & +                 & -                 & -     \\
      \varavgdrtaggedjetsAlt & average \(\Delta R\) between \cPqb-tagged jets   & -             & +             & +                 & -                 & -      \\
      \varavgdrtaggeduntaggedjetsAlt & average \(\Delta R\) between jets of which at least one is \cPqb-tagged   & -             & -             & -                 & +                 & -      \\
      \vardrbetweenlepandclosestjetAlt & \(\Delta R\) between lepton and closest jet         & +             & +             & -                 & -                 & -     \\
      \vardrbetweenlepandclosesttagAlt & \(\Delta R\) between lepton and closest \cPqb-tagged jet        & -             & +             & +                 & -                 & -     \\

      \varMlbAlt & mass of lepton and closest \cPqb-tagged jet                               & +             & +             & +                 & -                 & -     \\
      \varclosesttaggeddijetmassAlt & mass of closest \cPqb-tagged jets            & +             & +             & +                 & -                 & +     \\
      \varclosesttaggeduntaggeddijetmassAlt & mass of closest jets of which at least one is \cPqb-tagged            & -             & -             & -                 & +                 & -     \\
      \vartaggeddijetmassMaxMass & maximal mass of pairs of \cPqb-tagged jets       & -             & -             & -                 & +                 & +     \\
      \varclosesttaggeddijetpTAlt & combined \pt of closest \cPqb-tagged jets            & -             & -             & -                 & +                 & -     \\
      \varclosesttaggeduntaggeddijetpTAlt & combined \pt of closest jets of which at least one is \cPqb-tagged            & -             & -             & -                 & -                 & +     \\
      \varMjets & average mass of all jets                           & +             & +             & +                 & -                 &  -    \\
      \varMsquaredtaggedjets & average squared mass of all \cPqb-tagged jets                        & +             & -             & +                 & -                 & -     \\
      \vartaggeddijetmassclosesttoMHAlt & mass of pair of \cPqb-tagged jets closest to 125\GeV        & -             & +             & +                 & -                 & -     \\
      \varNjj  & number of pairs of jets (with at least one \cPqb-tagged jet) with an invariant mass within 110--140\GeV  & -             & -             & -                 & +                 & +     \\
      MEM  & matrix element method discriminant                                 & +             & +             & +                 & -                 & -     \\
      \hline
    \end{tabular}
  }
\end{table}

\clearpage

\begin{table}[!htp]
  \centering
  \topcaption{Continued from Table~\ref{tab:classifiers:inputs_part2}.}
  \label{tab:classifiers:inputs_part3}
  \cmsTable{
    \renewcommand{\arraystretch}{1.5}
    \begin{tabular}{cp{10cm}ccccc}
      \hline
      Variable & Definition & \rotatebox{90}{SL  \ljFourThreeIncl \xspace } &  \rotatebox{90}{SL  \ljFiveThreeIncl \xspace } &  \rotatebox{90}{SL  \ljSixThreeIncl \xspace } &  \rotatebox{90}{DL  \dlFourThree \xspace } &  \rotatebox{90}{DL  \dlFourFour \xspace } \\
      \hline
      \varHT & scalar sum of jet \pt \xspace                        & -             & +             & -                 & +                 & -     \\
      \varHTtagged & scalar sum of \cPqb-tagged jet \pt \xspace             & +             & +             & +                 & -                 & -     \\
      $A^{\text{j}}$ & $\frac{3}{2} \lambda_{3}$ where $\lambda_{i}$ are the eigenvalues of the momentum tensor built with jets ~\cite{PhysRevD.1.1416}  & -             & +             & +                 & -                 & -     \\
      $A^{\text{b}}$ & $\frac{3}{2} \lambda_{3}$ where $\lambda_{i}$ are the eigenvalues of the momentum tensor built with \cPqb-tagged jets ~\cite{PhysRevD.1.1416}                  & +             & +             & +                 & -                 & -     \\
      $C^{\text{j}}$ & \varHT divided by the sum of the energies of all jets & -             & -             & +                 & -                 & -     \\
      $C^{\text{b}}$ & \varHTtagged divided by the sum of the energies of all \cPqb-tagged jets & -             & -             & +                 & -                 & +     \\
      $S^{\text{j}}$ & $\frac{3}{2} (\lambda_{2} + \lambda_{3})$ where $\lambda_{i}$ are the eigenvalues of the momentum tensor built with jets ~\cite{PhysRevD.1.1416}                  & +             & +             & +                 & -                 & -     \\
      $S^{\text{b}}$ & $\frac{3}{2} (\lambda_{2} + \lambda_{3})$ where $\lambda_{i}$ are the eigenvalues of the momentum tensor built with \cPqb-tagged jets ~\cite{PhysRevD.1.1416}                 & -             & +             & +                 & -                 & -     \\
      $S^{\text{j}}_{\text{T}}$ & $\frac{2 \lambda_{2}}{\lambda_{2} + \lambda_{1}}$ where $\lambda_{i}$ are the eigenvalues of the momentum tensor built with jets ~\cite{PhysRevD.1.1416}         & +             & +             & +                 & -                 & -     \\
      $S^{\text{b}}_{\text{T}}$ & $\frac{2 \lambda_{2}}{\lambda_{2} + \lambda_{1}}$ where $\lambda_{i}$ are the eigenvalues of the momentum tensor built with \cPqb-tagged jets ~\cite{PhysRevD.1.1416}          & +             & +             & +                 & -                 & -     \\
      $I^{\text{b}}$ & a measure of how spherical or linear in $r-\phi$ space \cPqb-tagged jets are in the event                & -             & -             & -                 & +                 & -     \\
      \varhtwo & second Fox--Wolfram moment~\cite{tagkey1979543} & -             & +             & -                 & -                 & -     \\
      \varhthree & third Fox--Wolfram moment~\cite{tagkey1979543}    & +             & +             & -                 & -                 & -     \\
      \varhthreetagged & third Fox--Wolfram moment calculated with \cPqb-tagged jets ~\cite{tagkey1979543}   & -             & -             & -                 & -                 & +     \\
      \varRthree & ratio of Fox--Wolfram moments $\varhthree/\varhzero$ ~\cite{tagkey1979543}   & -             & -             & -                 & +                 & -     \\
      \varhfour & fourth Fox--Wolfram moment~\cite{tagkey1979543}   & +             & -             & +                 & -                 & -     \\

      \hline
    \end{tabular}
  }
\end{table}

The BDTs employed in the dilepton channel were trained using the stochastic gradient boost method~\cite{Friedman2002367,Hastie:ESL}, available as part of the TMVA package~\cite{TMVA}.
The number of trees ($N_{\text{trees}}$), the learning rate (shrinkage), the fraction of events used for the training of an individual tree (bagging fraction), the granularity of the cuts at each node splitting ($N_{\text{cuts}}$), and the number of node splittings per tree (depth) are listed in Table~\ref{tab:classifiers:configuration_BDT}.
\begin{table}[!htp]
  \centering
  \topcaption{
    Configuration of the BDTs used in the dilepton channel.
  }
  \label{tab:classifiers:configuration_BDT}
  \renewcommand{\arraystretch}{1.5}
  \begin{tabular}{cccccc}
    \hline
    Category & $N_{\text{trees}}$ & shrinkage & bagging fraction & $N_{\text{cuts}}$ & depth\\
    \hline
    \dlFourThree                & 955   & 0.022 & 0.42 & 30 & 2 \\
    \dlFourFour                 & 638   & 0.006 & 0.41 & 42 & 2 \\
    \hline
  \end{tabular}
\end{table}

The DNNs used in the single-lepton channel comprise two layers with 100 nodes each in each of the two network stages.
Overtraining is suppressed by random node dropout with a probability of 30\% and an L2 weight normalisation factor of $10^{-5}$.
All networks are optimised using the ADAM optimiser with a learning rate of $10^{-4}$, and the ELU activation function is used to add non-linearity to the response of the network~\cite{Goodfellow-et-al-2016}.

\clearpage

\section{Pre-fit discriminant shapes (single-lepton channel)}
\label{sec:appendix:prefitshapes}
\begin{figure}[!hbtp]
  \centering
  \begin{tabular}{c@{\hskip 0.05\textwidth}c}
    \includegraphics[width=0.35\textwidth]{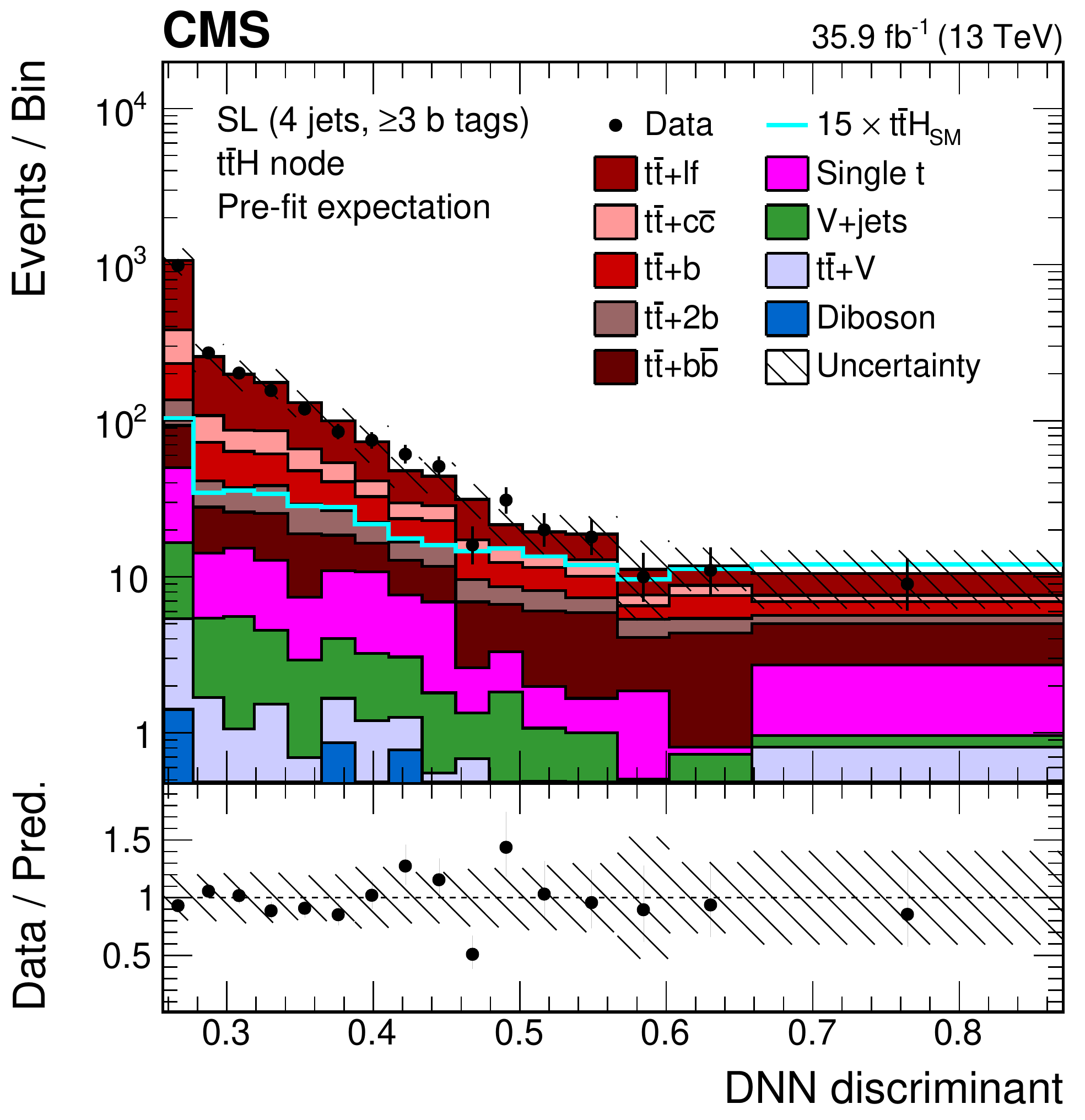} &
    \includegraphics[width=0.35\textwidth]{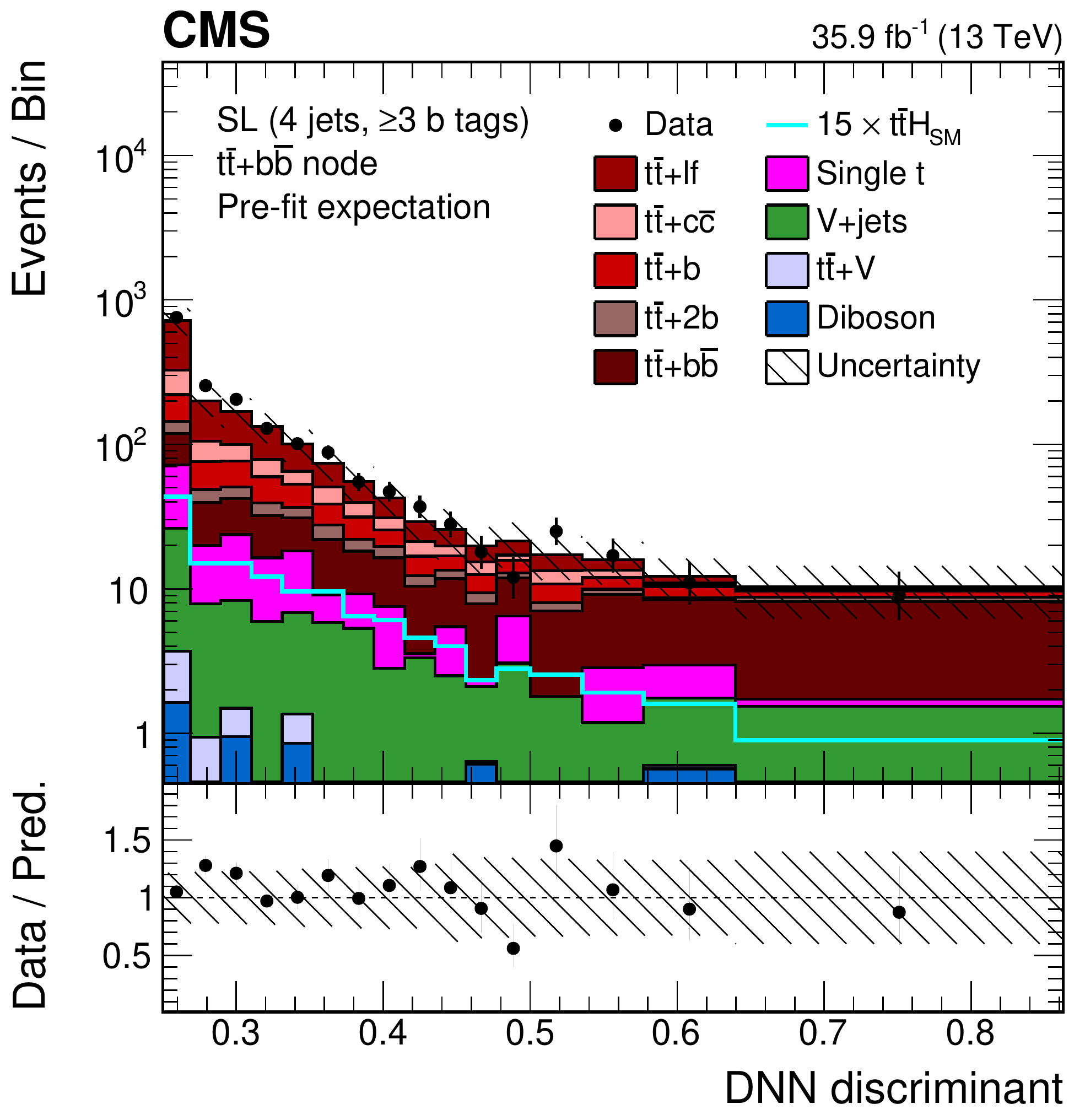}\\
    \includegraphics[width=0.35\textwidth]{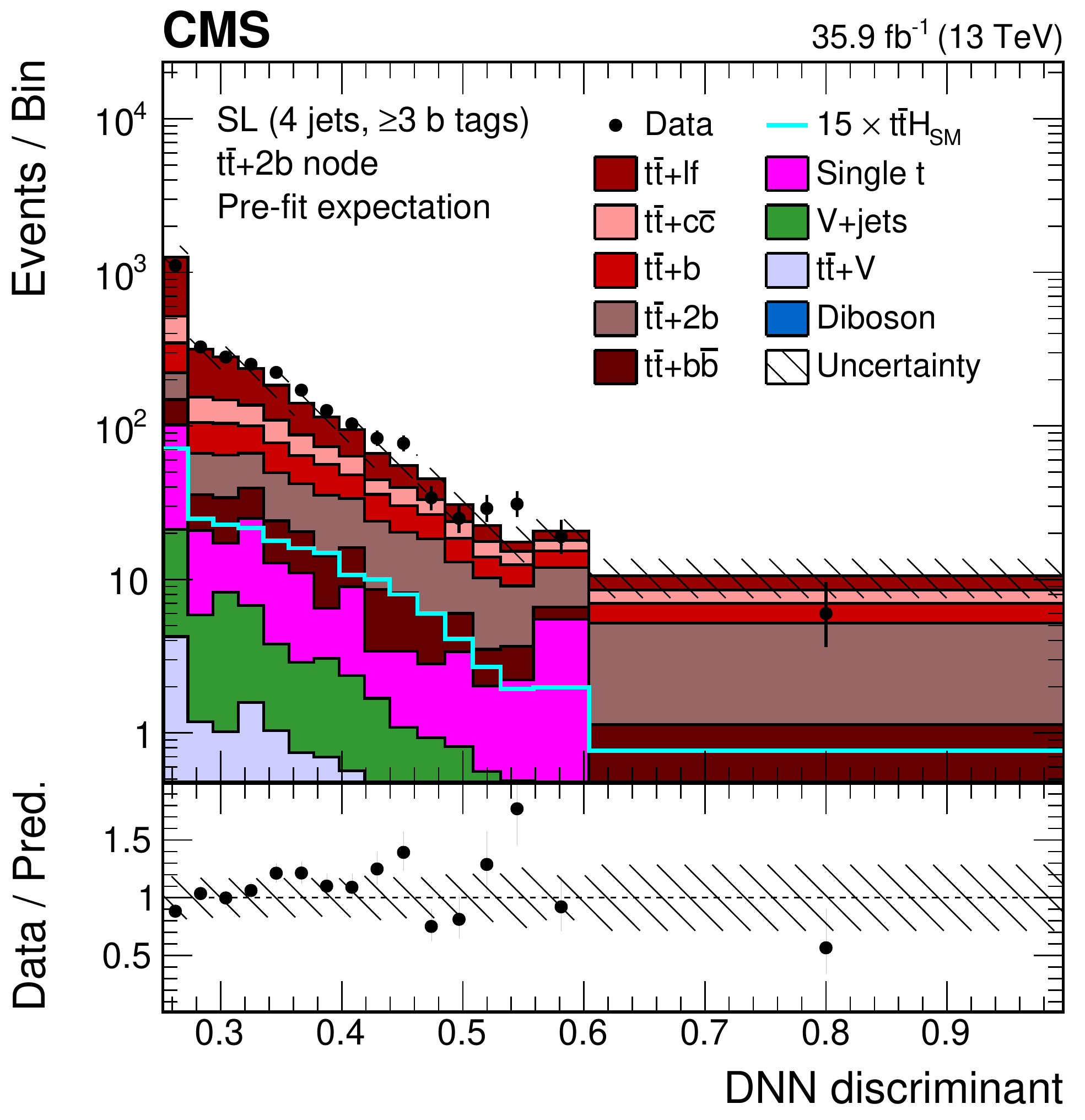} &
    \includegraphics[width=0.35\textwidth]{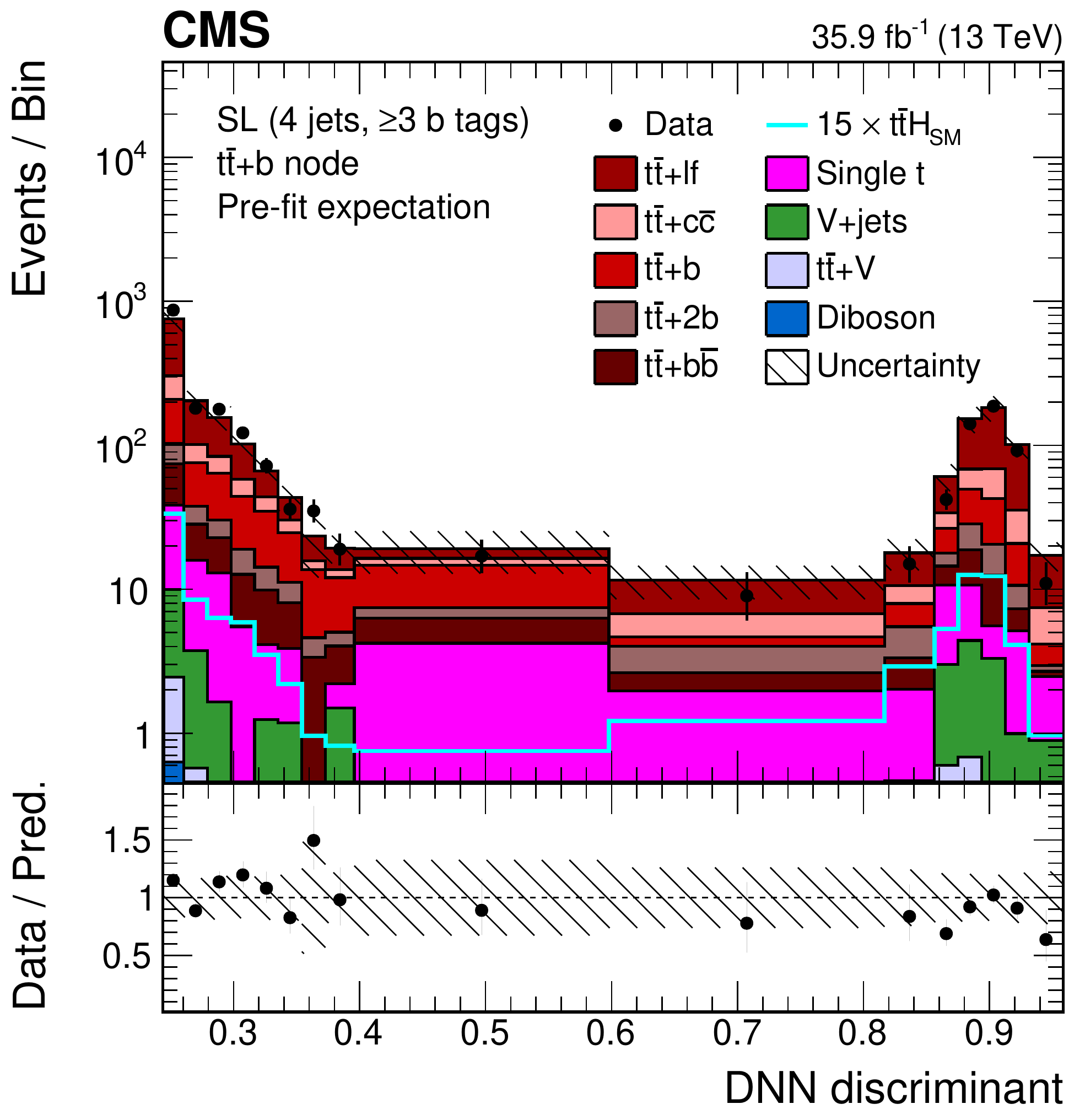}\\
    \includegraphics[width=0.35\textwidth]{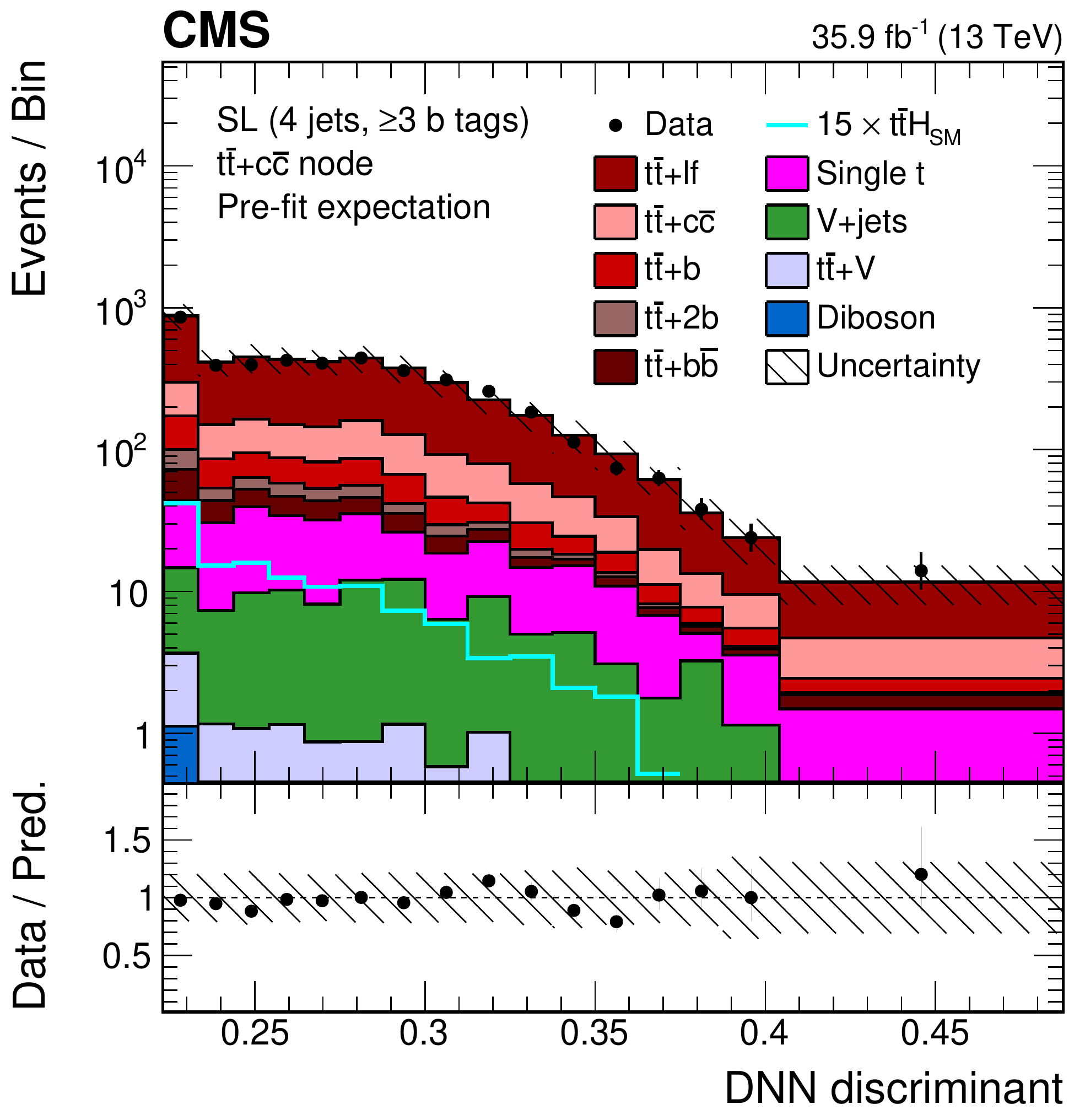} &
    \includegraphics[width=0.35\textwidth]{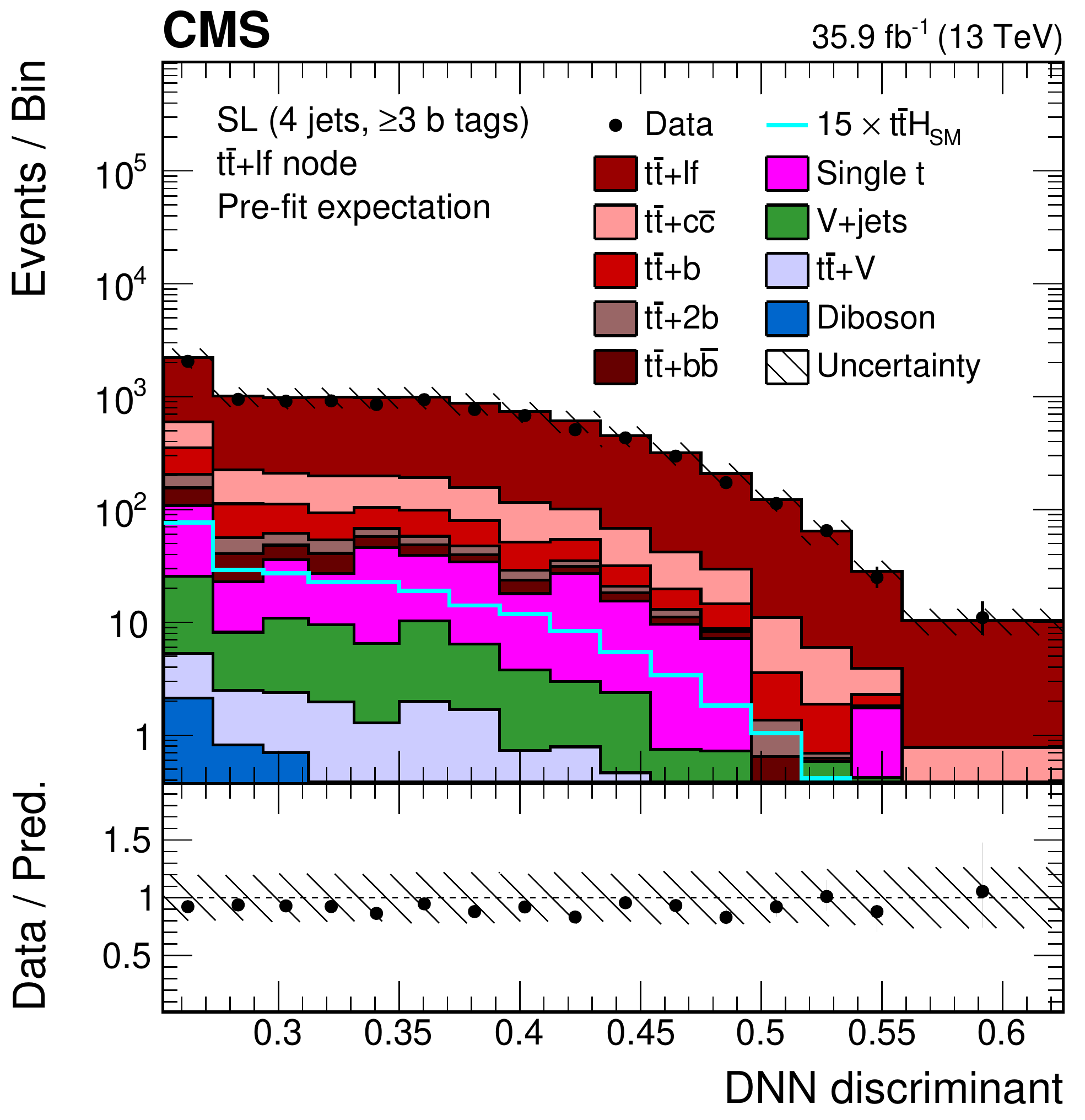}\\
    \end{tabular}
  \caption{Final discriminant (DNN) shapes in the single-lepton (SL) channel before the fit to data,
    in the jet-process categories with \ljFourThreeIncl and (from upper \cmsLeft to lower \cmsRight) \ttH, \ttbb, \tttwob, \ttb, \ttcc, and \ttlf .
    The expected background contributions (filled histograms) are stacked, and the expected signal
    distribution (line), which includes \Hbb and all other Higgs boson decay modes, is superimposed.
    Each contribution
    is normalised to an integrated luminosity of \lumivalue, and the
    signal distribution is additionally scaled by a factor of 15
    for better visibility.
    The hatched uncertainty bands include the total uncertainty of the fit model.
    The first and the last bins include underflow and overflow events, respectively.
    The lower plots show the ratio of the data to the background prediction.
  }
  \label{fig:appendix:prefit:ljdiscriminants_1}
\end{figure}

\begin{figure}[hbtp]
  \centering
  \begin{tabular}{c@{\hskip 0.05\textwidth}c}
    \includegraphics[width=0.35\textwidth]{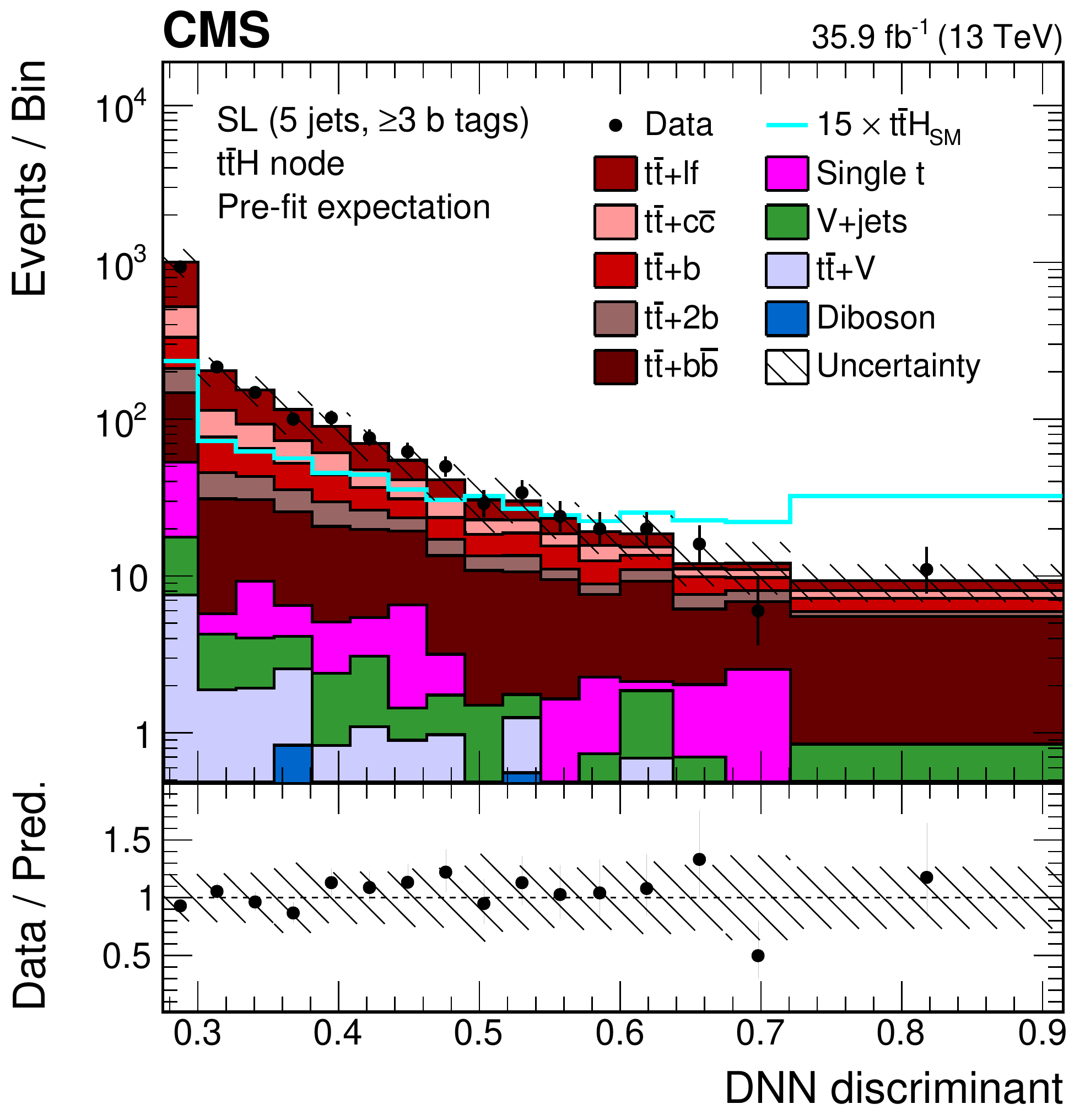} &
    \includegraphics[width=0.35\textwidth]{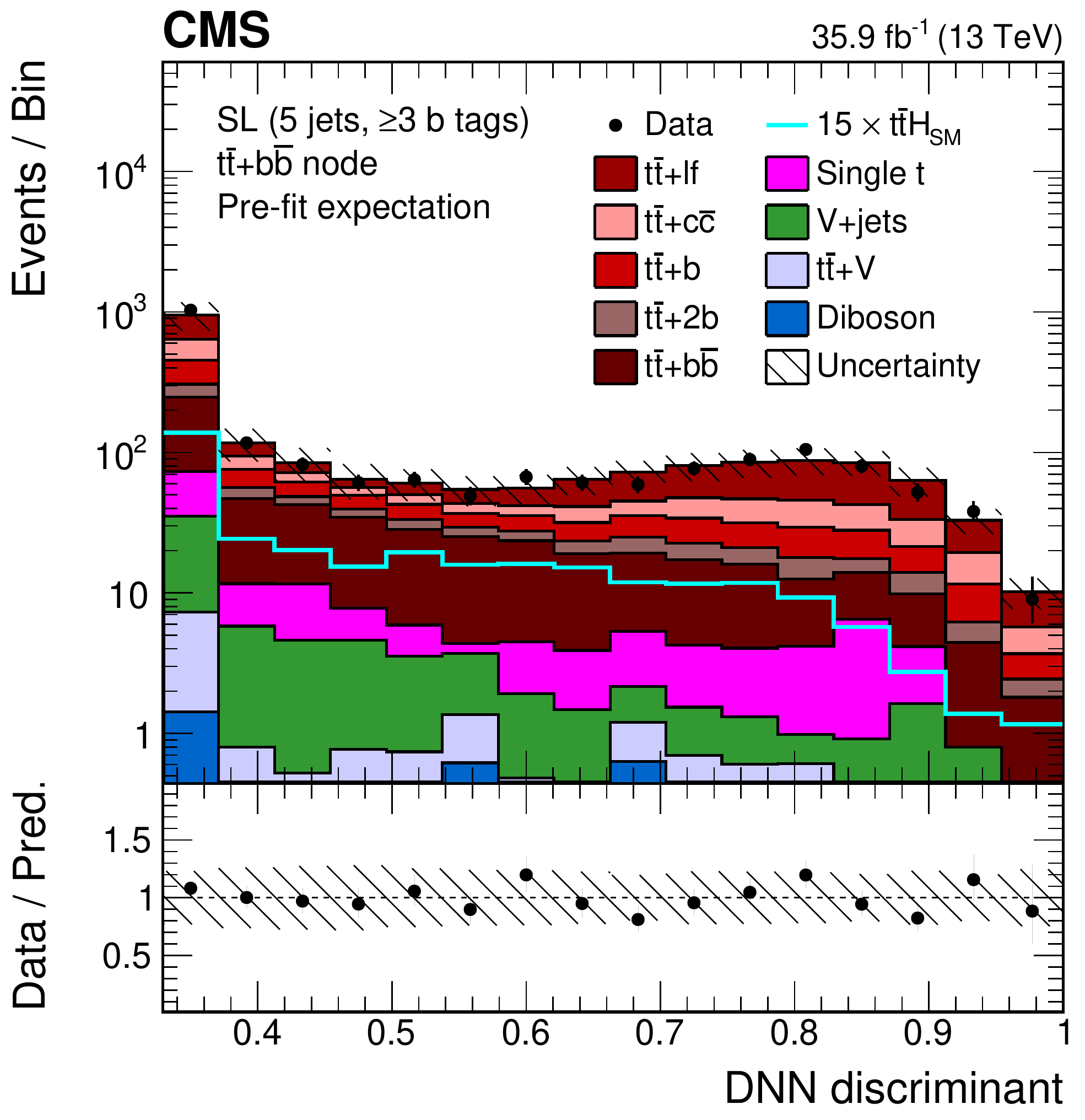}\\
    \includegraphics[width=0.35\textwidth]{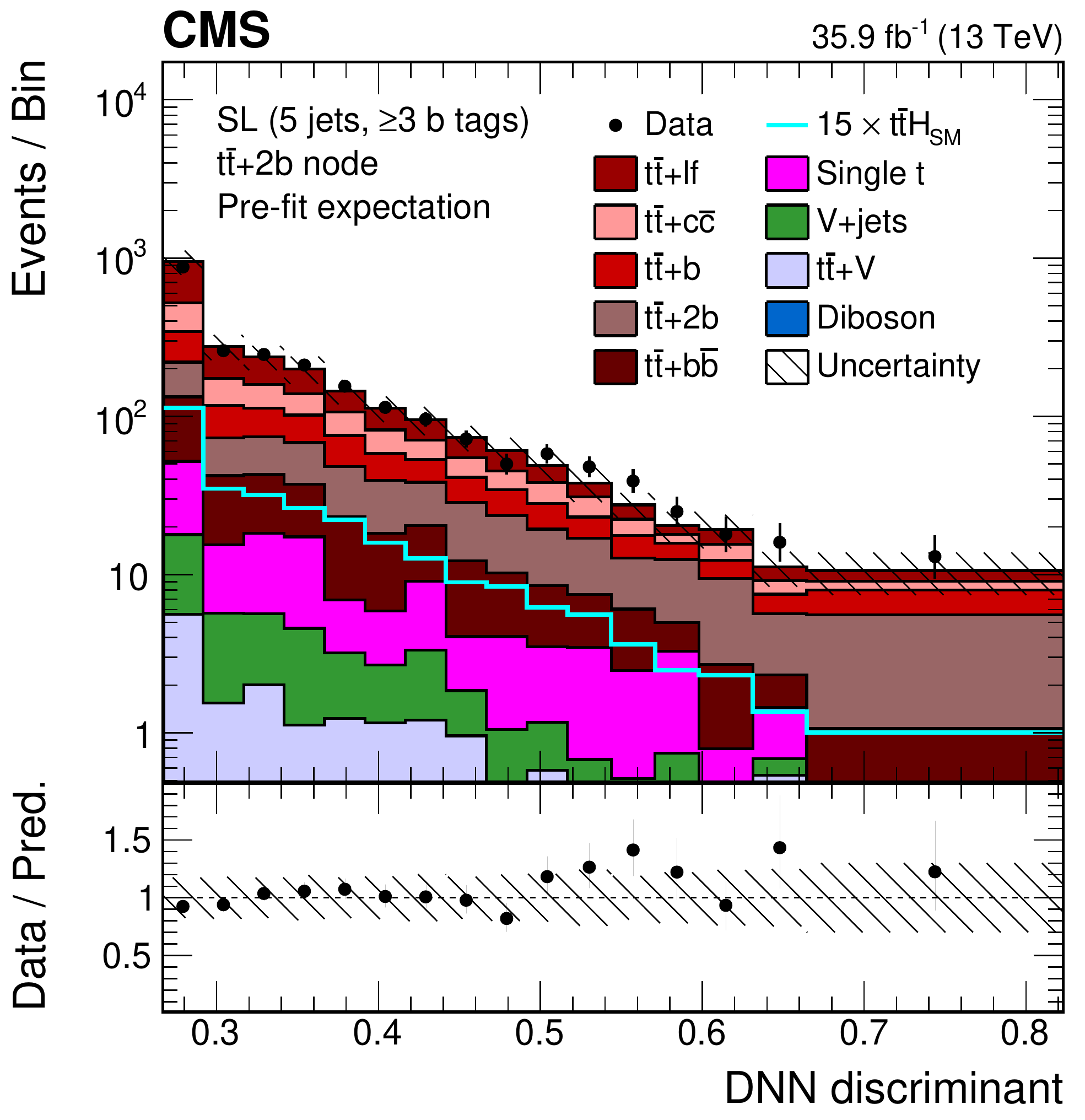} &
    \includegraphics[width=0.35\textwidth]{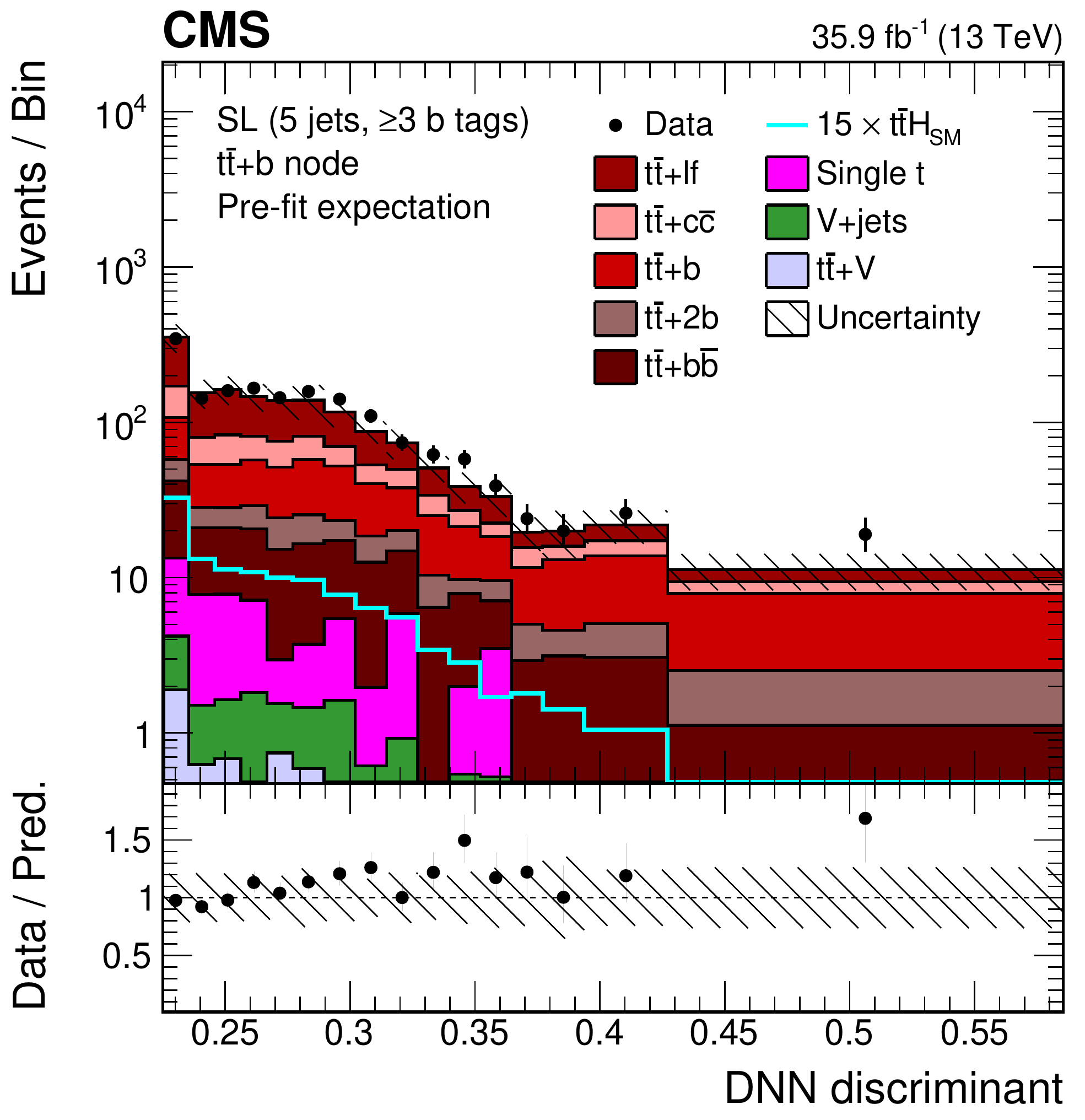}\\
    \includegraphics[width=0.35\textwidth]{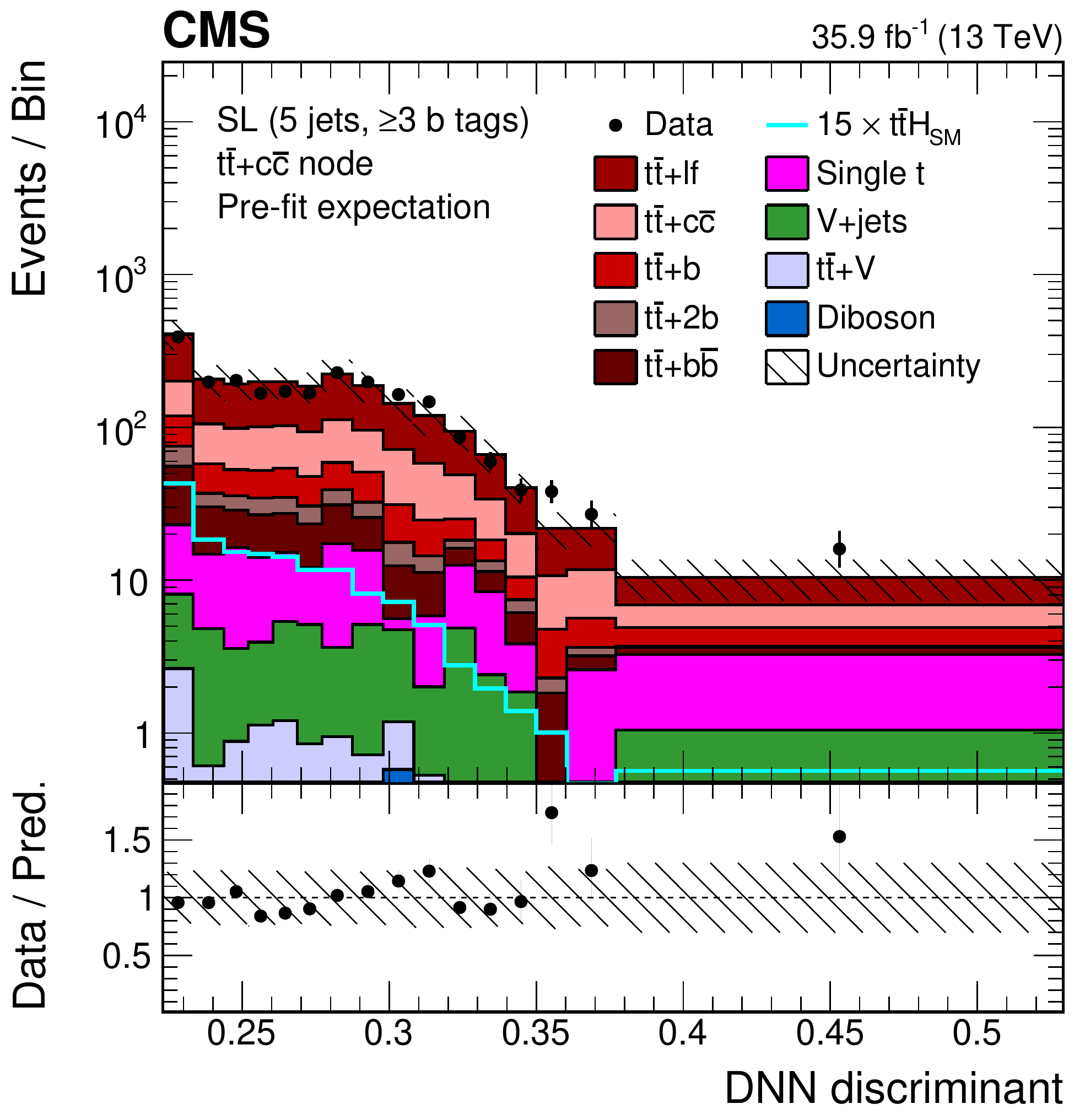} &
    \includegraphics[width=0.35\textwidth]{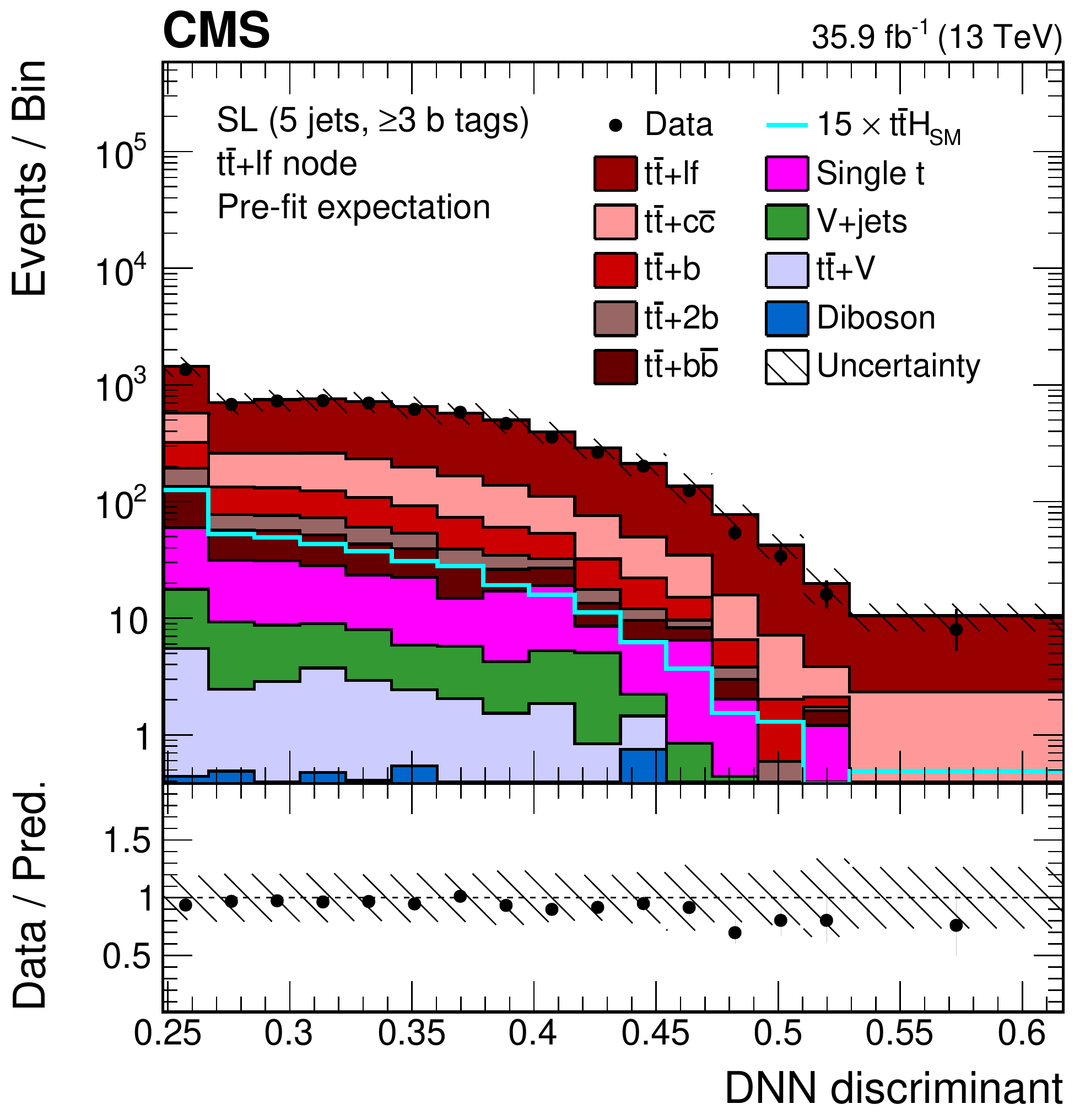}\\
    \end{tabular}
  \caption{Final discriminant (DNN) shapes in the single-lepton (SL) channel before the fit to data,
    in the jet-process categories with  \ljFiveThreeIncl and (from upper \cmsLeft to lower \cmsRight) \ttH, \ttbb, \tttwob, \ttb, \ttcc, and \ttlf .
    The expected background contributions (filled histograms) are stacked, and the expected signal
    distribution (line), which includes \Hbb and all other Higgs boson decay modes, is superimposed.
    Each contribution
    is normalised to an integrated luminosity of \lumivalue, and the
    signal distribution is additionally scaled by a factor of 15
    for better visibility.
    The hatched uncertainty bands include the total uncertainty of the fit model.
    The first and the last bins include underflow and overflow events, respectively.
    The lower plots show the ratio of the data to the background prediction.
  }
  \label{fig:appendix:prefit:ljdiscriminants_3}
\end{figure}

\begin{figure}[hbtp]
  \centering
  \begin{tabular}{c@{\hskip 0.05\textwidth}c}
    \includegraphics[width=0.35\textwidth]{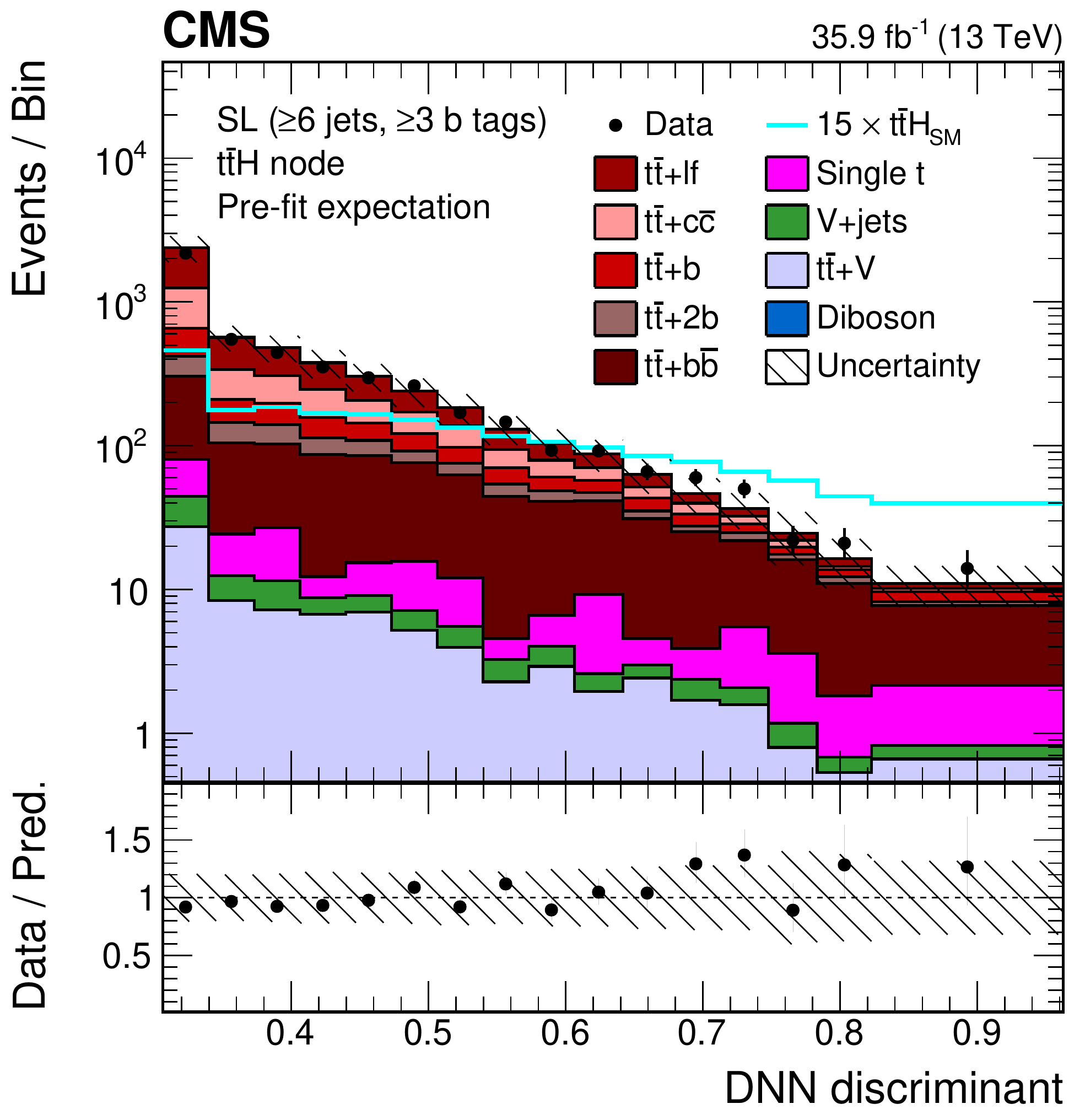} &
    \includegraphics[width=0.35\textwidth]{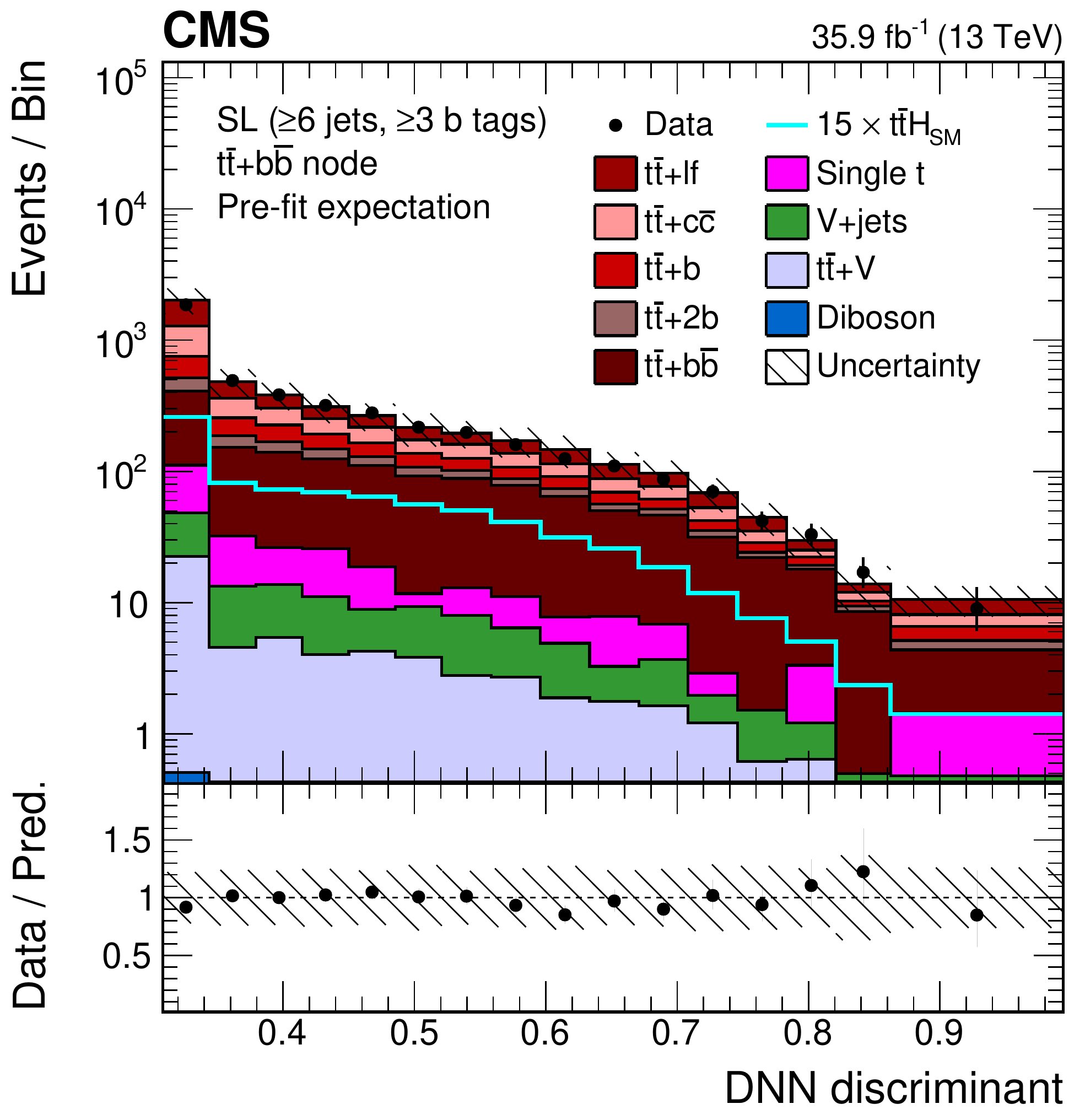}\\
    \includegraphics[width=0.35\textwidth]{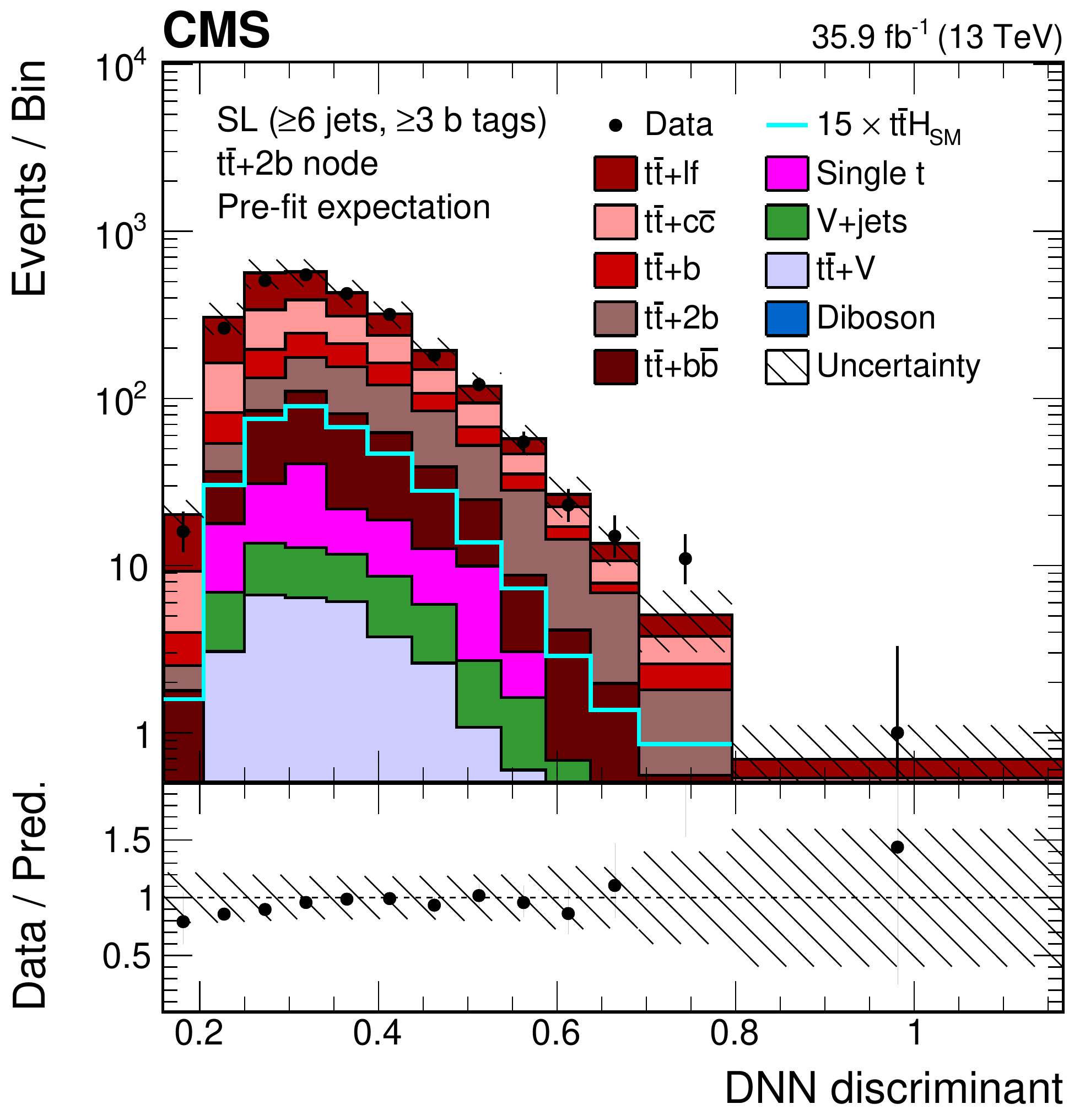} &
    \includegraphics[width=0.35\textwidth]{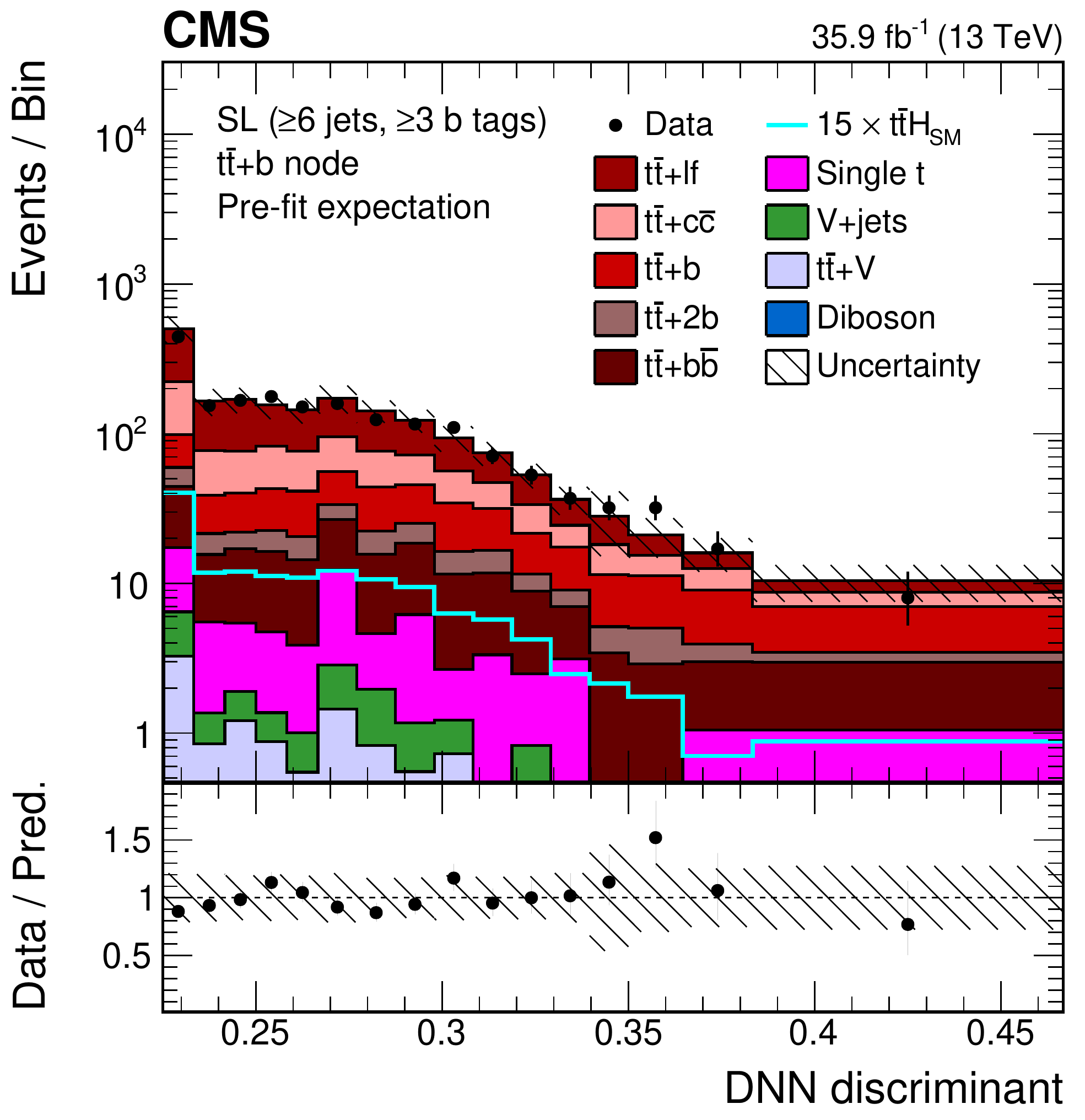}\\
    \includegraphics[width=0.35\textwidth]{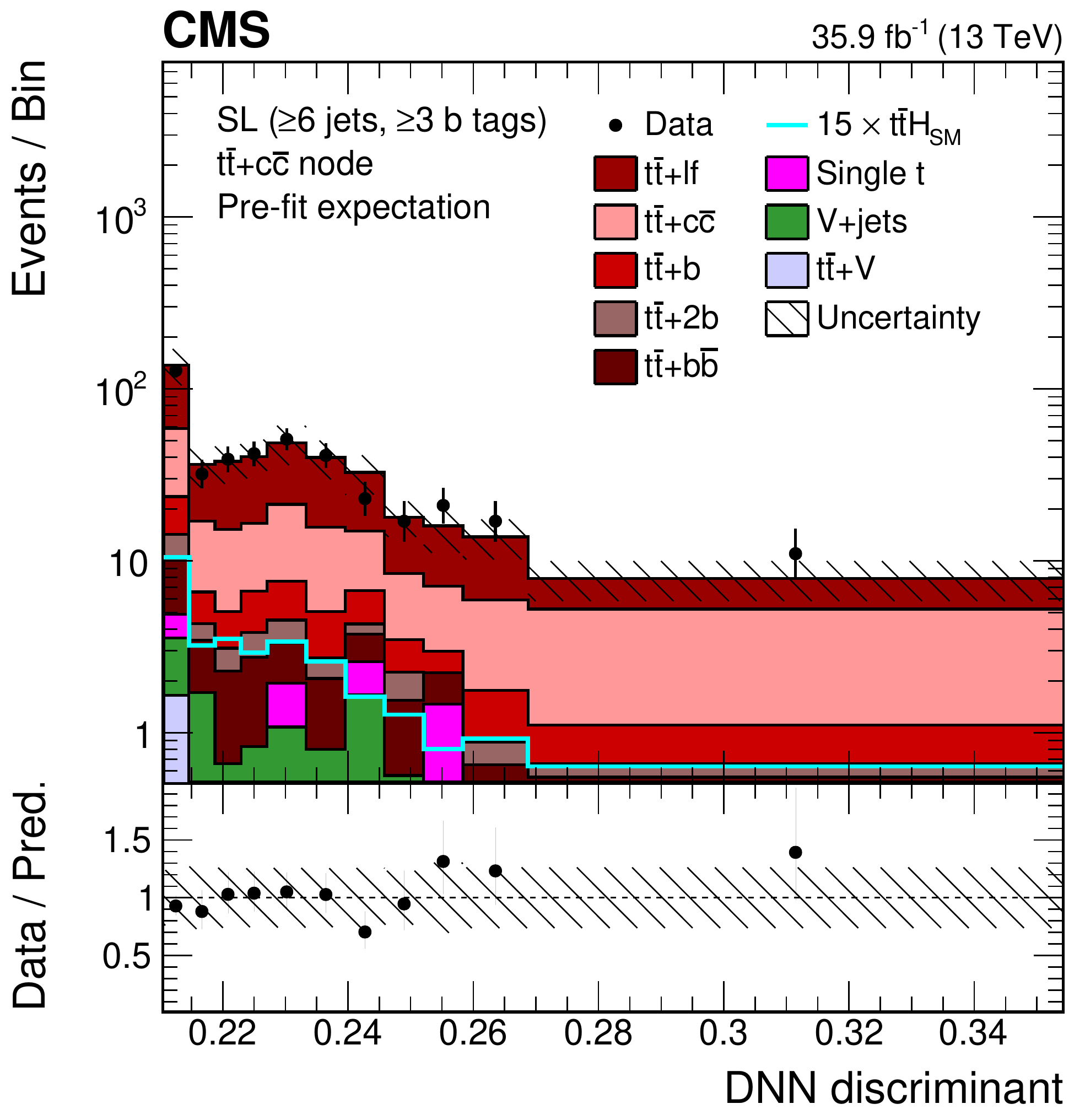} &
    \includegraphics[width=0.35\textwidth]{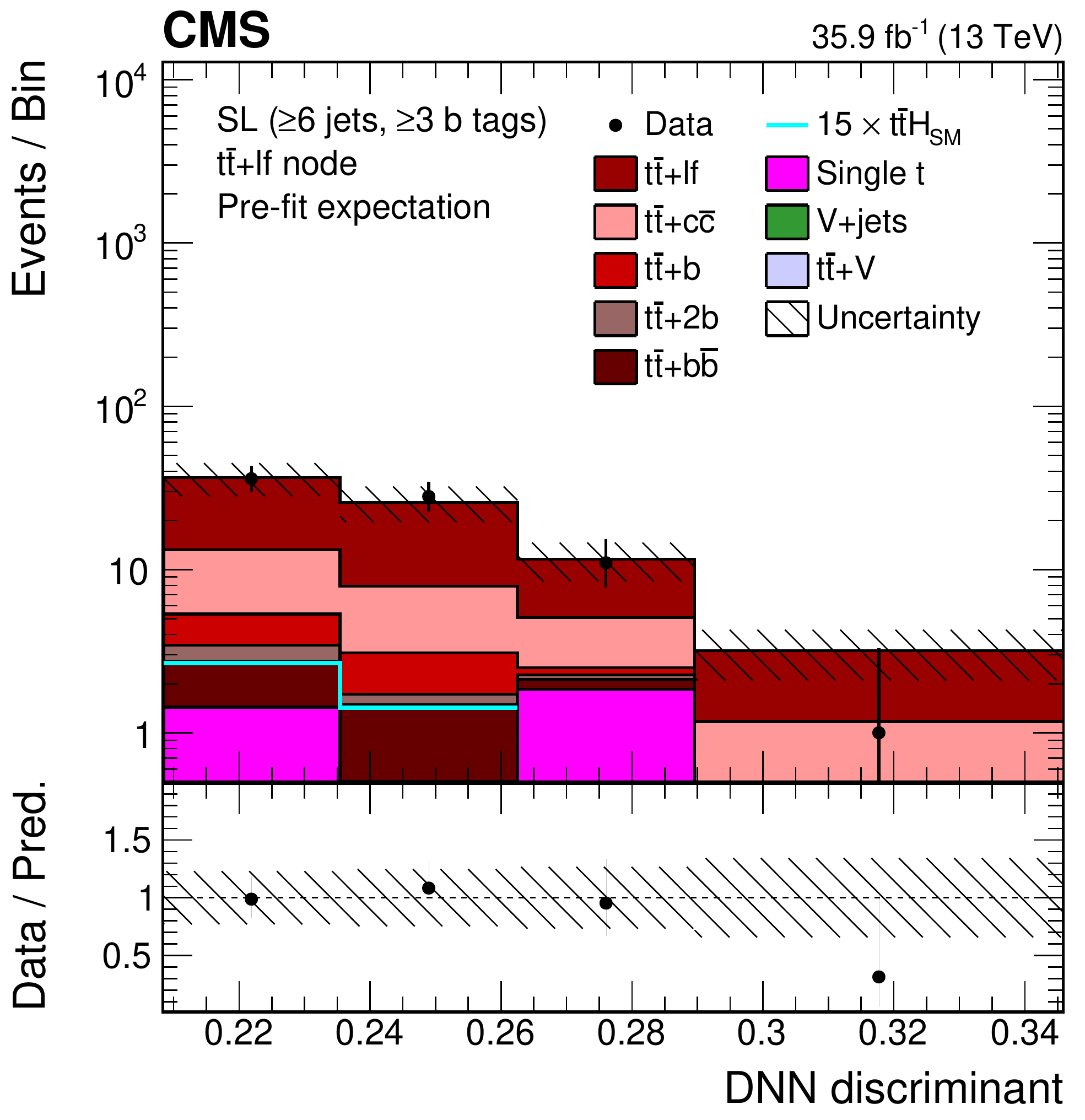}\\
    \end{tabular}
  \caption{Final discriminant (DNN) shapes in the single-lepton (SL) channel before the fit to data,
    in the jet-process categories with \ljSixThreeIncl and (from upper \cmsLeft to lower \cmsRight) \ttH, \ttbb, \tttwob, \ttb, \ttcc, and \ttlf .
    The expected background contributions (filled histograms) are stacked, and the expected signal
    distribution (line), which includes \Hbb and all other Higgs boson decay modes, is superimposed.
    Each contribution
    is normalised to an integrated luminosity of \lumivalue, and the
    signal distribution is additionally scaled by a factor of 15
    for better visibility.
    The hatched uncertainty bands include the total uncertainty of the fit model.
    The first and the last bins include underflow and overflow events, respectively.
    The lower plots show the ratio of the data to the background prediction.
  }
  \label{fig:appendix:prefit:ljdiscriminants_5}
\end{figure}

\clearpage

\section{Post-fit discriminant shapes (single-lepton channel)}
\label{sec:appendix:postfitshapes}
\begin{figure}[!!!!hbtp]
  \centering
  \begin{tabular}{c@{\hskip 0.05\textwidth}c}
    \includegraphics[width=0.35\textwidth]{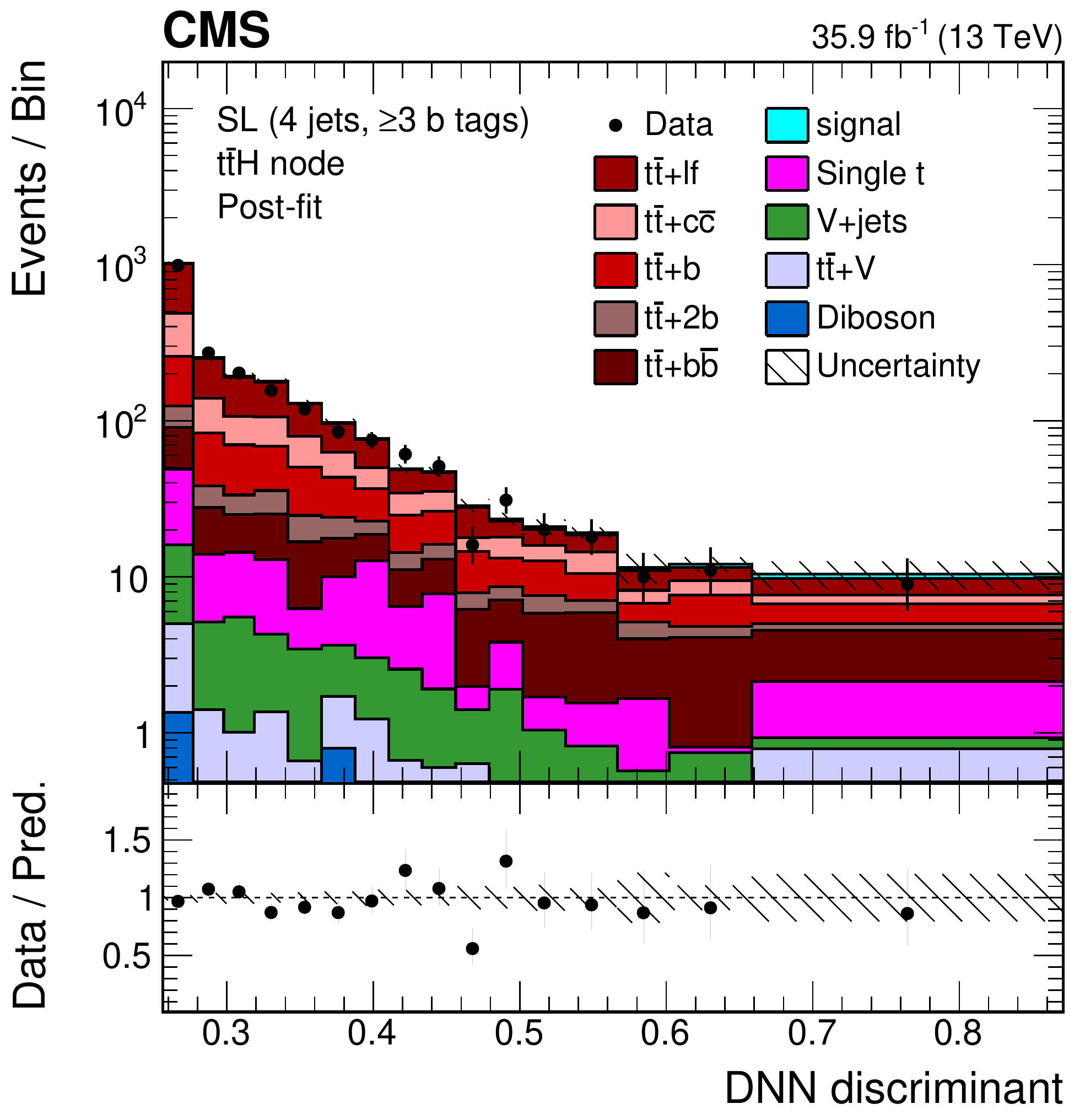} &
    \includegraphics[width=0.35\textwidth]{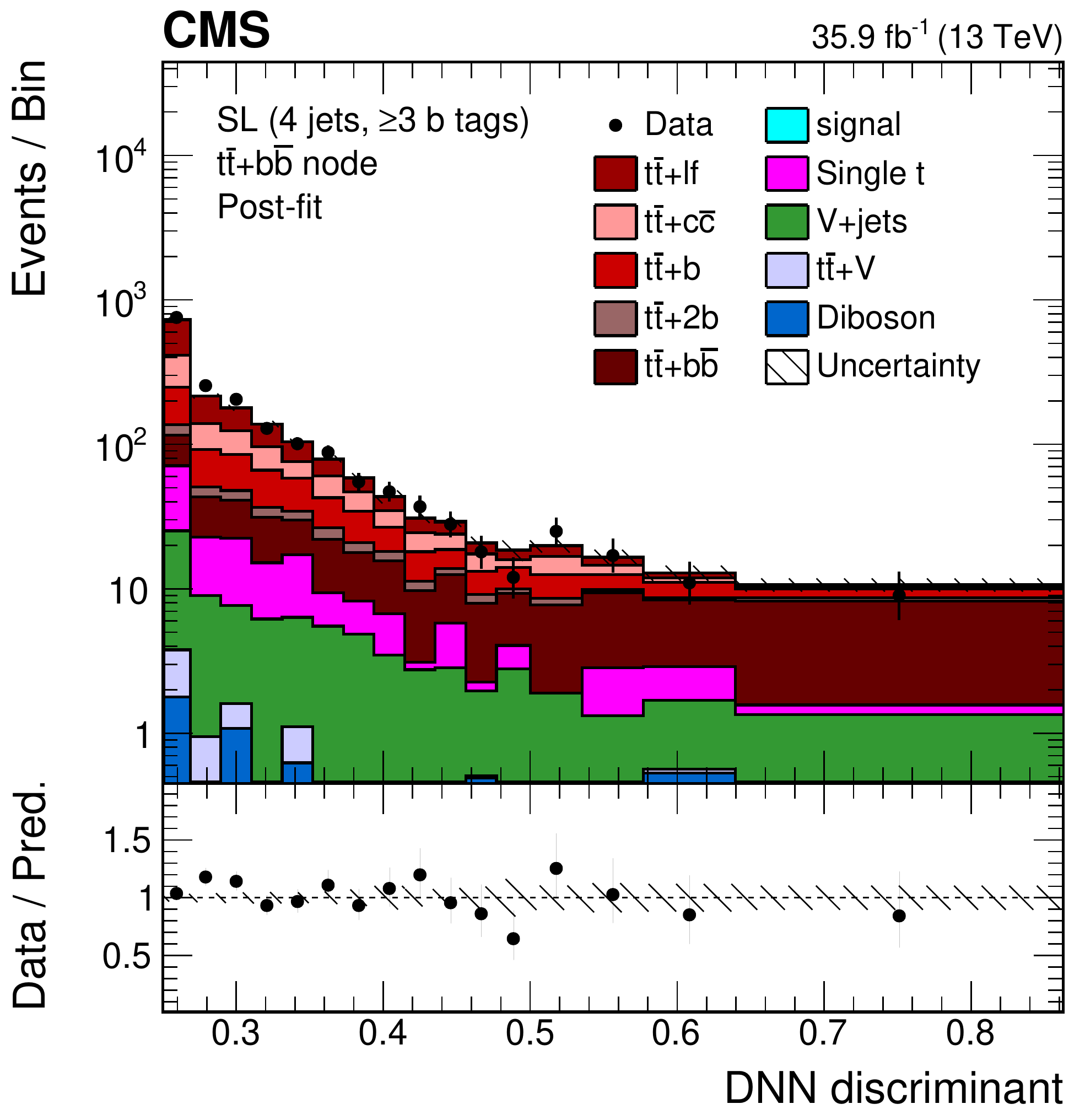}\\
    \includegraphics[width=0.35\textwidth]{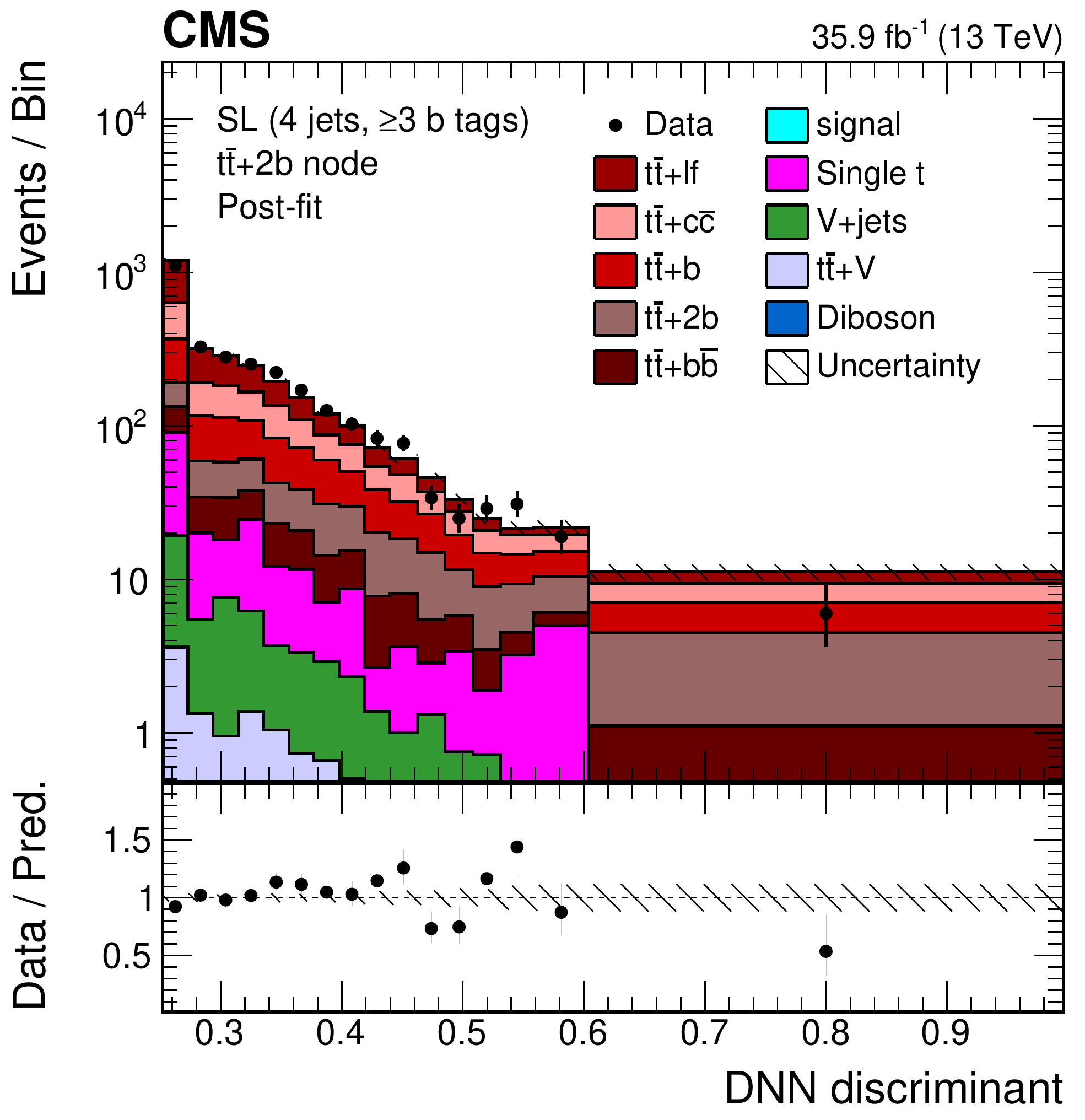} &
    \includegraphics[width=0.35\textwidth]{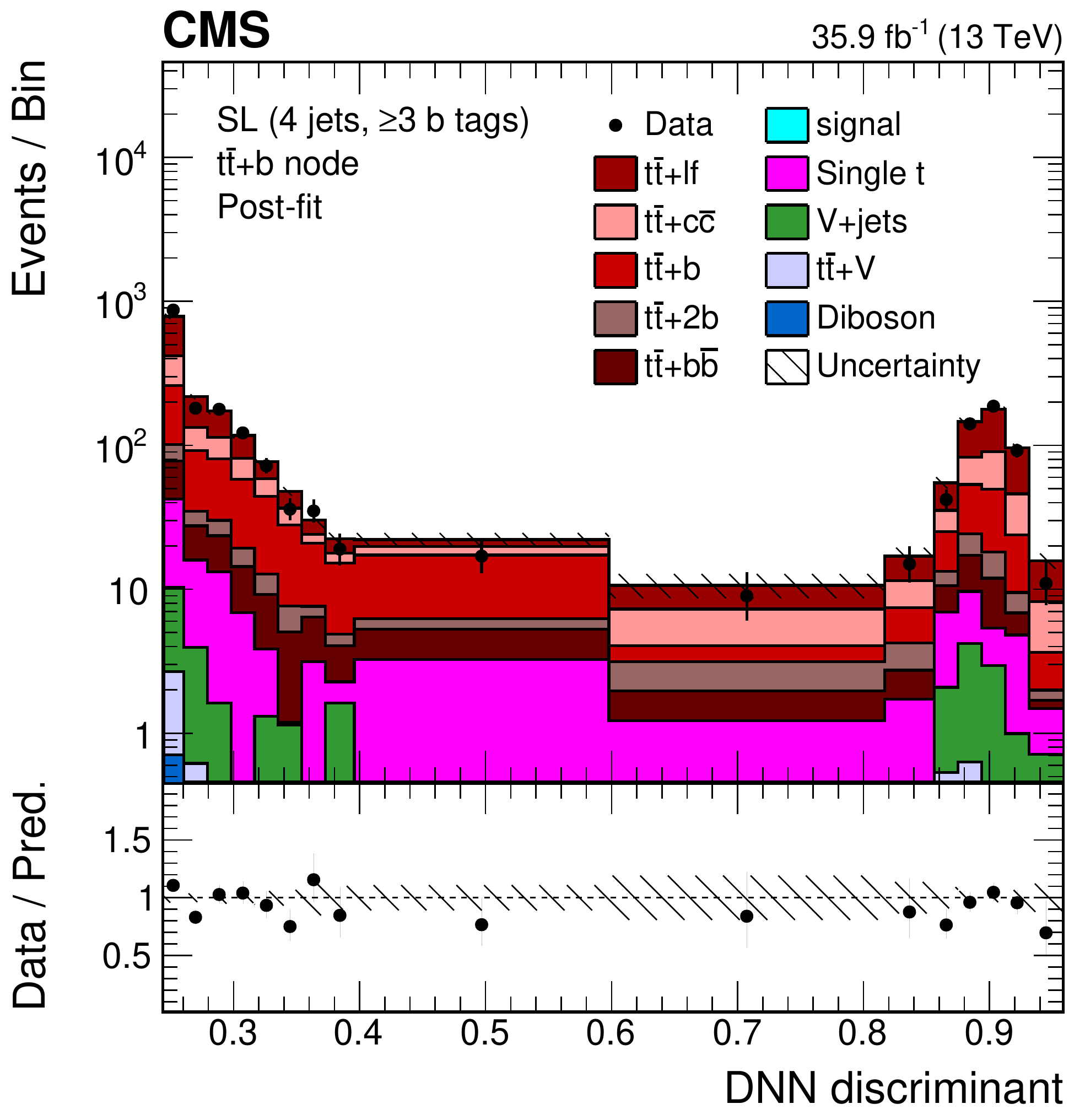}\\
    \includegraphics[width=0.35\textwidth]{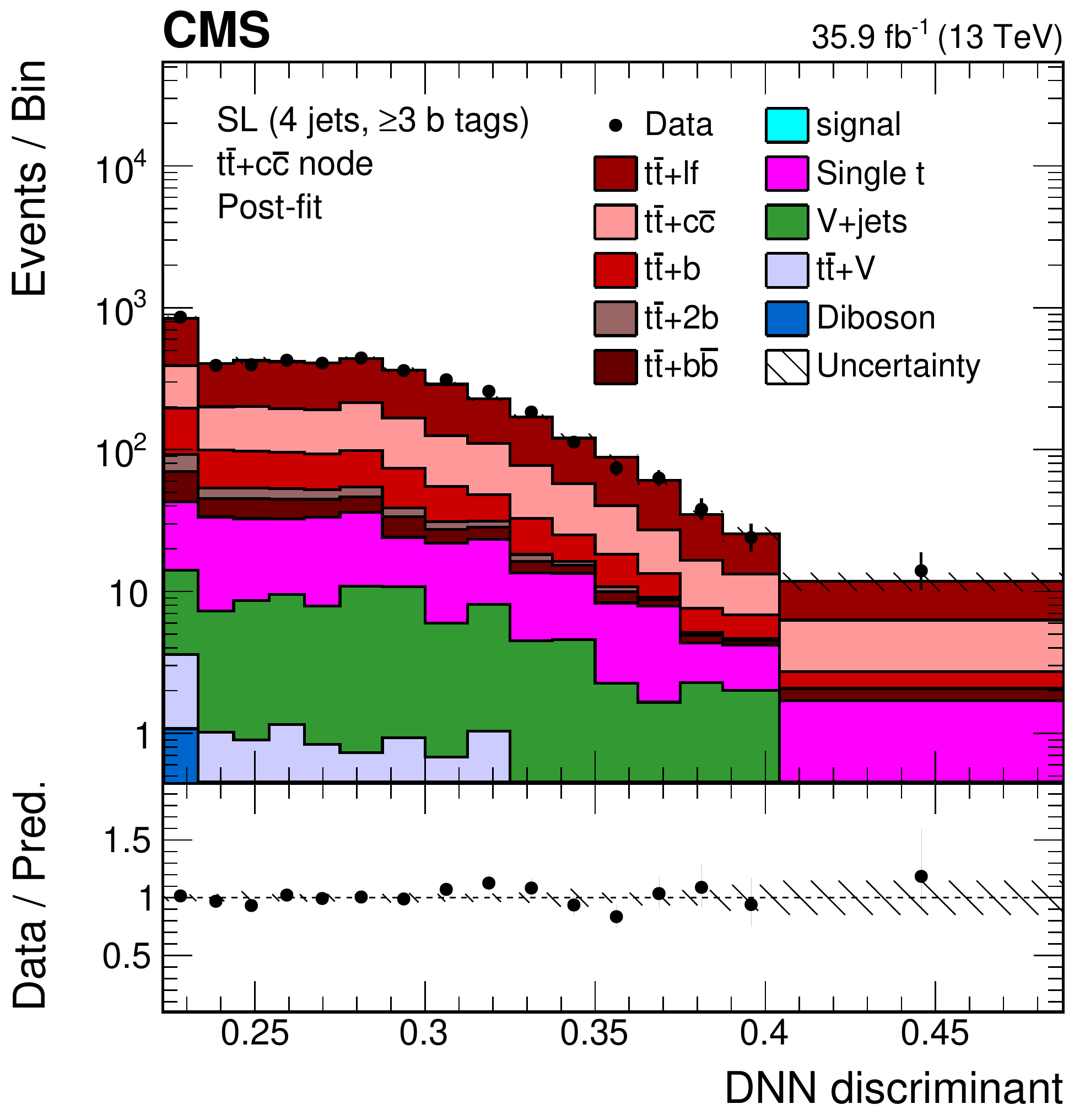} &
    \includegraphics[width=0.35\textwidth]{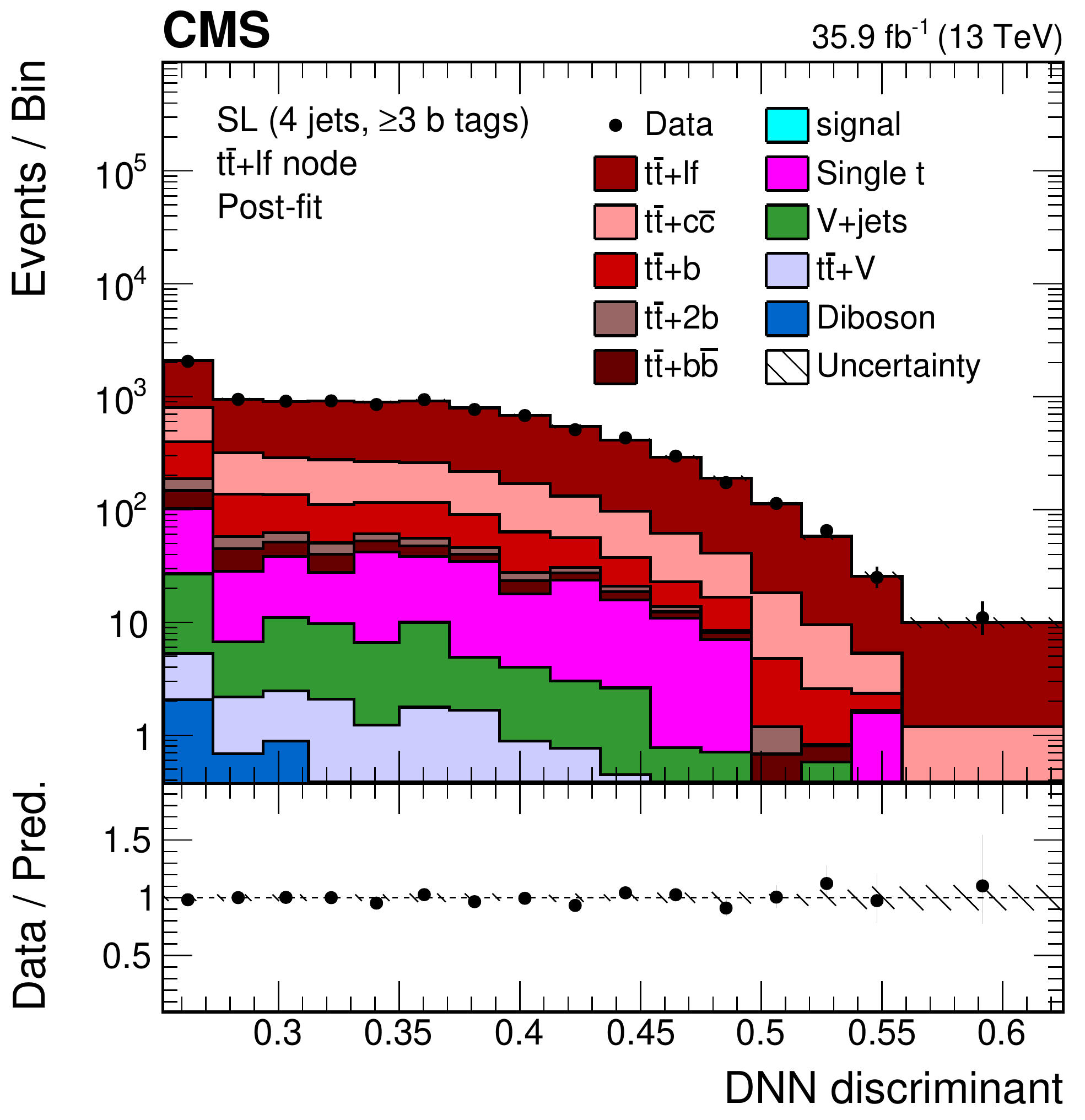}\\
    \end{tabular}
  \caption{Final discriminant (DNN) shapes in the single-lepton (SL) channel after the fit to data,
    in the jet-process categories with \ljFourThreeIncl and (from upper \cmsLeft to lower \cmsRight) \ttH, \ttbb, \tttwob, \ttb, \ttcc, and \ttlf .
    The error bands include the total uncertainty after the fit to data.
    The first and the last bins include underflow and overflow events, respectively.
    The lower plots show the ratio of the data to the post-fit background plus signal distribution.
  }
  \label{fig:appendix:postfit:ljdiscriminants_1}
\end{figure}

\begin{figure}[hbtp]
  \centering
  \begin{tabular}{c@{\hskip 0.05\textwidth}c}
    \includegraphics[width=0.35\textwidth]{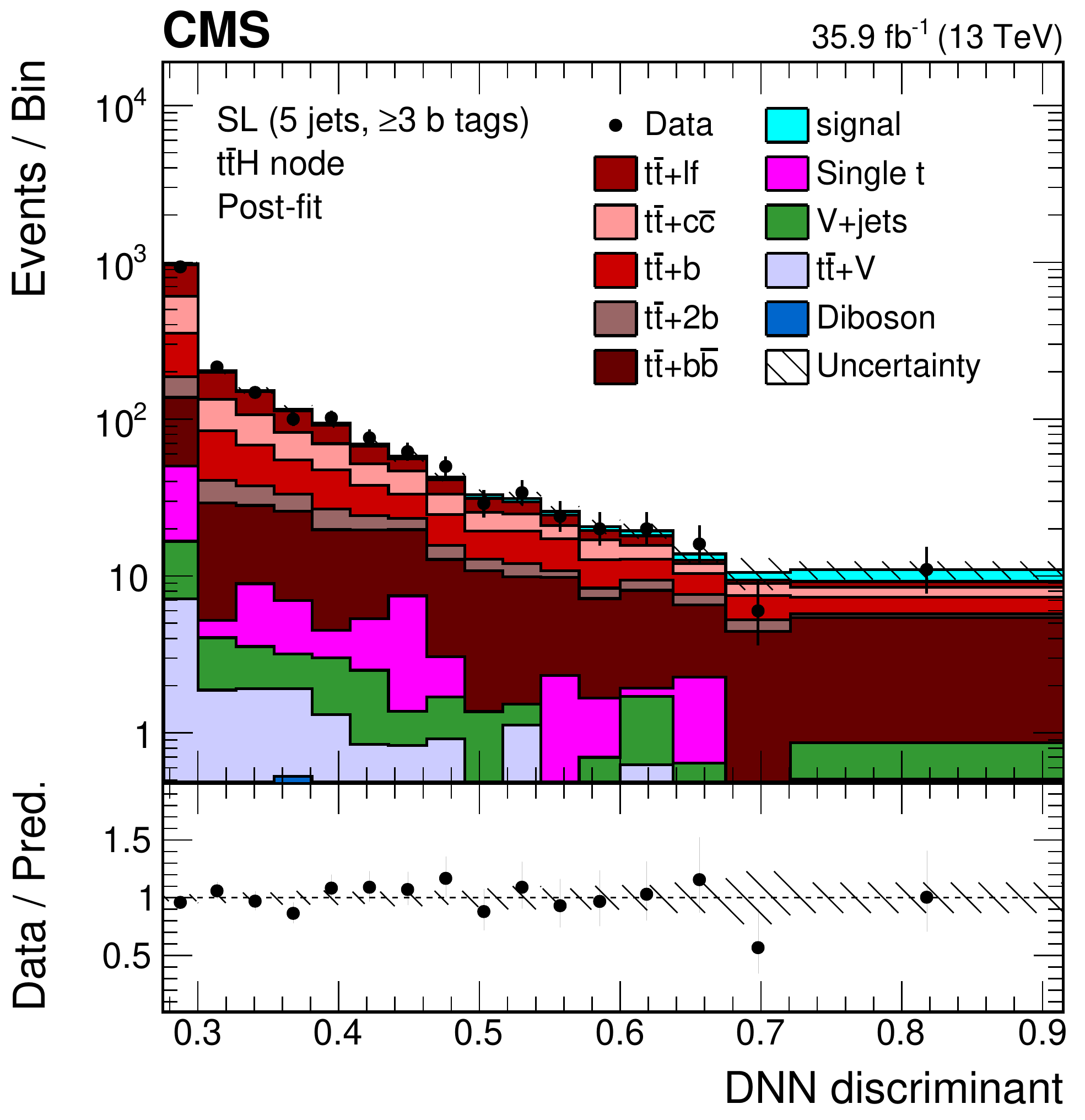} &
    \includegraphics[width=0.35\textwidth]{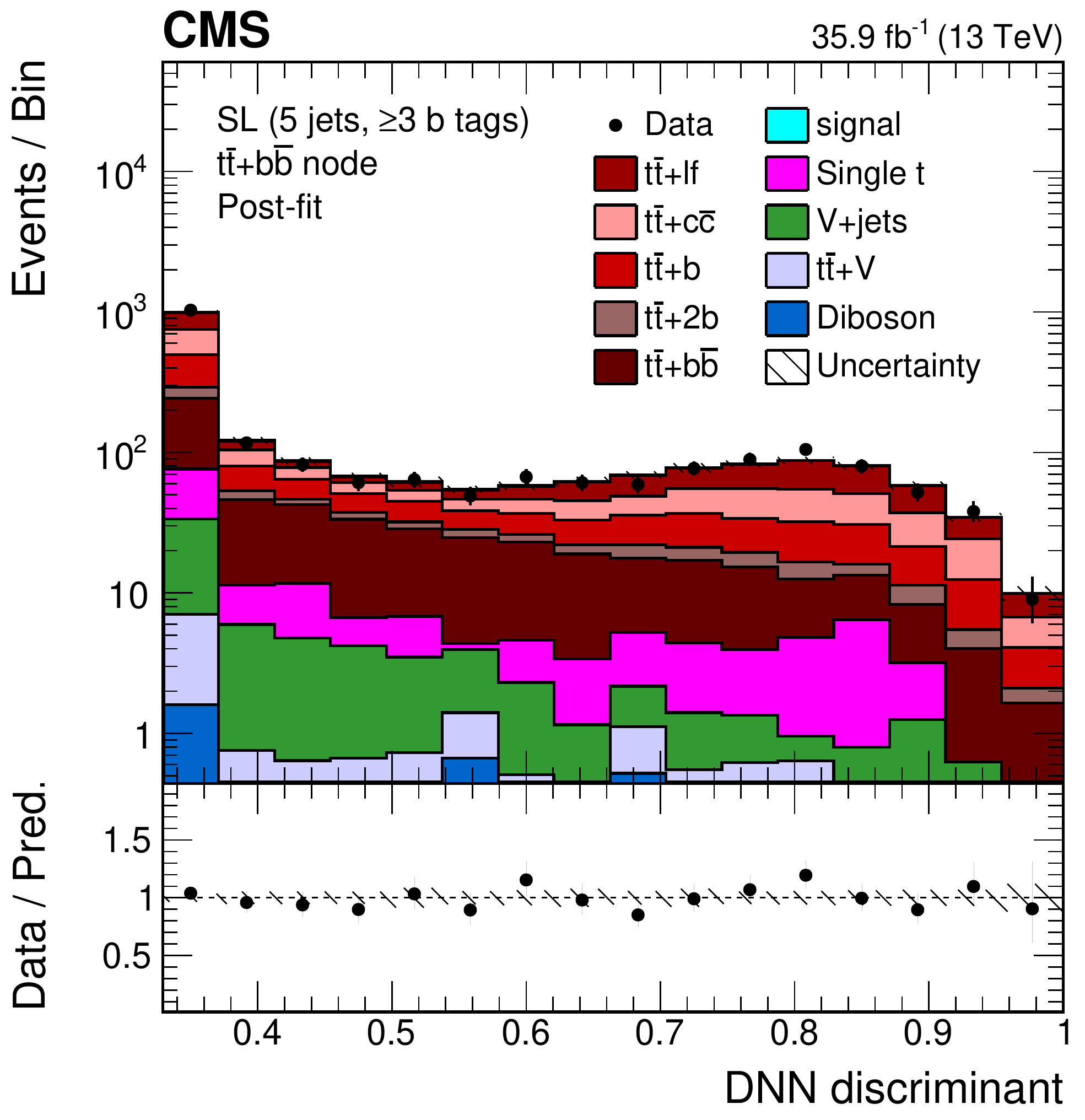}\\
    \includegraphics[width=0.35\textwidth]{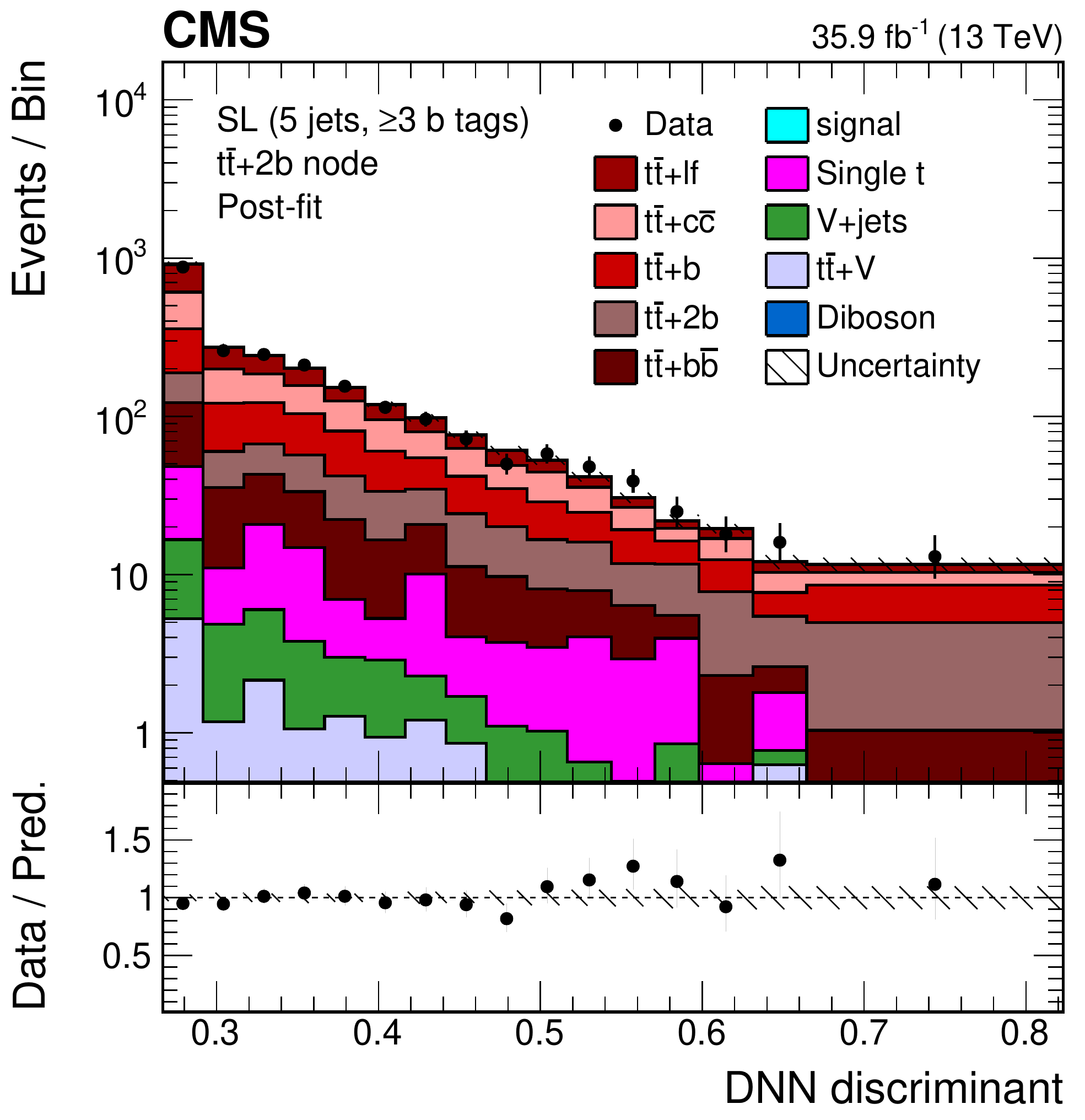} &
    \includegraphics[width=0.35\textwidth]{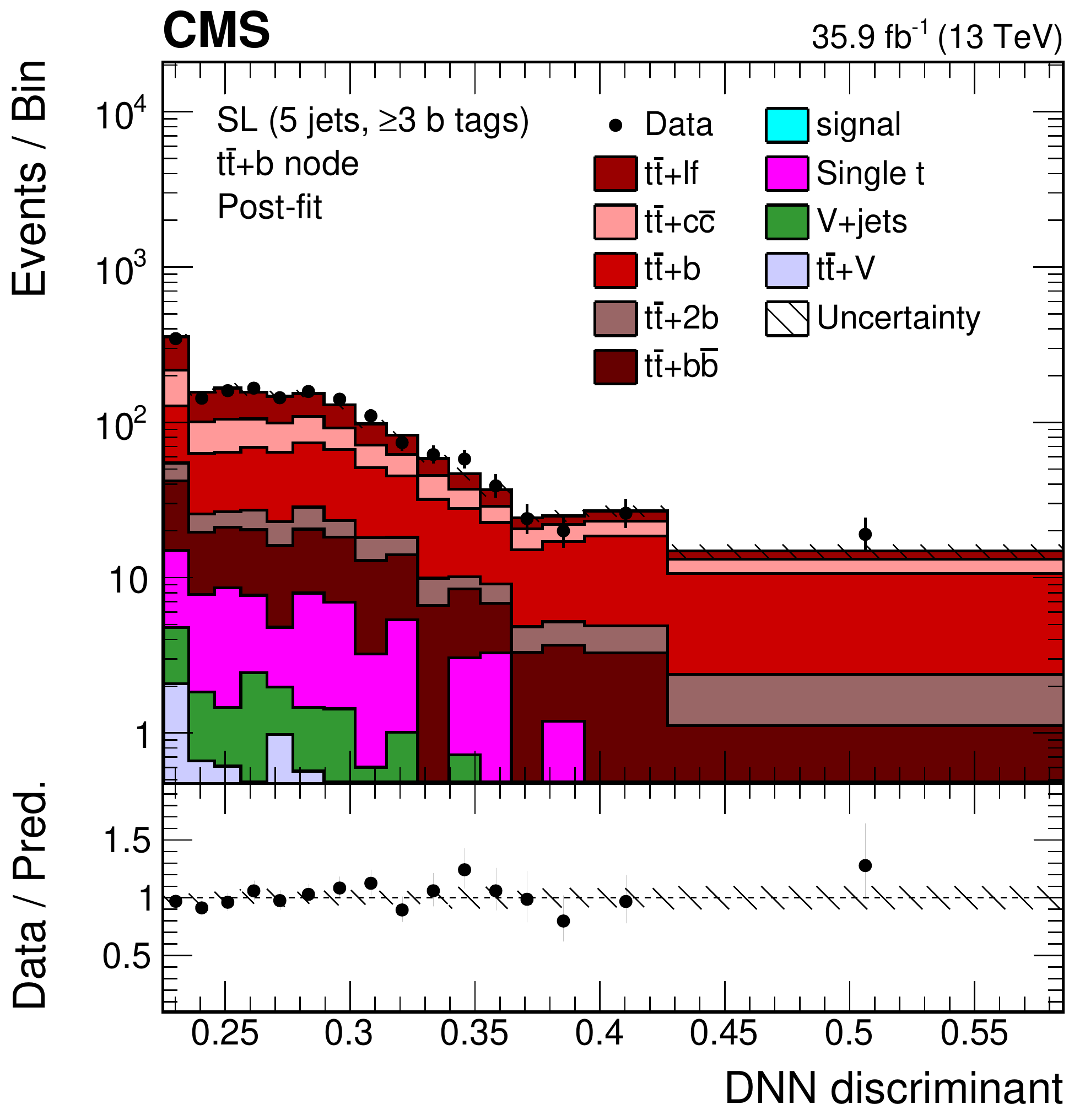}\\
    \includegraphics[width=0.35\textwidth]{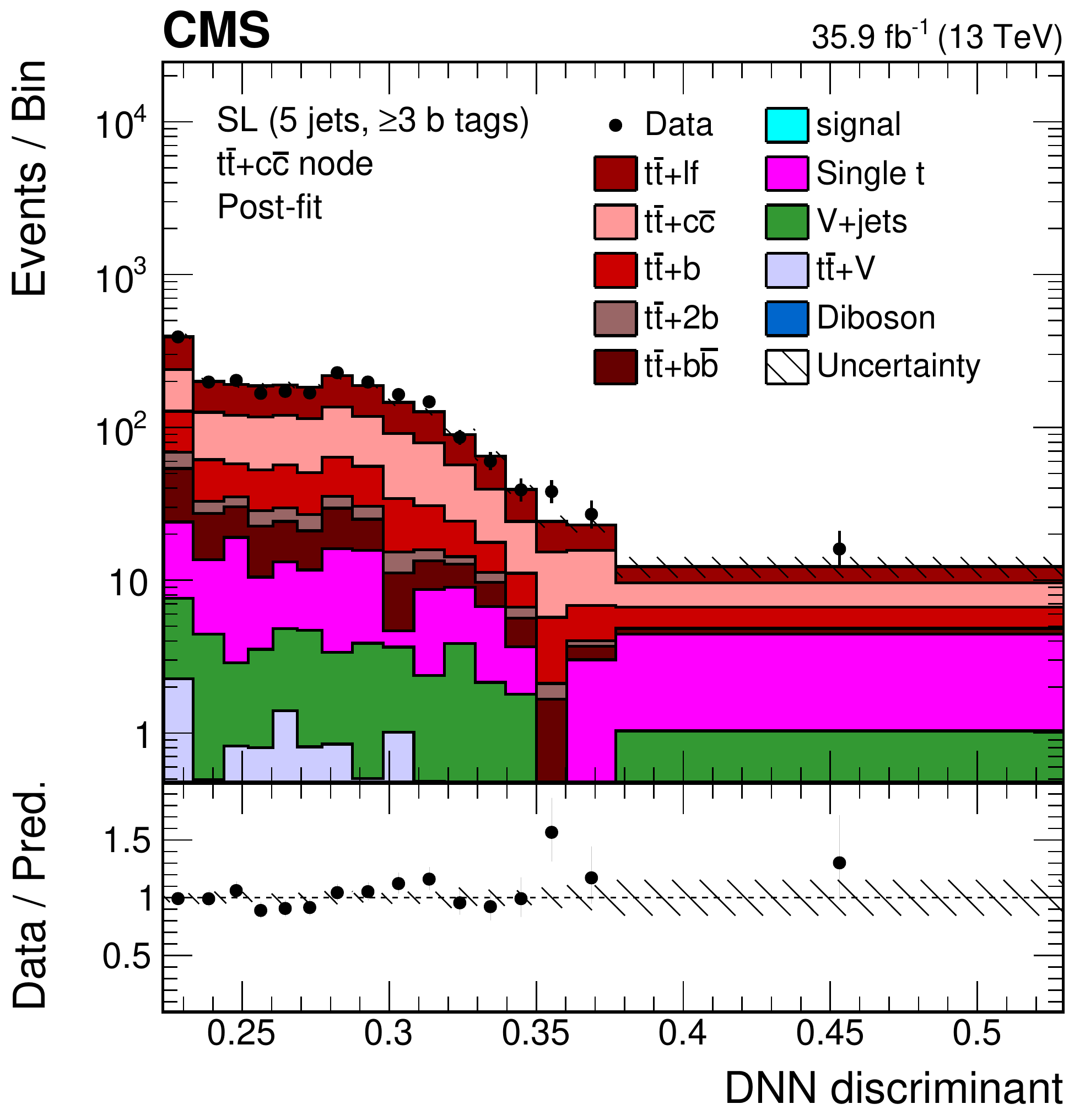} &
    \includegraphics[width=0.35\textwidth]{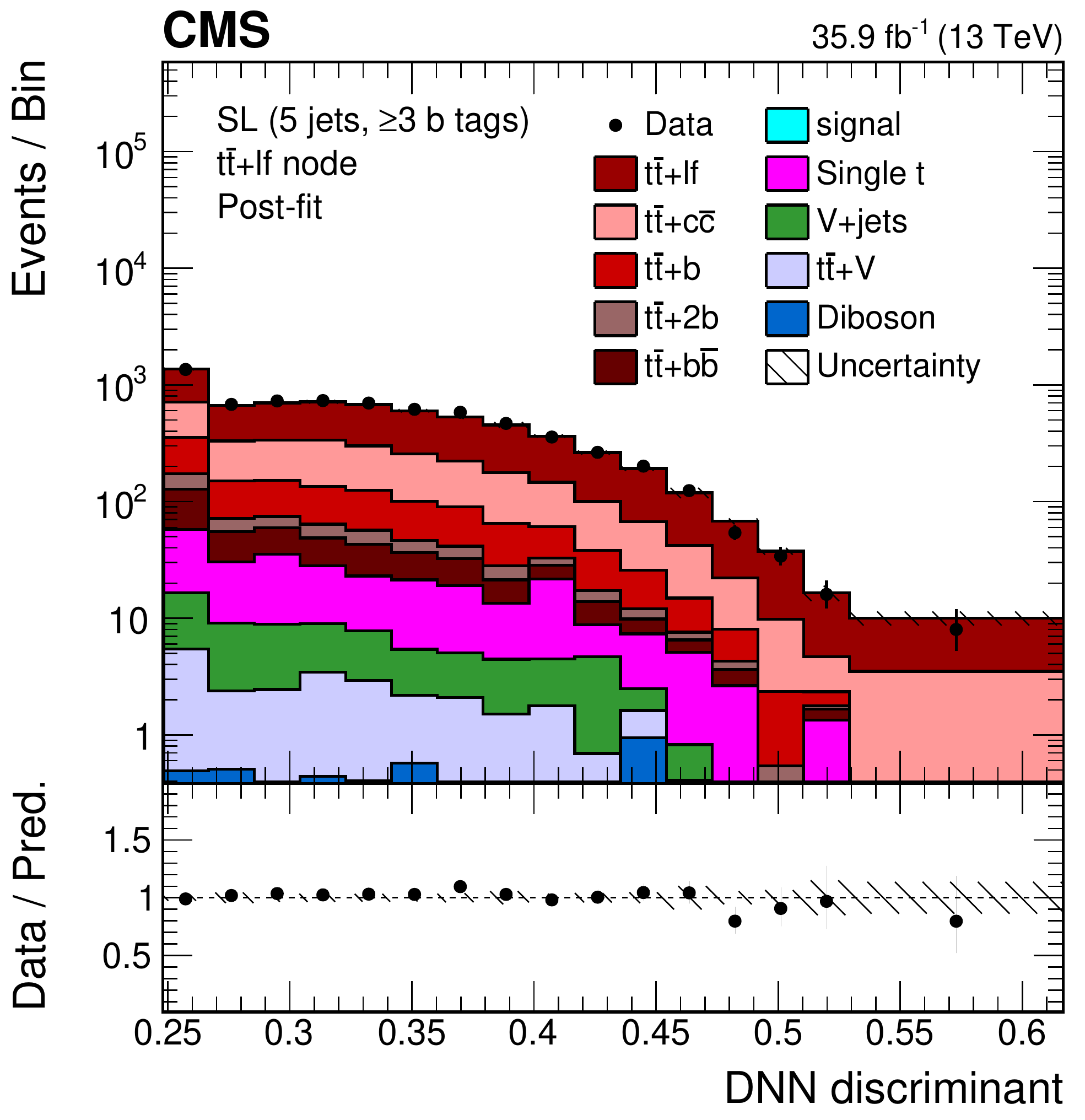}\\
    \end{tabular}
  \caption{Final discriminant (DNN) shapes in the single-lepton (SL) channel after the fit to data,
    in the jet-process categories with \ljFiveThreeIncl and (from upper \cmsLeft to lower \cmsRight) \ttH, \ttbb, \tttwob, \ttb, \ttcc, and \ttlf .
    The error bands include the total uncertainty after the fit to data.
    The first and the last bins include underflow and overflow events, respectively.
    The lower plots show the ratio of the data to the post-fit background plus signal distribution.
  }
  \label{fig:appendix:postfit:ljdiscriminants_3}
\end{figure}

\begin{figure}[hbtp]
  \centering
  \begin{tabular}{c@{\hskip 0.05\textwidth}c}
    \includegraphics[width=0.35\textwidth]{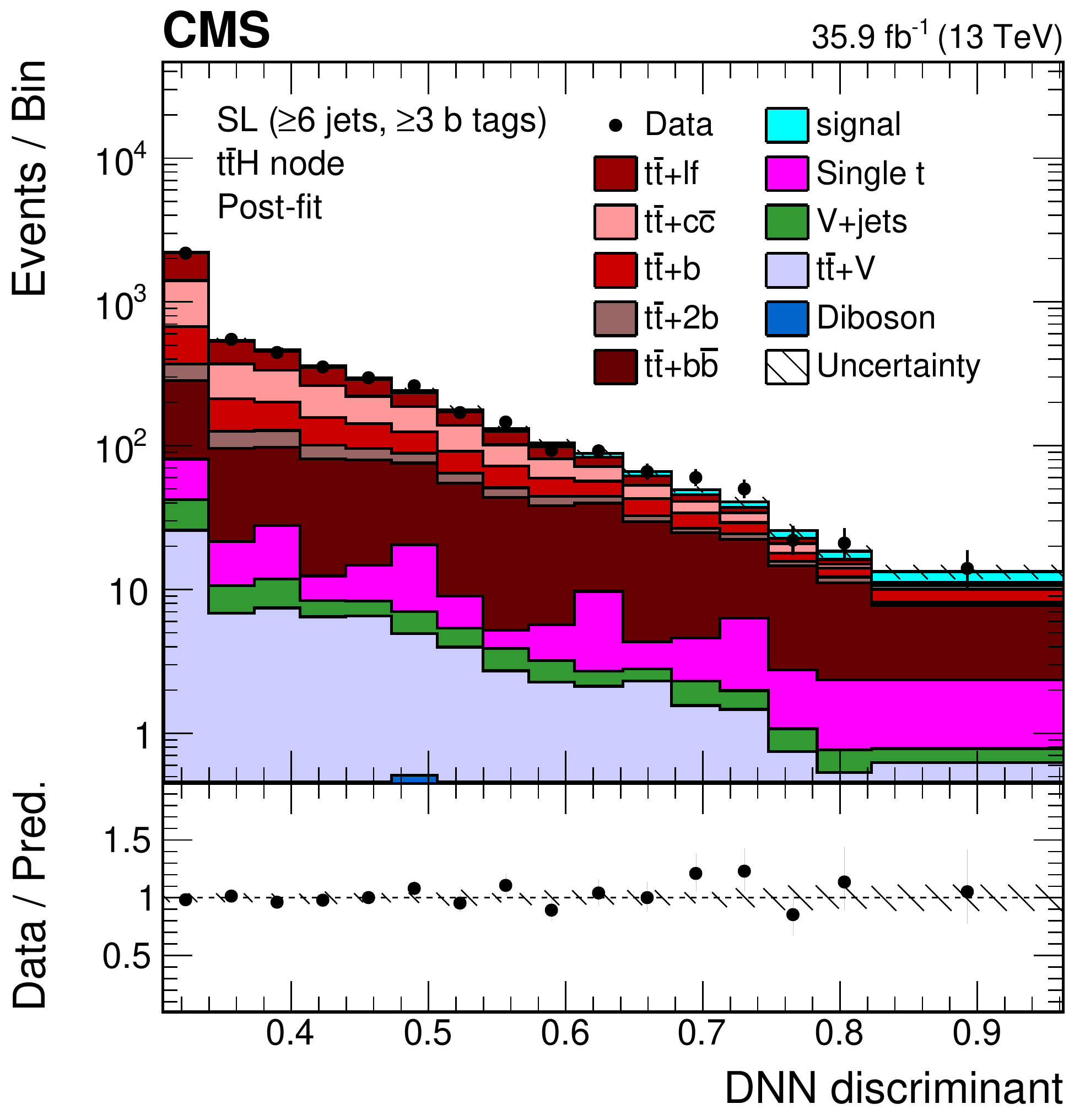} &
    \includegraphics[width=0.35\textwidth]{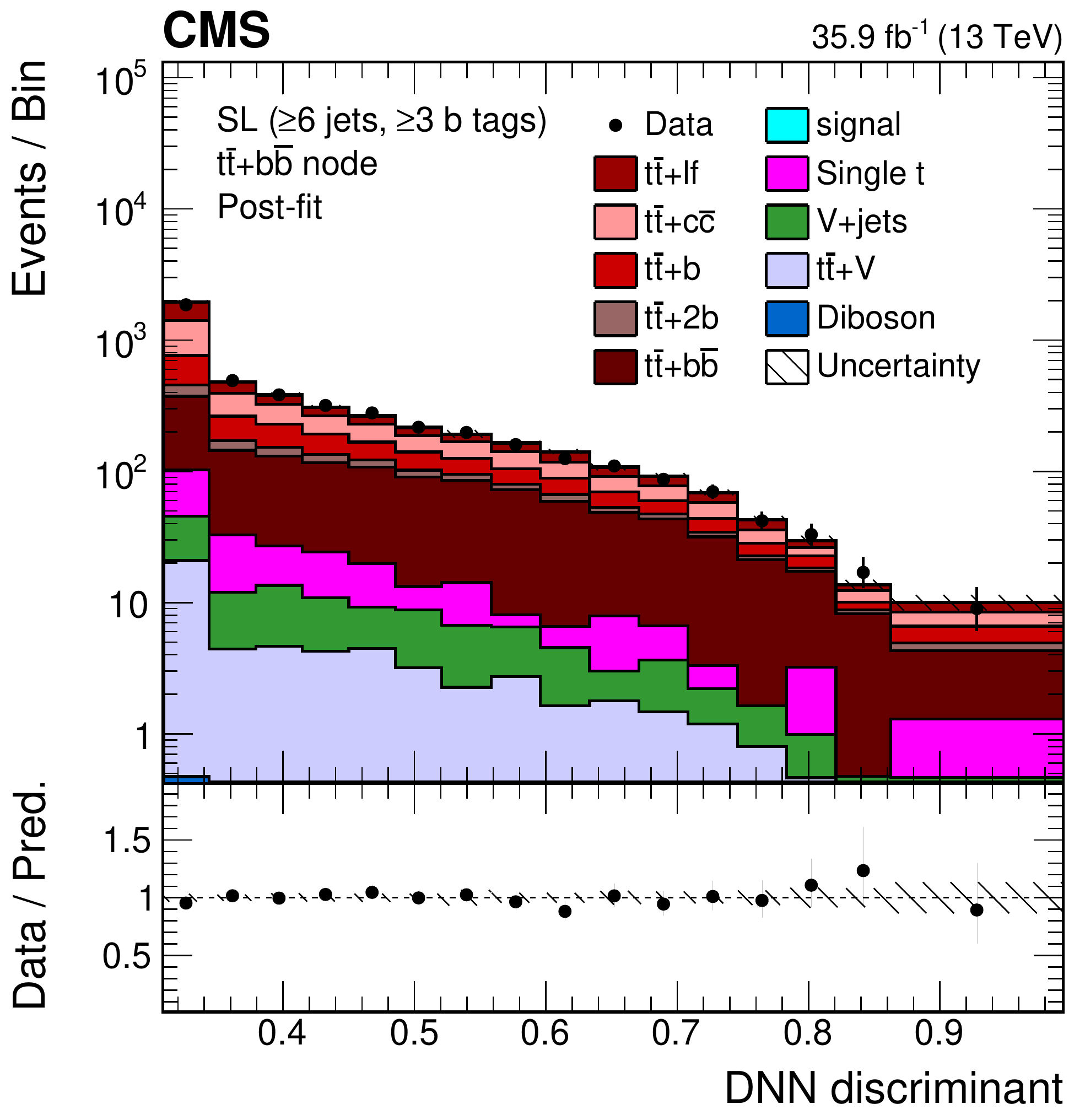}\\
    \includegraphics[width=0.35\textwidth]{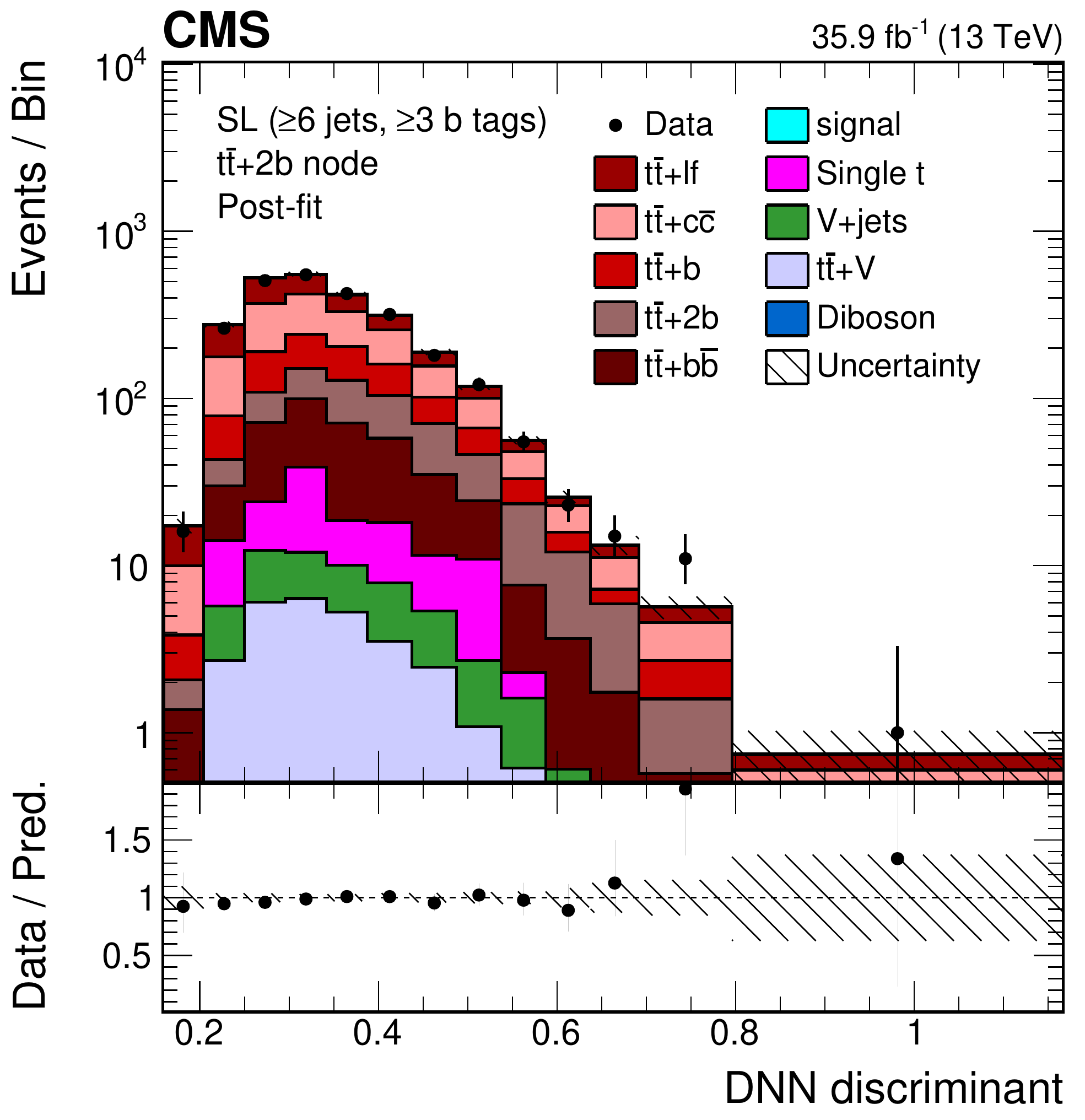} &
    \includegraphics[width=0.35\textwidth]{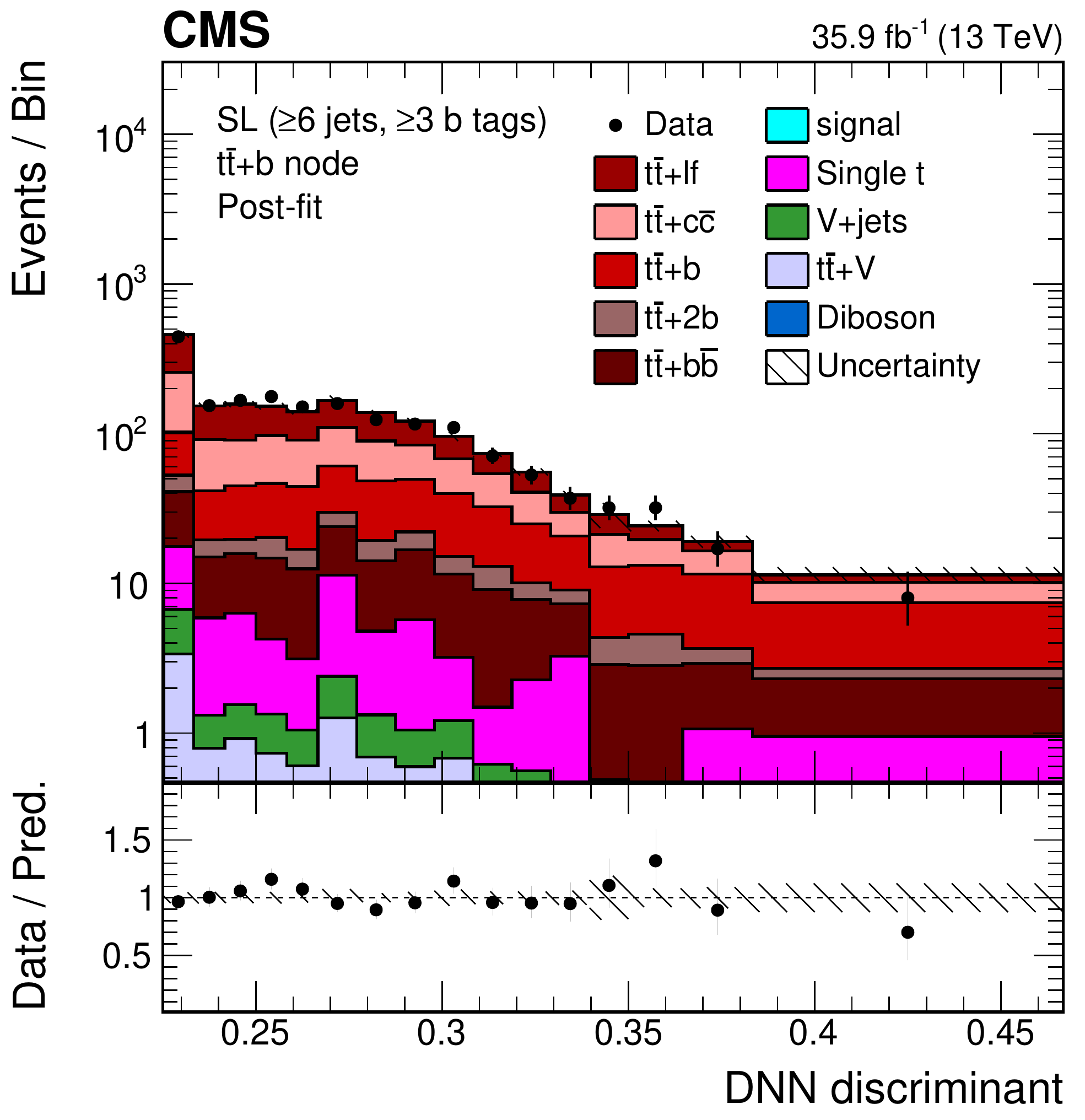}\\
    \includegraphics[width=0.35\textwidth]{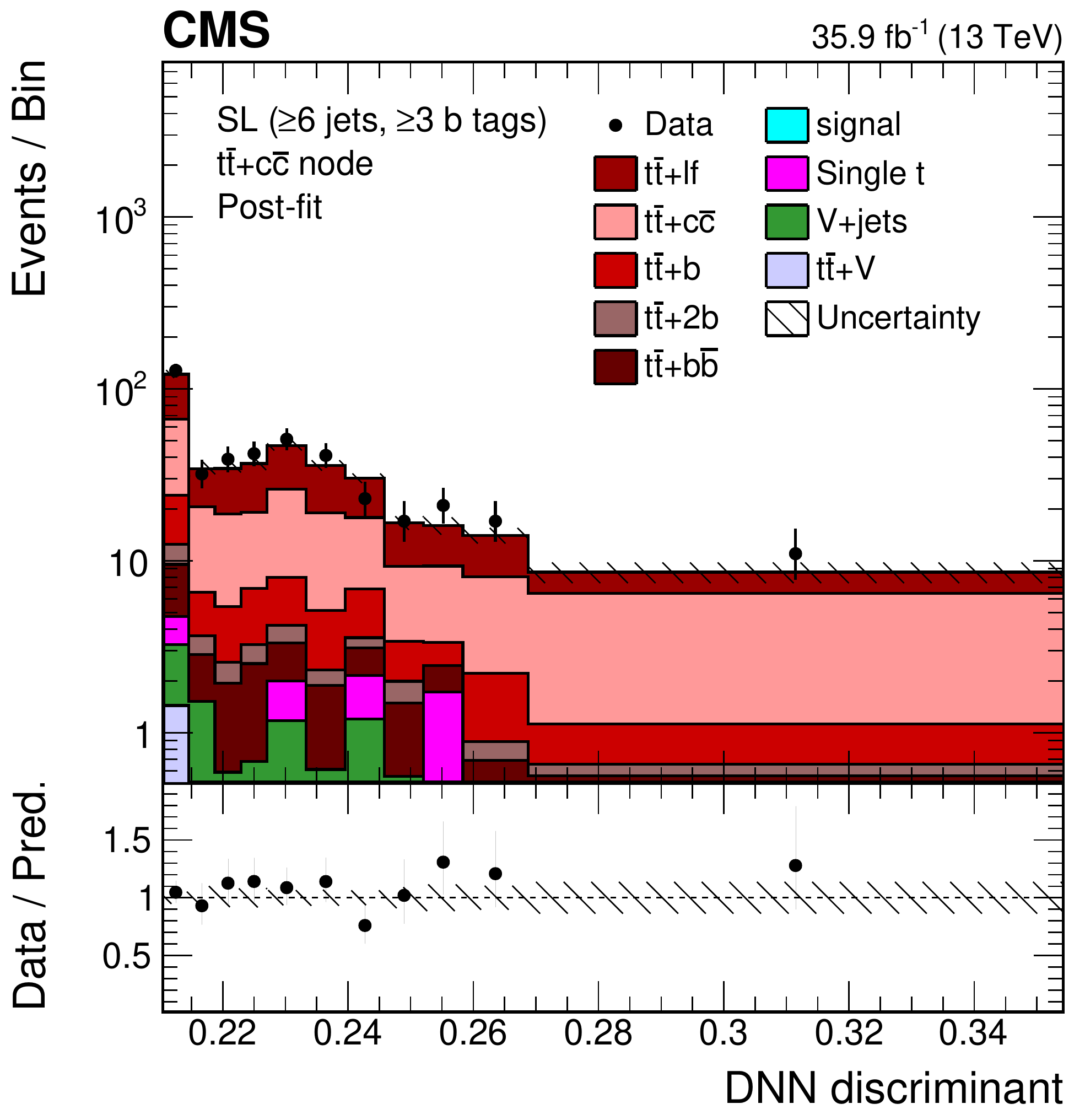} &
    \includegraphics[width=0.35\textwidth]{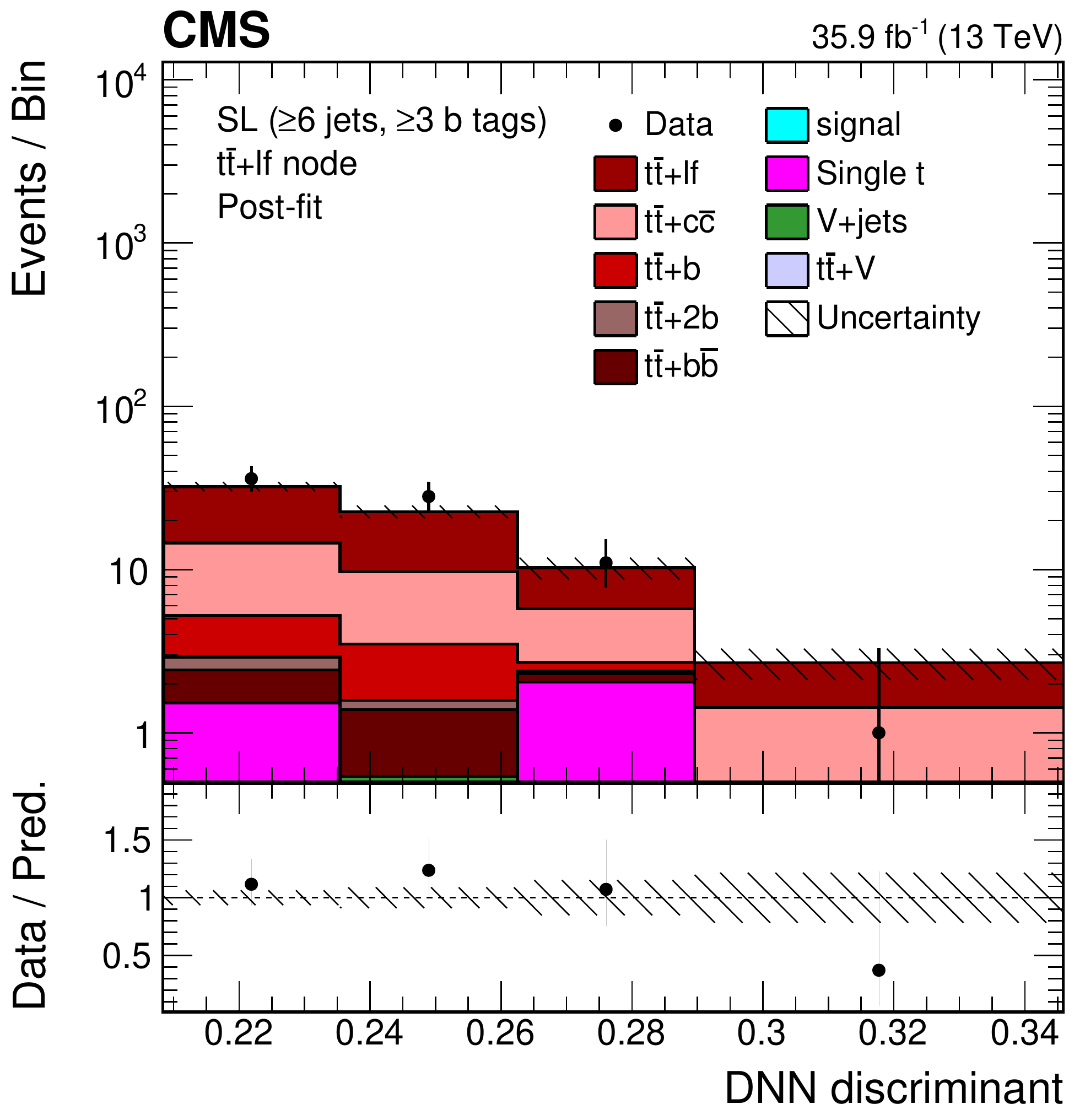}\\
    \end{tabular}
  \caption{Final discriminant (DNN) shapes in the single-lepton (SL) channel after the fit to data,
    in the jet-process categories with \ljSixThreeIncl and (from upper \cmsLeft to lower \cmsRight) \ttH, \ttbb, \tttwob, \ttb, \ttcc, and \ttlf .
    The error bands include the total uncertainty after the fit to data.
    The first and the last bins include underflow and overflow events, respectively.
    The lower plots show the ratio of the data to the post-fit background plus signal distribution.
  }
  \label{fig:appendix:postfit:ljdiscriminants_5}
\end{figure}
\cleardoublepage \section{The CMS Collaboration \label{app:collab}}\begin{sloppypar}\hyphenpenalty=5000\widowpenalty=500\clubpenalty=5000\vskip\cmsinstskip
\textbf{Yerevan Physics Institute, Yerevan, Armenia}\\*[0pt]
A.M.~Sirunyan, A.~Tumasyan
\vskip\cmsinstskip
\textbf{Institut f\"{u}r Hochenergiephysik, Wien, Austria}\\*[0pt]
W.~Adam, F.~Ambrogi, E.~Asilar, T.~Bergauer, J.~Brandstetter, M.~Dragicevic, J.~Er\"{o}, A.~Escalante~Del~Valle, M.~Flechl, R.~Fr\"{u}hwirth\cmsAuthorMark{1}, V.M.~Ghete, J.~Hrubec, M.~Jeitler\cmsAuthorMark{1}, N.~Krammer, I.~Kr\"{a}tschmer, D.~Liko, T.~Madlener, I.~Mikulec, N.~Rad, H.~Rohringer, J.~Schieck\cmsAuthorMark{1}, R.~Sch\"{o}fbeck, M.~Spanring, D.~Spitzbart, A.~Taurok, W.~Waltenberger, J.~Wittmann, C.-E.~Wulz\cmsAuthorMark{1}, M.~Zarucki
% ...
\vskip\cmsinstskip
% ...
1:  Also at Vienna University of Technology, Vienna, Austria\\
% ...
\end{sloppypar}
\end{document}